\documentclass[journal,article,moreauthors]{Definitions/mdpi} 
\firstpage{1} 
\pubvolume{1}
\issuenum{1}
\articlenumber{0}
\pubyear{2023}
\copyrightyear{2023}
\datereceived{ } 
\daterevised{ } 
\dateaccepted{ } 
\datepublished{ } 
\hreflink{https://doi.org/} 

\usepackage{natbib,graphicx}

\Title{A Survey on Blood Pressure Measurement Technologies: Addressing Potential Sources of Bias}

\TitleCitation{A Survey on Blood Pressure Measurement Technologies: Addressing Potential Sources of Bias}

\Author{Seyedeh~Somayyeh~Mousavi $^{1} $\orcidA{}, Matthew~A.~Reyna$^{1}$\orcidB{}
, Gari~D.~Clifford$^{1,2} $\orcidC{} and Reza Sameni $^{1}$* \orcidD{}}

\AuthorNames{Seyedeh~Somayyeh~Mousavi, Matthew~A.~Reyna, Gari~D.~Clifford and Reza Sameni}

\AuthorCitation{Mousavi, S.; Reyna, M.; Clifford, G.; Sameni, R}

\address{%
$^{1}$ \quad  Department of Biomedical Informatics, Emory
University, GA, USA;  seyedeh.somayyeh.mousavi@emory.edu, matthew@dbmi.emory.edu, gari@dbmi.emory.edu, rsameni@dbmi.emory.edu\\
$^{2}$ \quad  Biomedical Engineering Department, Georgia Institute of Technology, GA, USA}

\corres{Corresponding author: Reza Sameni; \url{rsameni@dbmi.emory.edu}}

\abstract{Regular blood pressure (BP) monitoring in clinical and ambulatory settings plays a crucial role in the prevention, diagnosis, treatment, and management of cardiovascular diseases. Recently, the widespread adoption of ambulatory BP measurement devices has been driven predominantly by the increased prevalence of hypertension and its associated risks and clinical conditions. Recent guidelines advocate for regular BP monitoring as part of regular clinical visits or even at home. This increased utilization of BP measurement technologies has brought up significant concerns, regarding the accuracy of reported BP values across settings. In this survey, focusing mainly on cuff-based BP monitoring technologies, we highlight how BP measurements can demonstrate substantial biases and variances due to factors such as measurement and device errors, demographics, and body habitus. With these inherent biases, the development of a new generation of cuff-based BP devices which use artificial-intelligence (AI) has significant potential. We present future avenues where AI-assisted technologies can leverage the extensive clinical literature on BP-related studies together with the large collections of BP records available in electronic health records. These resources can be combined with machine learning approaches, including deep learning and Bayesian inference, to remove BP measurement biases and to provide individualized BP-related cardiovascular risk indexes. }


\begin{document}

\section{Introduction}\label{sec:introduction}
In 2021, the World Health Organization (WHO) reported that 32\% of the world’s mortality is related to cardiovascular diseases (CVDs) \citep{who}. In 2020, CVDs were the leading cause of death in the United States, surpassing cancer and COVID-19 \citep{ahmad2021leading}. Strokes and heart attacks are the leading causes of CVD-related mortalities \citep{who,LeonardiBee2002, Lawes2004}, and hypertension is the most significant risk factor for CVDs \citep{Vasan2001, Zhou2017, Guyenet2006}. Hypertension, rarely shows early symptoms before causing severe damage to organs such as the heart, blood vessels, brain, eyes, and kidneys \citep{Brenner1988}. Therefore, it is known as the ``silent killer'' \citep{mukkamala2015toward, kalehoff2020story}. Monitoring the blood pressure (BP) is one of the effective and widely accessible methods for diagnosing and reducing CVD prevalence \citep{rastegar2020non}. Abnormal BP is even more critical and life endangering for vulnerable populations, including the elderly and pregnant women.

BP is measured manually and automatically in medical centers. Most commercial BP devices use a \textit{cuff} --- a non-elastic fabric commonly wrapped around the arm --- to apply sufficient external pressure on the artery wall to temporarily block the blood flow and to measure the BP when the arterial BP and the monitored external pressure are balanced. Over the decades, cuff-based BP devices have evolved from manual mercury-based devices to the current ones based on pressure sensors and automatic electronic measurements \citep{Tholl2004}. 

Cuff-based BP devices have passed the test of time, due to their simple operation, low-cost, availability and ease of interpretation \citep{worldtechnical}. Automatic and portable BP devices have also enabled out-of-clinic ambulatory BP monitoring by patients and their families. Although ambulatory BP is not always as accurate as in-clinic measurements, it can be acquired more frequently, reduces clinic visits and costs, increases patients' satisfaction and overcomes their clinical environment stress, which leads to the so-called \textit{white-coat} hypertension \citep{van2019validation, Pickering2006}. Specifically, the latest guidelines advise patients with gestational and chronic hypertension to repeat BP measuring at home \citep{van2019validation}. 

Hypertension diagnosis and the treatment of many other diseases are based on the accurate measurement of the BP. It is therefore critical to assess the accuracy of BP values reported in ambulatory and in-clinic settings \citep{ding2016continuous, Sewell2016-lp}. However, most users are unaware or neglectful of the standard BP measurement protocols that should be followed during BP acquisition to acquire accurate BP values. Therefore, BP measurements --- even in clinical settings --- can be significantly biased and variant due to the measurement circumstances (beyond the patient's physiological factors). This results in misinterpretations of BP readings and hampers the reliability of this vital sign for clinical diagnosis.



In recent years, many studies have focused on the notion of bias and its significance in different areas of biomedical research, including bias in false beliefs about the biological differences between various races \citep{Hoffman2016-km}, pain assessment and treatment recommendations \citep{Lee2019-fm}, medical equipment \citep{Valbuena2022-gw}, racial biases in algorithmic diagnosis \citep{Obermeyer2019-cb}, performance metrics in algorithmic diagnosis \citep{Reyna2022-on}, and reducing bias in machine learning (ML) for medical applications \citep{Vokinger2021-td}.

In this survey, we focus on potential sources of biases in BP measurement, which can influence BP-based diagnosis of hypertensive and hypotensive patients. 
We will focus on the most popular commercial cuff-based devices, which are currently the most accurate and popular BP devices used for in- and out-patients. They are also used for calibration of cuff-less BP devices. We have conducted a broad literature survey in terms of the various factors that can potentially impact BP measurements, including patient-related factors, BP acquisition session circumstances and device-related factors. The paper is organized as follows: Section~\ref{sec: BP_review} reviews the biophysics of the BP. Section~\ref{sec: BP_techs} classifies BP measurement methods. Section~\ref{sec: cuff-based} investigates different validation standards and reviews various commercial BP technologies and their operation principles. 
Section~\ref{sec: BP_tech_biases} presents potential sources of bias in BP technologies from different perspectives. Section~\ref{sec: Future} details future perspectives for using machine learning techniques for individualized BP assessment and bias correction. Section~\ref{sec: conclusion} concluded this study and discusses the impact and limitations of the study.

\section{A Review of the blood pressure physiology}
\label{sec: BP_review}
Vital signs and physiological measurements are proxies for assessing the fundamental body functions. The BP is one of the important clinical parameters measured from the body, together with the vital signs \citep{Magder2018-vy}. A regulated BP guarantees timely and adequate supply of blood \citep{M9}, which is essential for the blood functions: 1) transportation of nutrients, waste, hormones, oxygen and carbon dioxide; 2) regulation of osmotic pressures, temperature, and pH; and 3) protection against infections via white blood cells, antibodies and clots (to prevent excessive blood loss during injuries).



\subsection{Blood pressure definition}
\begin{figure}[tb]
\centering
\includegraphics[trim={3cm 2cm 3cm 1cm},clip,width=1\textwidth]{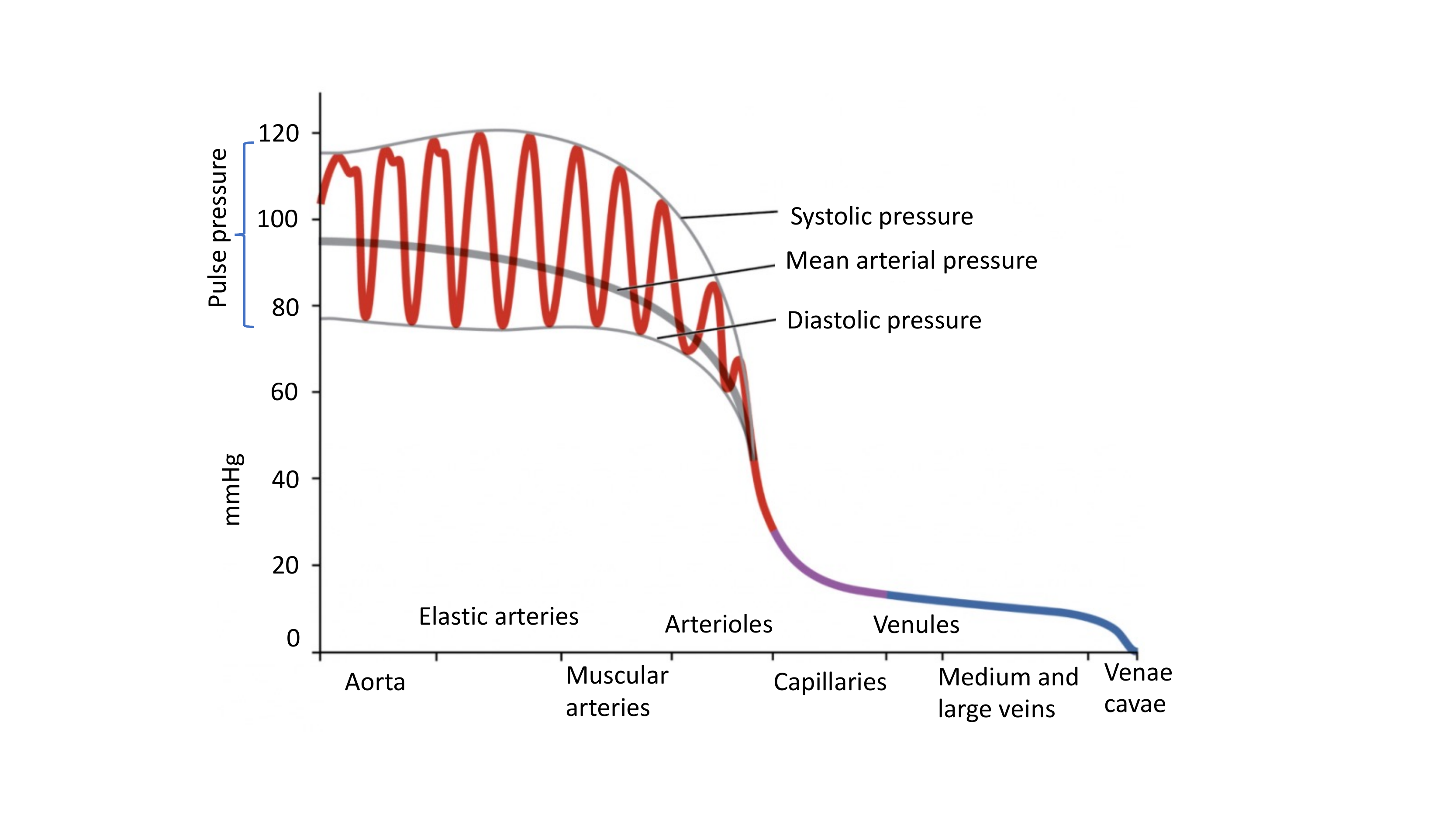}
\caption{The systolic, diastolic, mean arterial, pulse pressures, and the overall blood pressure at different blood vessel points; adopted from \citep{bettsanatomy}, by OpenStax College (CC-BY-3.0)}‌\label{Fig: BP_meaning}
\end{figure}

Blood flows across the body due to the pressure difference in the arterial system \citep{mousavi2018designing}. The BP assesses the mechanical function of the heart (as a pump). It is the force per unit area of the arterial system, commonly measured in millimeters of mercury (mmHg). In healthy subjects, the heart contracts between 60 to 100 times per minute, resulting in a pulsatile and almost periodic pressure wave in the arterial system. The pressure wave's maximum is during \textit{systole}, when cardiac contraction exerts its maximum pressure to the blood and arterial walls. This is when blood is pumped from the heart into the arteries. The lowest pressure corresponds to \textit{diastole}, when the heart is at rest \citep{M5}. 
Therefore, the arterial BP fluctuates between the maximum and minimum levels and damp down to zero through the end of the arterial circulation system, Fig.~\ref{Fig: BP_meaning} \citep{bettsanatomy}. 
Throughout this work we will study the following BP parameters:
\begin{itemize}
\item\textit{Systolic blood pressure (SBP)}, or the maximum pressure inside the arteries during cardiac contraction;
\item\textit{Diastolic blood pressure (DBP)}, or the minimum atrial pressure during cardiac rest;
\item\textit{Mean Arterial Pressure (MAP)}, which is an empirical weighted average of SBP and DBP, to approximate the average BP over the cardiac cycle using only its maximum and minimum values \citep{Vlachopoulos2011-kp}:
\begin{equation}
\text{MAP}=\frac{\text{SBP}+2\times\text{DBP}}{3}
\label{Eq: Map}
\end{equation}
where the higher weight of the DBP empirically accounts for the asymmetry of the continuous BP (see Fig.~\ref{Fig: KORTOKOFF}).
\end{itemize}

\subsection{Factors impacting the blood pressure}
  
Arterial BP values are subjective, and depend on many factors, including physics and physiology of the body; body position; brain activities; digestive activities; muscle activities; nerval stimulations; environmental factors (air temperature and audio noise level); smoking; alcohol and coffee consumption; and medications \citep{M10, M1}. There are also various biological factors that impact the BP \citep{desaix2013anatomy}, including:
\subsubsection{Cardiac output} Cardiac output is the amount of blood that is pumped into the ventricles by the heart. BP and flow increase when factors such as sympathetic stimulation, epinephrine and norepinephrine, thyroid hormones, and increased calcium ion levels increase cardiac output (heart rate, stroke volume, or both). Conversely, factors that decrease heart rate, stroke volume, or both, such as parasympathetic stimulation, increased or decreased levels of potassium ions, decreased calcium levels, anoxia, and acidosis, will decrease cardiac output \citep{desaix2013anatomy, M10}.

\subsubsection{Compliance} Compliance is the ratio of the change in volume to the change in the pressure applied to a vessel \citep{Glasser1997-ll}. Arterial compliance has a direct relationship with its efficiency. There is an inverse relationship between BP and compliance of blood vessels. When vascular diseases cause artery stiffening, compliance decreases and the heart works harder to push blood through the stiffened arteries, resulting in an increase in the BP
\citep{desaix2013anatomy}.

\subsubsection{Blood volume} The total amount of blood in the body directly affects blood flow and pressure. If blood volume decreases due to bleeding, dehydration, vomiting, severe burns, or certain medications, BP and flow also decrease. However, the body's regulatory mechanisms are efficient in controlling BP, and symptoms may not appear until 10-20\% of blood volume is lost. Hypovolemia can be treated with intravenous fluid replacement, but the underlying cause must also be addressed. Intravenous fluid replacement is typically part of the treatment. Hypovolemia can be treated with intravenous fluid replacement, but the underlying cause must also be addressed to restore homeostasis in these patients \citep{desaix2013anatomy, M12}.

\subsubsection{Blood viscosity} Blood viscosity is the fluids' thickness (resistance to blood flow). Blood viscosity is inversely proportional to flow and directly proportional to resistance. As a result, any factor that increases the blood viscosity raises the resistance and lowers the flow. In contrast, factors that decrease viscosity increases flow and lowers the resistance. Blood viscosity typically does not vary over short time intervals. Plasma proteins and the formed elements are the two primary factors influencing blood viscosity. Any condition that affects the number of plasma constituents, such as red blood cells, can change viscosity \citep{desaix2013anatomy}. Since the liver produces most plasma proteins, liver impairments or dysfunctions such as hepatitis, cirrhosis, alcohol damage and drug toxicity can also alter the viscosity and decrease the blood flow \citep{Letcher1981-qq}.

\subsubsection{Blood vessel length and diameter} The vessels' resistance and lengths are directly related. Longer vessels have more resistance and a lower flow. A higher surface area of the vessel makes it harder for blood to flow through it. Similarly, shorter vessels have a smaller resistance, resulting in a higher flow. The length of blood vessels grows with age, but they tend to stabilize and remain constant in length during adulthood under normal physiological conditions \citep{M10, desaix2013anatomy}.
The diameter of blood vessels differs depending on their type and can change throughout the day in response to chemical and neural signals. Unlike vessel length, vessel diameter is inversely related to resistance. Intuitively, a vessel with a larger diameter allows blood to flow with less friction and resistance (even with the same blood volume), because the blood has less contact with the vessel walls. \citep{desaix2013anatomy, Jeppesen2007-az}.
\subsection{Blood pressure norms}
In 2018, the Journal of the American Heart Association (AHA) published guidelines for the prevention, detection, evaluation, and management of BP \citep{Whelton2018}. Table \ref{Tab: BPcategory}
lists the BP ranges for adults in four categories. BP values for children and adolescents are generally lower than for adults, and they gradually rise with age \citep{azegami2021blood}. Hypertension (abnormally high BP) is characterized by a continuous elevation of BP values above the normal range. On the other hand, hypotension (abnormally low BP) occurs when BP values are below normal ranges. Hypotension can happen due to a sudden blood loss or a decrease in blood volume, and hypertension is linked to an increased risk of various forms of CVDs \citep{Whelton2018}: 
  
\begin{table}[tb]
    \begin{adjustwidth}{-\extralength}{0cm}
    \caption{Blood pressure norms based on the health status of an adult \citep{Whelton2018}}\label{Tab: BPcategory}
    \begin{center}
    \begin{tabularx}{\textwidth}{p{5cm}CCC}
        \toprule
        \textbf{BP category} & \textbf{SBP (mmHg)} & 
        &
        \textbf{DBP (mmHg)} \\
        \midrule
        \textbf{Normal BP} & $<$120 & {AND} & $<$80 \\
        \hline
        \textbf{Elevated} & 120--129 & {AND} & $<$80 \\
        \hline
        \textbf{High BP (Hypertension) Stage1} & 130--139 & {OR} & 80--89 \\
        \hline
        \textbf{High BP (Hypertension) Stage2} & $\geq$140 & {OR} & $\geq$90 \\
        \bottomrule
    \end{tabularx}
    \end{center}
    \end{adjustwidth}
\end{table}

\section{Blood pressure measurement methods}\label{sec: BP_techs}
BP readings are also impacted by the measurement technology, measurement setup, patient conditions and time. We will study these factors in the sequel.

\subsection{Invasive blood pressure measurement}
\label{sec:invasive_BP_measurement}
In this type of measurement, a catheter (a thin tube utilized to administer drugs, fluids, or gases into or out of a patient's body) is inserted into a vessel and measures the arterial BP using a pressure transducer consisting of a delicate and sensitive diaphragm. The transducer's resistance varies with the slightest pressure changes, allowing for the detection of BP fluctuations \citep{ding2016continuous}. This technique is utilized to record and monitor changes in BP \citep{Cole2007-rx}.

In modern healthcare facilities, disposable pressure transducers are commonly utilized for more precise and continuous measurement of BP in specialized settings, such as cardiac catheterization labs, intensive care units (ICUs), and operating rooms. Although pressure transducers can measure intracranial and intra-abdominal pressures, they are most frequently used for invasive monitoring of arterial and venous BP. Invasive methods of monitoring BP can be generally classified into two categories \citep{M15}: \textit{Intravascular:} where the pressure sensor is inserted into the vessel at the tip of the catheter; \textit{Extra vascular:} where the pressure sensor is located outside the vessel and along the catheter's end.

\subsection{Noninvasive blood pressure measurement}
Noninvasive BP measurement techniques determine the BP without any physical injury to the body. This technique is classified into two groups: cuff-based and cuff-less methods. Cuff-less methods are still in the developmental stage and are commonly not common in clinical studies. Therefore, the scope of the current survey is on cuff-based methods.


All commercially available cuff-based BP measurement technologies consist of a manometer (digital or analog), a pressure pump (manual or automatic), and a cuff. The basic principle of these devices is to apply sufficient pressure to the extremities (arm, wrist or leg) to temporarily block the blood flow through the artery. The cuff is then slowly deflated and the pressure is reduced until the blood begins to flow through the artery. At this point, the pressure in the cuff is close to the SBP --- the peak of BP  (Fig.~\ref{Fig: KORTOKOFF}). Since the BP oscillates through the cardiac cycle, it repeatedly falls below the external pressure, which results in repeated obstruction of blood flow (which can be heard through a stethoscope or sensed via a measurement device). The cuff pressure is then further reduced until blood flows through the artery with no obstruction, where the pressure in the cuff drops below the DBP, and no further sounds are heard by the physician (or sensed by the automatic analysis hardware/software). In Fig.~\ref{Fig: KORTOKOFF}, we can see how the actual and reported BP values can deviate, especially for the DBP.


There are various methods to identify the moments when the cuff pressure is equal to SBP and DBP. The main difference between these methods lies in their ability to detect the external-internal pressure balance points accurately \citep{mousavi2018designing}. These methods are:
\subsubsection{Auscultatory} 
\begin{figure}[tb]
\centering
\includegraphics[trim={2in 1in 2.5in 1.5in},clip,width=\columnwidth]{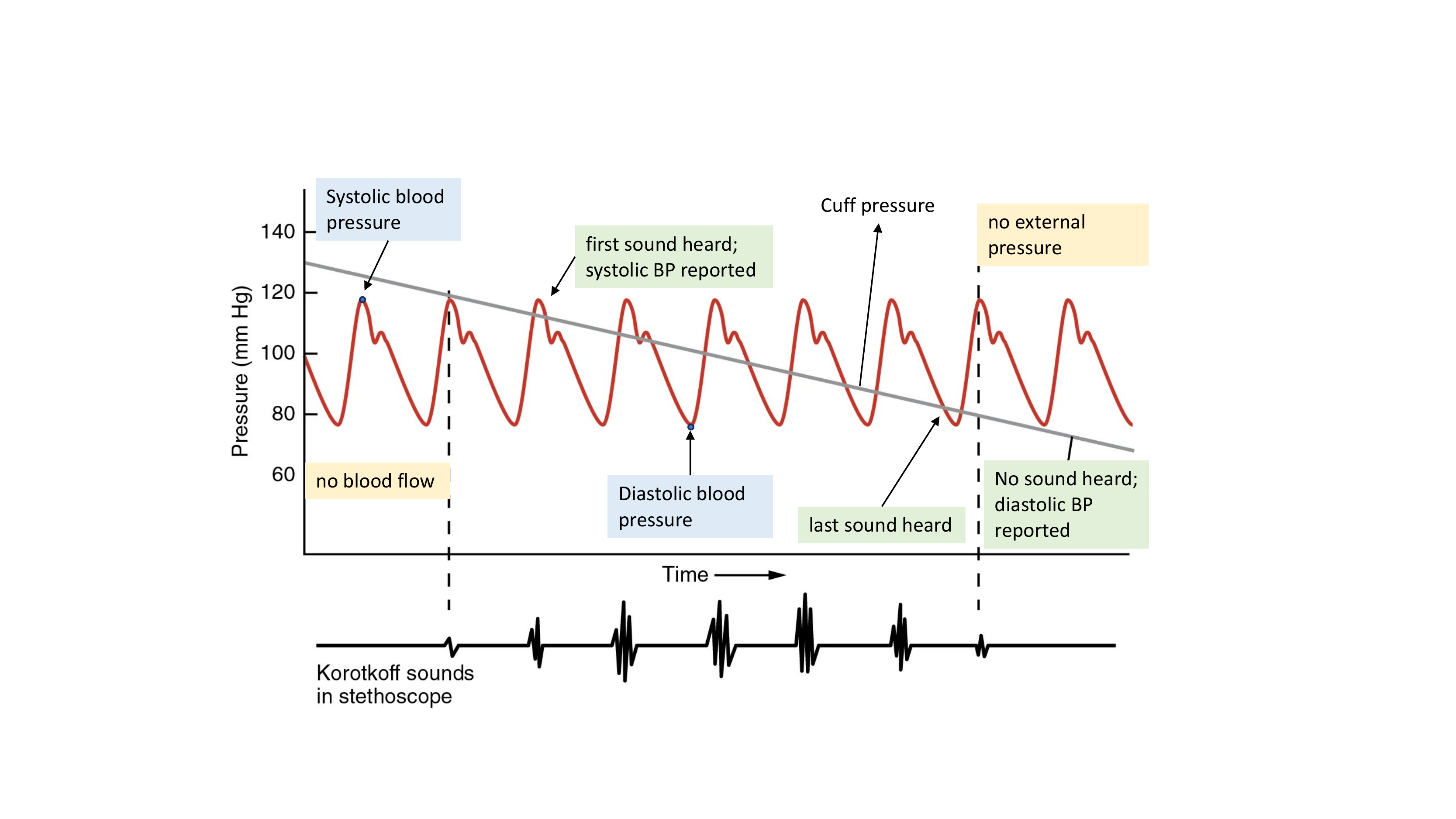}
\caption{Auscultatory method measures blood pressure based on Korotkoff sounds. Notice the difference between the actual and reported systolic/diastolic blood pressures; adapted from \citep{desaix2013anatomy} (CC-BY-3.0)}\label{Fig: KORTOKOFF}
\end{figure}

This method dates back to the late 18th century and remains the gold standard for validating novel BP measurement methods \citep{Kumar2021-em}. It is based on Korotkoff sounds, which are produced by the turbulent flow of blood through the compressed artery. As the cuff pressure is slowly released, the artery begins to open and its pressure exceeds the cuff pressure. The pressure that the manometer displays when the first Korotkoff sound is heard corresponds to the systolic BP. As the cuff pressure reduces, the blood flows turbulently in the artery, and the sound continues until the pressure of the cuff reaches the lowest arterial pressure. At this moment, the Korotkoff sound disappears, and the corresponding manometer pressure is the distolic BP, as shown in Fig. \ref{Fig: KORTOKOFF} \citep{Kumar2021-em,peter2014review}. 

Initially, BP was measured using mercury sphygmomanometers, but the toxicity of mercury led to the adoption of aneroid sphygmomanometers, which use a mechanical pressure gauge that is calibrated to display the pressure readings. The precision of the aneroid sphygmomanometer depends on the operator's proficiency in auscultation and use of a stethoscope, and visual acuity \citep{Kumar2021-em}. More recently, hybrid sphygmomanometers automate the detection of Korotkoff sounds and the display of the SBP, DSP and pulse rate on digital monitors \citep{lim2022blood}. To note, aneroid sphygmomanometers have varying accuracy across manufacturers. Over the past decade, surveys have been conducted to assess the accuracy of these devices. The results demonstrate that the BP readings from various manufacturers have significant deviations ranging from 1 to 44\,\% \citep{Yarows2001, Canzanello2001, Mion1998, Kumar2021-em}. Additionally, using a small gauge to read the pressure is another potential source of bias in these devices \citep{Kumar2021-em}. Therefore, mercury sphygmomanometers still remain popular in clinical settings.

\subsubsection{Oscillometric} This method is currently the most popular technology for automated BP measurement devices \citep{worldtechnical}. Using a pressure sensor, it measures the pulsatile BP in the artery during cuff inflation and deflation \citep{rastegar2020non}. In this technology, the transducer detects the small variations in arterial pressure oscillations or the intra-cuff pressure, produced by the changes in pulse volume due to the heartbeats \citep{worldtechnical}. The oscillations begin when the cuff pressure exceeds the systolic BP and continues until it is lower than the diastolic BP. A microcontroller is used to control the inflating process, reading the analog output signal of the pressure sensor and its digitization. The micro-variations of the sensed pressure is filtered and pre-processed to obtain the mean arterial pressure as defined in \eqref{Eq: Map} \citep{worldtechnical}. Due to the indirect nature of this approach, the measured pressure value requires calibration to be mapped to the actual systolic and diastolic BPs. The microcontroller displays the measured BPs on a local screen \citep{Kumar2021-em}.

Some of the advantages of the oscillometric method include \citep{Kumar2021-em, worldtechnical}: 1) ease of use for patients (placement and removal of the cuff); 2) ease of calibration (commonly via a button on the device); 3) portability; 4) use with minimal training. The negative aspect of this method is that commercial oscillometric devices use different (and commonly proprietary) algorithms to estimate BP from the measurements \citep{ogedegbe2010principles}. This results in an intrinsic source of bias in BP measurement from different manufacturers.

\subsubsection{Ultrasound} This technique is based on the Doppler effect \citep{worldtechnical}. Similar to the previous methods, cuff inflation blocks the blood flow in the artery. Then, as the cuff deflates, the arterial wall starts to move at the systolic BP, and produces a Doppler phase shift in the reflected ultrasound. As the arterial motion decreases and reaches its endpoint), the cuff pressure at that moment is considered as the diastolic BP \citep{Kumar2021-em}. The detection of the onset and offset of arterial wall motions is performed by processing the Doppler reflection by a local (digital) processor.

\section{Cuff-based blood pressure measurement technologies}\label{sec: cuff-based}
As with all medical equipment, BP measurement devices must meet regulatory requirements and standards. In this section, the essential BP measurement and validation standards of commercial cuff-based BP monitors are reviewed.

\subsection{BP measurement device standards}
There are different standards for validating BP measurement devices. In 1987, the Association for the Advancement of Medical Instrumentation (AAMI) published the first standard for non-invasive BP medical devices \citep{association1987american}. In 1990, the British Hypertension Society (BHS) set another clinical protocol for validating these devices, which includes many of the AAMI standards \citep{gervsak2009procedure}, \citep{stergiou2018universal}. These parallel standards continued until 2018 when the AAMI, the European Society of Hypertension (ECH) and the
International Organization for Standardization (IOS) published a universal protocol named ``single universal standard'', also  referred by the ISO 81060-2:2018/Amd 1:2020 on ``non-invasive sphygmomanometers'', \citep{ISO81060-2:2018/Amd.1:2020, worldtechnical}. The unified standard facilitated the validation and comparison of measurements made by BP devices manufactured globally.

As part of this standard, manufacturers are required to collect a database of measurements from at least 85 individuals at least three times per individual. Therefore, at least 255 records are required to validate BP devices \citep{mousavi2018designing}. Moreover, the mean absolute error between the recorded BP values and the reference technology should be less than 5\,mmHg, and the standard deviation of the multiple measurements should be less than 8\,mmHg. The standard also identifies whether the BP measurement devices have cumulative absolute error into three groups of less than 5\,mmHg, between at least 5\,mmHg but less than 10\,mmHg, and at least 10\,mmHg but less 15\,mmHg. Devices will pass the certification test if at least 85 percent of the reported results based on the noted criteria are less than 10\,mmHg \citep{worldtechnical}. The reference BP measurement technology is the invasive BP measurement techniques (Section \ref{sec:invasive_BP_measurement}), but it is also acceptable to compare the BP results with any non-invasive measurement method with a maximum error of 1\,mmHg \citep{worldtechnical}.

\subsection{Commercial cuff-based BP measurement devices}

Commercial BP devices may be categorized into three groups: ambulatory BP monitors (ABPM), office BP (OBP) monitors, and home BP monitors (HBPM).

The ABPM or BP Holter is a portable monitor that is carried by individuals with hypertension (or those who are at a higher risk of developing hypertension) for a period of 24 or 48 hours, while engaging in their regular daily activities and during sleep. Based on the physician's required settings, the device measures the patient's BP at specific time intervals, e.g. 15 or 30 minutes. After the required period, the patient returns to the clinical center, the device is taken off the patient, and the BP data is transferred to the computer or cloud and analyzed by software \citep{mousavi2018designing}. ABPM is most commonly used for detecting non-dipping BP patterns, which refer to a phenomenon where an individual's BP fails to exhibit the normal nocturnal decrease during sleep, potentially indicating an elevated risk of cardiovascular issues \citep{hassler2005circadian}.

OBP is the most common type of BP measurement devices for clinical use. Its accuracy is critical (especially in emergency and surgical units), where physicians make essential real-time decisions from the BP and other vital signs values. In these situations, the physician may not have adequate time to repeat BP measurement (as advised by BP reading standards). These devices have two types. These BP devices are either integrated into bedside monitors or are used as discrete devices similar to the HBPM type.

\subsection{Standard blood pressure measurement conditions}
\begin{figure}[tb]
\centering
\includegraphics[trim={1cm 2cm 5cm 0.5cm},clip,width=0.7\textwidth]{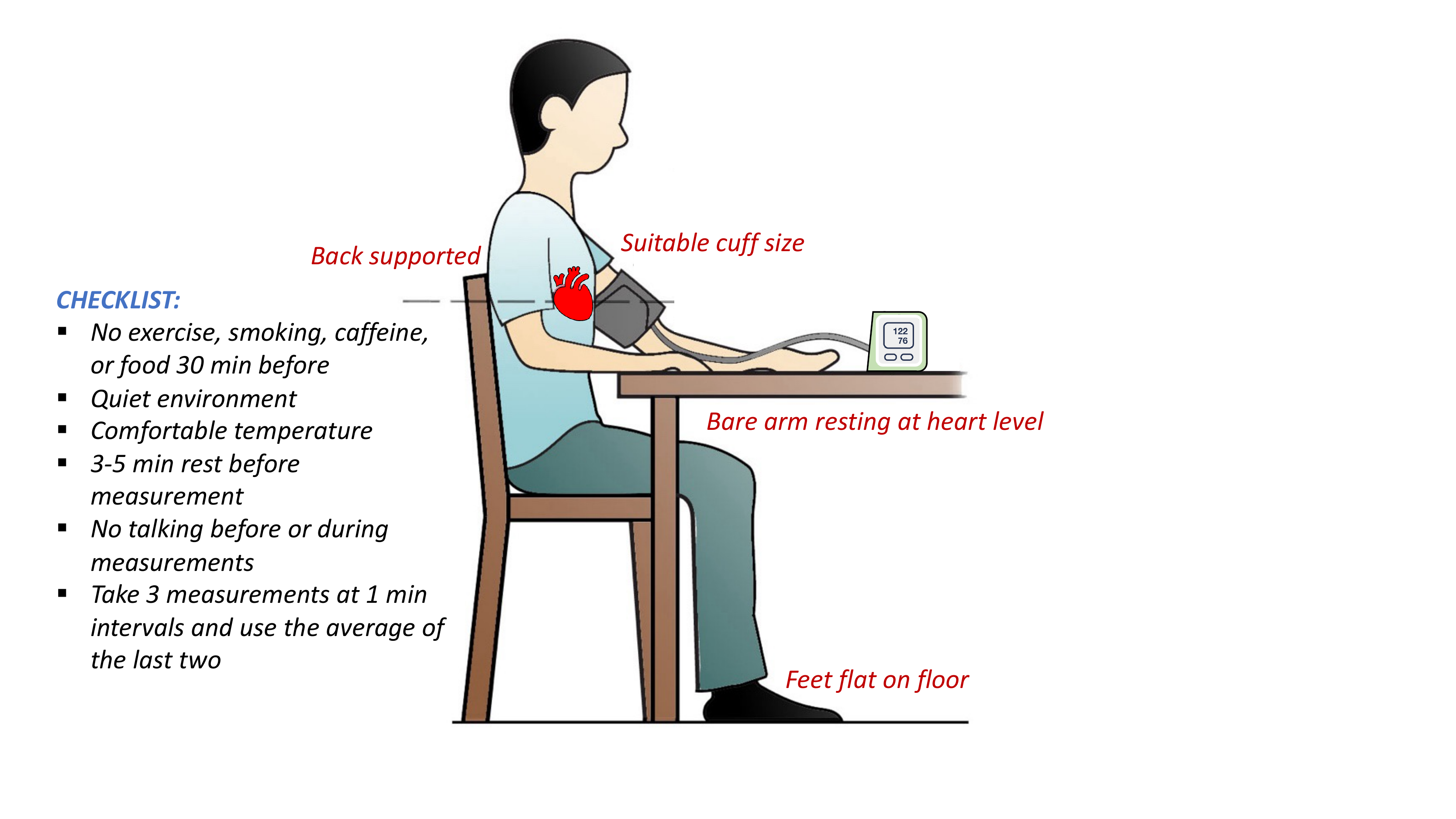}
\caption{The basic principles and important standards of blood pressure measurement; adapted from \citep{Stergiou2021-fl}. See STRIDE-BP (\url{https://stridebp.org/}) for validated electronic cuff-based blood pressure devices.}\label{Fig: principle}
\end{figure}

Several guidelines have been published to improve the accuracy of BP measurement devices by standardizing BP acquisition procedures. Although there are different recommendations in different countries and organizations, they typically address the same fundamental issues \citep{Wagner2012-id}. Fig.~\ref{Fig: principle} illustrates the basic principles of BP measurement \citep{Stergiou2021-fl}. The following are some of the common items for BP measurement guidelines: 
\begin{itemize}
    \item To maintain a stable BP measurement environment, it is recommended to refrain from opening and closing windows and doors. \citep{Kumar2021-em}.
    \item The temperature and relative humidity of the BP measurement environment should be in the range of 15–25\,$^{\circ}$C and 20–85\,\%, respectively \citep{Kumar2021-em}. 
    \item BP should be measured in a quiet environment \citep{Liu2022-xk, Stergiou2021-fl}.
    \item The patient should not smoke, eat or drink at least 30 minutes before measuring \citep{Stergiou2021-fl}. 
    \item The patient should have adequate rest time before the measurement to stabilize the BP.
    \item The patient should not speak and should remain quiet during the measurement \citep{Stergiou2021-fl}.
    \item The patient should sit on a chair with back and arm supports and without crossing legs \citep{Liu2022-xk}.
    \item The patient's arm should be placed and remain at the same level as the heart throughout BP measurement \citep{Kumar2021-em}.
    \item An appropriate cuff should be used for measuring according to AHA guidelines (Table~\ref{Tab:cuffsizeAHA}) \citep{Kumar2021-em}.
    \item The antecubital fossa (the area between the anatomical arm and the forearm) should be 2-3\,cm above the lower end of the cuff \citep{muntner2019measurement}. 
    \item During the measurement, the patient's leg should remain flat on the floor \citep{Stergiou2021-fl}.
    \item Measuring BP should be done using direct contact of the cuff with the upper part of the arm (not over sleeves) \citep{Liu2022-xk}. 
    \item It is recommended to take three BP measurements with one-minute intervals in-between. The average of the results should be reported as the BP values \citep{Stergiou2021-fl, Liu2022-xk}.
\end{itemize}

\newcolumntype{C}[1]{>{\centering\arraybackslash}p{#1}}

\begin{table}[tb]
\caption{AHA Recommendation for the Appropriate Cuff Size per Patient \citep{Kumar2021-em}}\label{Tab:cuffsizeAHA}
\centering
\begin{tabularx}{\textwidth}{r C{3.5cm} C{3.5cm} C{3.5cm}}
\toprule
\textbf{Cuff} & \textbf{Arm Circumference (cm)} & \textbf{Bladder Width (cm)} & \textbf{Bladder Length (cm)} \\
\midrule
\textbf{Newborn} & $<$ 6 & 3 & 6 \\\hline
\textbf{Infant}  & 6-15 & 5 & 15 \\\hline
\textbf{Child} & 16-21 & 8 & 21 \\\hline
\textbf{Small Adult} & 22-26 & 10 & 24 \\\hline
\textbf{Adult} & 27-34 & 13 & 30 \\\hline
\textbf{Large Adult} & 35-52 & 20 & 42 \\\hline
\textbf{Adult Thigh} & 45-52 & 20 & 42 \\
\bottomrule
\end{tabularx}
\end{table}

\section{Potential sources of bias in blood pressure technologies} \label{sec: BP_tech_biases}
Patient positions or acquisition circumstances that do not meet the measurement guidelines may potentially result in BP measurement biases (over or under-estimation) and misdiagnosis \citep{Wagner2012-id,kallioinen2017sources, ogedegbe2010principles}. We will study the potential sources of biases under three categories \citep{mancia20132013, ogedegbe2010principles}.

\subsection{Biases related to blood pressure measurement devices}
The BP measurement device is the first source of measurement bias. BP devices comprise of the main measurement unit and the consumable parts, as detailed below.

\subsubsection{Main blood pressure measurement unit} 
Biases related to the main BP measurement unit are known as systematic errors. Manufacturers report the acceptable range of measurement errors in the user's manual of BP measurement devices. Measurements falling outside of this range are considered unacceptable, indicating that the device needs calibration. Therefore, they require maintenance and regular calibration to identify and reduce measurement uncertainties to an acceptable level. Medical instruments comprise numerous electro-mechanical elements that undergo natural aging and wearing and are impacted by microscopic airborne contaminants that accumulate on their sensitive electronic elements and sensors. These effects change the electro-mechanical characteristics of the devices (even if they are not used), and result in a gradual drift from their nominal operating point. At a system's level, the deviations of the device elements contribute to measurement biases. It should be highlighted that the effect of device aging is not identical to the failure of the elements or the device. Therefore, even though a BP device may appear new or be fully functional, it may deviate from the original calibration point. Hence, the regular calibration of medical devices is an essential requirement that should not be compromised. 
To note, systematic errors, such as the changes in the cuff deflation rate \citep{Yong1987-gn}, are not mitigated by averaging, as they tend to drift the reported values by a constant bias (unknown to the user) \citep{speechly2007sphygmomanometer}. The identification and correction of systematic errors is performed by reference instruments that have been well-maintained and calibrated by medical instrumentation experts.

Another source of systematic error is the software/algorithm used for BP calculation from the pressure sensor measurements. The lack of calculation calibration for the electronic pressure sensors can result in systemic errors \citep{speechly2007sphygmomanometer, turner2004effects}. For example, in automatic BP measurement devices, the rate of cuff inflation and deflation is significant, because the isometric exercise involved in inflating the cuff causes a temporary elevation of about 10\,mmHg. Although this only takes around 20\,s, if the cuff is deflated too quickly, the pressure may not have returned to baseline, resulting in a falsely high systolic pressure \citep{ogedegbe2010principles}. Other biasing factors include sensor accuracy and the software/firmware logic \citep{Kumar2021-em}. Moreover, the cuff's deflation or `bleed rate' is assumed to be a consistent rate when the algorithm calculates the systolic and diastolic pressures \citep{Kumar2021-em}, which may not hold in practice.

The WHO regularly publishes and updates guidelines for standard practices to ensure that healthcare providers are cognizant of calibration requirements and potential biases to provide quality healthcare \citep{yayan2020key}. The process of calibrating medical equipment is different for each device. Most devices have built-in mechanisms for calibration. 
Manufacturers also advise regular maintenance based on the rate of utilization. 
Automatically calibrated devices can be used within predefined tolerances, beyond which they should be returned to the manufacturer for technical inspection or be disposed.

\subsubsection{Consumables of blood pressure measurement devices}
All BP devices have consumable components that require regular replacement. This includes batteries, rubber tubing, hoses, fittings, and the cuffs \citep{worldtechnical}. For example, since the BP device consumes variable power throughout the inflation-deflation cycle, weak batteries may result in erroneous BP values. Worn cuffs and rubber tubing with weak elasticity also impact the reported BP values. Consumable parts of BP devices should therefore be replaced according to the manufacturers' recommended timelines.

\subsection{Subject-specific biases} BP values can significantly fluctuate across subjects \citep{pickering2005recommendations}. In the following section, we list some of the most important sources of subject-specific bias.

\subsubsection{Demographic features} 
Demographic features such as sex, age, race, and genetic background have physiological and anatomical impacts that influence BP measurements. 

Sex is a prominent factor that influences the BP \citep{Reckelhoff2018, Sandberg2012}. Generally, males tend to have higher BP levels, which can be associated to differences in hormonal activities and anatomical differences between male and female bodies \citep{Sandberg2012, Reckelhoff2001-td}. A notable study involving 32,833 individuals (54\% women) was conducted over a span of four decades, ranging from ages 5 to 98 years \citep{Ji2020}. The study showed that women exhibited a steeper increase in BP measurements compared to men, starting as early as the third decade of life and continuing throughout their lifespan \citep{Ji2020}. Table~\ref{Tab: sex} presents a compilation of studies reporting BP values based on sex. 

\begin{table}[tb]
\caption{The results of studies reporting blood pressure values based on sex. N is the number of patients, and summaries of SBP and DBP include mean $\pm$ standard deviation.}\label{Tab: sex}
\begin{center}
\begin{tabular}{rccccccc}
\toprule
\textbf{Ref.} & \textbf{Total N} & \multicolumn{3}{c}{\begin{tabular}{@{}c@{}} \textbf{Male} 
\end{tabular} }  
&  \multicolumn{3}{c}{\textbf{Female} } 
\\\hline
 &   &  N & SBP & DBP  & N & SBP & DBP
\\\hline
\citep{Somani2018-cr} & 20 & 10 & 126.0$\pm$8.0 & 73.0$\pm$5.0 & 10 & 122.0$\pm$5.0 & 73.0$\pm$5.0
 \\\hline
\citep{Somani2018-cr} & 26 & 13 & 117.0$\pm$5.0 & 65.0$\pm$7.0 & 13 & 103.0$\pm$6.0 & 62.0$\pm$8.0
\\\hline
\citep{olatunji2011water} & 37 & 22 & 121.2$\pm$9.7 & 73.7$\pm$8.5 & 15 & 117.4$\pm$13.9 & 74.8$\pm$12.2
\\\hline
\citep{kho2006acute} & 39 & 24 & 122.9$\pm$13.2 & 82.6$\pm$10.1 & 15 & 110.5$\pm$8.8 & 74.5$\pm$7.3
\\\hline
\citep{Papakonstantinou2016-mj} & 40 & 20 & 128.2$\pm$12.3 & 83.3$\pm$5.8 & 20 & 117.0$\pm$14.4 & 75.5$\pm$12.3
\\\hline
\citep{Monnard2017-je} & 45 & 23 & 119.0$\pm$9.5 & 76.0$\pm$4.7 & 22 & 111.0$\pm$4.6  & 72.0$\pm$4.6
\\\hline
\citep{helfer2001does} & 55 &  26 & 129.1$\pm$9.1 & 64.2$\pm$8.3 & 29 & 108.0$\pm$9.8 & 61.7$\pm$6.7
\\\hline
\citep{harshfield1989race} & 92 & 55 & 105.6$\pm$10.3 & 58.5$\pm$9.3 & 37 & 103.4$\pm$11.8  & 56.9$\pm$8.9
\\\hline
\citep{harshfield1989race} & 107 & 42  & 114.3$\pm$12.2 & 62.5$\pm$13.3 & 65 & 100.3$\pm$10.0  & 63.6$\pm$10.9
\\\hline
\citep{costa2018gender} & 122 & 52 & 137.0$\pm$20.0 & 86.0$\pm$12.0 & 70 & 145.0$\pm$26.0  & 87.0$\pm$15.0
\\\hline
\citep{Ki2013-ai} & 141 & 117 & 128.8$\pm$10.4 & 81.3$\pm$5.3 & 24 & 126.0$\pm$11.8  & 77.6$\pm$7.4
\\\hline
\citep{Wang2006-jk} & 312 & 142 & 116.3$\pm$9.9 & 66.4$\pm$7.1 & 170 & 112.3$\pm$8.3 & 66.5$\pm$6.8
\\\hline
\citep{Wang2006-jk} & 351 & 184 & 113.7$\pm$9.0 & 64.5$\pm$6.6 & 167 & 109.8$\pm$7.5 & 64.1$\pm$5.9
\\\hline
\citep{Song2016-ho} & 806 & 237 & 120.6$\pm$12.9 & 77.9$\pm$8.7 & 569 & 112.7$\pm$12.3 & 71.7$\pm$8.4
\\\hline
\citep{lan2012prevalence} & 1030 & 614  & 123.3$\pm$12.3 & 77.3$\pm$8.2  & 416 & 117.1$\pm$10.6 & 73.9$\pm$7.1
\\\hline
\citep{privvsek2018epidemiological} & 1298 & 638  & 127.4$\pm$14.0 & 77.7$\pm$10.5 & 660 & 124.4$\pm$15.7  & 
74.5$\pm$9.7
\\\hline
\citep{cui2002genes} & 1378 & 664 & 122.0$\pm$10.5 & 72.0$\pm$9.4 & 714 & 113.0$\pm$9.9 & 68.2$\pm$8.6
\\\hline
\citep{cui2002genes} & 1534 & 767  & 132.0$\pm$16.4 & 83.1$\pm$9.3  & 767 & 126.0$\pm$16.1 & 78.9$\pm$9.2
\\\hline
\citep{vallee2019relationship} & 2105 & 945  & 132.0$\pm$18.0 & 79.0$\pm$11.0 & 1160 & 122.0$\pm$18.0  & 
75.0$\pm$10.0
\\\hline
\citep{Giggey2011-ri} & 2442 & 1577 & 129.7$\pm$19.2 & 80.9$\pm$10.6 & 865 & 123.2$\pm$20.9 & 76.5$\pm$10.3
\\\hline
\citep{bourgeois2017associations} & 2849 & 1505  & 124.6$\pm$15.5 & 73.8$\pm$15.5 & 1344 & 120.0$\pm$18.3  & 71.2$\pm$14.6
\\\hline
\citep{bourgeois2017associations} & 3654 & 1915  & 120.0$\pm$21.8 & 70.5$\pm$17.5 & 1739 & 115.0$\pm$20.8  & 68.2$\pm$16.6
\\\hline
\citep{bourgeois2017associations} & 6485 & 3379 & 121.7$\pm$23.2 & 72.9$\pm$17.4 & 3106 & 117.8$\pm$22.2  & 70.4$\pm$16.7
\\\hline
\citep{pan1986role} & 33599 & 19704  & 138.7$\pm$18.4 & 81.3 $\pm$11.5 & 13895 & 132.1$\pm$19.3 & 77.4 $\pm$11.6
\\
\bottomrule
\end{tabular}
\end{center}
\end{table}

Age is another significant demographic feature. Arterial stiffness is known to increase with age, arterial compliance reduces, and pulse pressure increases \citep{muntner2019measurement}. Elderly individuals may experience systolic hypertension, characterized by elevated SBP without a rise in DBP. This condition, commonly known as ``pseudohypertension'', has been linked to a decline in arterial distensibility, which can lead to bias and inaccurately high BP readings, as the external cuff pressure reduces on the artery \citep{muntner2019measurement}. Carrico et al.\ have examined the relationship between age, sex, and BP in a group comprising 965 men and 1114 women \citep{carrico2013predictive}. Their findings indicate that overall BP values increase with age, but decrease again after approximately 70 years of age. In most age groups, BP is higher in males than in females, but both sexes have similar BP until their teenage years. However, after about 70 years of age, females' SBP values surpass those of males, while their DBP values remain approximately the same.

There has also been a significant interest in examining health measures across racial groups. Numerous studies have examined the relationship between BP and race \citep{Jones2006, hardy2021racial}. These studies have found that specific diseases are more common in certain racial groups, suggesting a potential link between race and hypertension prevalence \citep{Cooper1998, fryar2017hypertension}. For instance, Staessen et al. \citep{staessen1997nocturnal} showed that Asian populations have a higher rate of non-dipping than European populations, indicating possible variations in BP patterns among different races. Furthermore, individuals of African American descent have been observed to have higher rates of hypertension, cardiovascular, and cerebrovascular morbidity and mortality compared to those of European descent \citep{mayet1998ethnic}. This suggests that race (or more accurately genetic factors) may play a significant role in hypertension susceptibility. Another possible explanation is that the experiences of some social groups, including but not limited to the medical care that the groups receive, affect their BP. Most likely, the higher prevalence of hypertension in African Americans has multiple sources including genetic, socioeconomic, systemic, and other factors. Table~\ref{Tab: Race} summarizes a group of studies that have compared BP values based on race. 

\begin{table}
\caption{The summarized results of studies reporting blood pressure values across race. N is the number of patients, and summaries of SBP and DBP include mean $\pm$ standard deviation.}\label{Tab: Race}

\begin{center}
\begin{tabular}{rcccc}
\toprule
\textbf{Ref.} & \textbf{N}   &\textbf{Race} & \textbf{SBP} & \textbf{DBP} \\
\hline
\citep{harshfield1989race} & 199 & Black & 105.7$\pm$10.9 & 63.2$\pm$11.9 \\
 &  & White & 104.7$\pm$10.9 & 57.9$\pm$9.1 \\
\hline
\citep{mayet1998ethnic} & 245 & White hypertensive & 145.0$\pm$18.3 & 92.0$\pm$10.7 \\
 &  & Black hypertensive & 142.0$\pm$14.9 & 93.0$\pm$10.8
 \\
\hline
\citep{Wang2006-jk} & 663 & European Americans & 111.8$\pm$8.3 & 64.3$\pm$6.2 \\
 &  & African Americans & 114.1$\pm$9.0 & 66.4$\pm$6.9 \\
\hline
\citep{bourgeois2017associations} & 6503 & Non-Hispanic Black & 122.4$\pm$16.9 & 72.5$\pm$21.3 \\
 &  & Mexican American & 117.6$\pm$30.2 & 69.4$\pm$24.1 \\
\hline
\citep{bourgeois2017associations} & 9334 & Non-Hispanic White & 119.8$\pm$32.2 & 71.7$\pm$24.1 \\
 &  & Non-Hispanic Black & 122.4$\pm$16.9 & 72.5$\pm$21.3 \\\hline
\citep{bourgeois2017associations} & 10139 & Non-Hispanic White & 119.8$\pm$32.2 & 71.7$\pm$24.1 \\
 &  & Mexican American & 117.6$\pm$30.2 & 69.4$\pm$24.1 \\
 \bottomrule
\end{tabular}
\end{center}
\end{table}

\subsubsection{Subject-wise factors}
Height and weight are two essential factors influencing BP values. Studies have shown that taller individuals tend to have lower SBP and higher DBP than shorter individuals \citep{bourgeois2017associations}. The body mass index (BMI) --- weight divided by the squared height (kg/m$^2$) --- can be used as a compound factor for BP assessment \citep{he2000blood, jena2018relationship}. Table \ref{Tab: BMI} presents the findings from various studies that have reported BP values based on BMI, overall showing a positive correlation between BP and BMI \citep{neter2003influence, Neter2003}.

\begin{table}[tb]
\centering
\caption{The results of studies reporting blood pressure values based on BMI}\label{Tab: BMI}
\small{
\begin{tabular}{rcccc}
\toprule
\textbf{Ref.} & \textbf{N} & \textbf{BMI} &\textbf{SBP} &\textbf{DBP} \\
\hline
\citep{Tibana2013-ve} & 13 & 30.7$\pm$4.2 & 124.7$\pm$13.0 & 82.4$\pm$10.1
\\\hline
\citep{Karatzi2013-yr} & 17  & 24.3$\pm$2.4 & 115.4$\pm$6.2 & 68.5$\pm$5.4
\\\hline
\citep{Fantin2016-bt} & 21  & 23.9$\pm$3.3 & 115.4$\pm$13.5 & 71.2$\pm$9.4
\\\hline
\citep{Papakonstantinou2016-mj} & 40 & 23.6$\pm$3.5 & 122.6$\pm$14.4 & 79.4$\pm$10.3
\\\hline
\citep{Kayrak2010-lk} & 45 & 29.8$\pm$4.7 & 174.0$\pm$14.1 & 95.8$\pm$11.5
\\\hline
\citep{Monnard2017-je} & 45 & 22.6$\pm$2.6 & 115.0$\pm$6.7 & 74.0$\pm$6.7
\\\hline
\citep{cunha2017acute} & 50 & 28.6$\pm$3.9 & 133.9$\pm$12.3 & 66.4$\pm$9.7
\\\hline
\citep{Netea2003} & 57 & 25.7$\pm$4.4 & 135.7$\pm$24.8 & 79.5$\pm$9.7
\\\hline
\citep{Kayrak2010-lk} & 70 & 30.6$\pm$5.6 & 168.3$\pm$18.4 & 83.4$\pm$9.4
\\\hline
\citep{talukder2016effect} & 88 & 22.0$\pm$4.4 & 108.0$\pm$10.0 & 65.0$\pm$9.0
\\\hline
\citep{Xu2019-rx} & 100 & 23.7$\pm$2.9 & 132.9$\pm$16.5 & 80.0$\pm$10.4
\\\hline
\citep{Azar2016-eg} & 165 & 21.3$\pm$4.1 & 112.0$\pm$10.0 & 67.0$\pm$9.0
\\\hline
\citep{talukder2016effect} & 194 & 26.0$\pm$5.0 & 120.3$\pm$15.8 & 76.4$\pm$11.3
\\\hline
\citep{krzesinski2016diagnostic} & 280 & 28.7$\pm$4.2 & 143.8$\pm$14.3 & 92.4$\pm$9.5
\\\hline
\citep{Li2019-ax} & 287 & 25.0$\pm$3.9 & 139.2$\pm$16.9 & 74.6$\pm$12.0
\\\hline
\citep{Wang2006-jk} & 312 & 24.0$\pm$7.0 & 114.1$\pm$9.0 & 66.4$\pm$6.9
\\\hline
\citep{Wang2006-jk} & 351 & 22.0$\pm$5.0 & 111.8$\pm$8.3 & 64.3$\pm$6.2
\\\hline
\citep{Walker2019-oe} & 389  & 29.4$\pm$5.7 & 121.1$\pm$16.3 & 53.8$\pm$4.8
\\\hline
\citep{Widlansky2007-ih} & 500 & 27.9$\pm$5.3 & 123.0$\pm$17.0 & 70.0$\pm$11.0
\\\hline
\citep{Widlansky2007-ih} & 599 & 28.1$\pm$5.1 & 128.0$\pm$18.0 & 72.0$\pm$12.0
\\\hline
\citep{privvsek2018epidemiological} & 638 & 27.5$\pm$3.5 & 127.4$\pm$14.0 & 77.7$\pm$10.5
\\\hline
\citep{privvsek2018epidemiological} & 660 & 27.3$\pm$5.2 & 124.4$\pm$15.7 & 74.5$\pm$9.7
\\\hline
\citep{Widlansky2007-ih} & 733 & 28.0$\pm$5.2 & 122.0$\pm$17.0 & 69.0$\pm$11.0
\\\hline
\citep{Widlansky2007-ih} & 735 & 28.2$\pm$5.5 & 124.0$\pm$18.0 & 71.0$\pm$11.0
\\\hline
\citep{Song2016-ho} & 806  & 23.7$\pm$3.0 & 115.0$\pm$13.0 & 73.5$\pm$8.9
\\\hline
\citep{Walker2019-oe} & 833 & 27.5$\pm$4.7 & 124.3$\pm$9.5 & 68.5$\pm$6.1
\\\hline
\citep{Walker2019-oe} & 927 & 27.3$\pm$5.6 & 117.1$\pm$14.3 & 53.9$\pm$4.6
\\\hline
\citep{vallee2019relationship} & 945 & 26.1$\pm$4.4 & 132.0$\pm$18.0 & 79.0$\pm$11.0 \\
\hline
\citep{Walker2019-oe} & 1030 & 30.8$\pm$6.3 & 138.1$\pm$18.6 & 71.9$\pm$8.6
\\\hline
\citep{vallee2019relationship} & \text{1160} & 25.7$\pm$5.2 & 122.0$\pm$18.0& 75.0$\pm$10.0 \\
\hline
\citep{privvsek2018epidemiological} & 1298 & 27.4$\pm$4.5 & 125.9$\pm$14.9 & 76.1$\pm$10.2
\\\hline
\citep{bourgeois2017associations} & 1344 & 30.0$\pm$7.3 & 120.0$\pm$18.3 & 71.2$\pm$14.6
\\\hline
\citep{cui2002genes} & 1378 & 23.4$\pm$3.5 & 112.5$\pm$10.1 & 70.0$\pm$8.9
\\\hline
\citep{bourgeois2017associations} & 1505 & 27.1$\pm$7.7 & 124.6$\pm$15.5 & 73.8$\pm$15.5
\\\hline
\citep{cui2002genes} & 1534 & 26.5$\pm$3.9 & 129.0$\pm$16.2 & 81.0$\pm$9.2
\\\hline
\citep{Walker2019-oe} & 1559 & 28.8$\pm$5.2 & 137.2$\pm$16.4 & 71.8$\pm$8.3
\\\hline
\citep{bourgeois2017associations} & 1739 & 28.6$\pm$8.3 & 115.0$\pm$20.8 & 68.2$\pm$16.6
\\\hline
\citep{bourgeois2017associations} & 1915 & 27.7$\pm$8.7 & 120.0$\pm$21.8 & 70.5$\pm$17.5
\\\hline
\citep{barba2006body} & 1937 & 19.2$\pm$3.8 & 96.5$\pm$13.3 & 60.6$\pm$9.4
\\\hline
\citep{barba2006body} & 1968 & 19.5$\pm$3.9 & 97.5$\pm$13.2 & 61.3$\pm$9.0
\\\hline
\citep{vallee2019relationship} & \text{2105} & \text{25.9$\pm$5.1} & 127.0$\pm$19.0 & 77.0$\pm$11.0 \\
\hline
\citep{Sano2020-zk} & 2423  & 24.3$\pm$3.3 & 154.7$\pm$16.2 & 90.1$\pm$11.9
\\\hline
\citep{Giggey2011-ri} & 2442  & 24.9$\pm$3.6 & 127.4$\pm$20.1 & 79.4$\pm$10.7
\\\hline
\citep{bourgeois2017associations} & 3106 & 26.9$\pm$11.1 & 117.8$\pm$22.2 & 70.4$\pm$16.7
\\\hline
\citep{bourgeois2017associations} & 3379 & 27.5$\pm$5.8 & 121.7$\pm$23.2 & 72.9$\pm$17.4
\\\hline
\citep{Li2019-ax} & 6887  &  25.7$\pm$4.4 & 134.3$\pm$20.2 & 79.6$\pm$11.6
\\\hline
\citep{Li2019-ax} & 12624  & 25.5$\pm$4.4 & 131.9$\pm$23.1 & 79.7$\pm$11.9
\\\hline
\citep{Li2019-ax} & 17921  & 25.6$\pm$4.4 & 133.1$\pm$22.4 & 79.9$\pm$11.8
\\\hline
\citep{Zheng2021-tq} & 32710 & 23.6$\pm$3.3 & 123.6$\pm$19.8 & 78.9$\pm$12.4
\\\hline
\citep{Kang2020-mq} & 417907 & 23.8$\pm$3.6 & 128.1$\pm$19.0 & 76.1$\pm$10.4
\\\hline
\citep{Lewington2012-iv} & 506673 & 23.7$\pm$3.4 & 131.0$\pm$21.0 & 78.0$\pm$11.0
\\
\bottomrule
\end{tabular}}
\end{table}

\subsubsection{Individual-wise background medical conditions}
Accurate interpretation of BP measurements should consider patient-specific conditions and comorbid factors, which can directly or indirectly influence BP readings. Often, the inherent biological rhythms in the disease process and their potential clinical implications are overlooked or inadequately considered when treating hypertensive patients \citep{hassler2005circadian}. In the following section, we discuss several situations where these considerations play a significant role.

Healthy individuals typically show BP variations throughout the day, rising during daytime and dropping at night \citep{Ohkubo2002}. Most people have nighttime BP readings 10--20\% lower than daytime. However, some experience a non-dipping pattern where nighttime BP drops by less than 10\% \citep{routledge2007nondipping}. About 25\% of hypertensive individuals with unknown causes exhibit this non-dipping pattern \citep{Pickering2001}.

The \textit{``white-coat effect''}, caused by the medical environment and physician presence, can elevate a patient's clinic BP \citep{ogedegbe2010principles}. This effect often results in higher SBP and DBP compared to baseline ambulatory BP \citep{kallioinen2017sources}. Table~\ref{Tab: enviro} summarizes studies comparing BP measurements in settings like home and office.

\begin{table*}[tb]
\caption{The results of studies reporting blood pressure values based on the measuring environment}\label{Tab: enviro}
\begin{center}
\begin{tabular}{rccccccc}
\toprule
\textbf{Ref.} & \textbf{N} & \multicolumn{3}{c}{\begin{tabular}{@{}c@{}} \textbf{Home} 
\end{tabular} }  
&  \multicolumn{3}{c}{\textbf{Clinic} } 
\\\hline
 &   &  N & SBP & DBP  & N & SBP & DBP
\\\hline
\citep{Ragot2000-ym} & 454 & 199  & 144.0$\pm$18.0 & 88.6$\pm$10.0 & 255 & 160.0$\pm$13.0 & 99.7$\pm$4.0
\\\hline
\citep{Asayama2022-ft} & 574 & 287 & 125.7$\pm$8.4 & 72.9$\pm$8.6 & 287 & 139.2$\pm$16.9 & 74.6$\pm$12.0
\\\hline
\citep{Sano2020-zk} & 4846 & 2423  & 152.4$\pm$3.1 & 89.7$\pm$9.3  & 2423 & 154.7$\pm$16.2 & 90.1$\pm$11.9
\\\hline
\citep{Li2019-ax} & 13774 & 6887 & 127.3$\pm$18.1 & 76.2$\pm$9.9 & 6887 & 134.3$\pm$20.2 & 79.6$\pm$11.6 
\\\hline
\citep{Li2019-ax} & 35842 & 17921 & 129.1$\pm$18.6 & 76.9$\pm$9.8 & 17921 & 133.1$\pm$22.4 & 79.9$\pm$11.8
\\
\bottomrule
\end{tabular}
\end{center}
\end{table*}

Pregnancy-related hemodynamic changes can affect BP readings. While many automated BP device types exist, few are calibrated for pregnant women, including those with hypertensive disorders \citep{van2019validation}. Nearly 10\% of pregnant women face high BP, risking both the fetus(es) and the mother. Those with gestational diabetes or preeclampsia are especially at risk \citep{Wagner2012-id, Katebi2022, Katebi2023}.

Obesity correlates with high BP in both children and adults. For each BMI unit increase, SBP and DBP rise by 0.56 and 0.54\,mmHg, respectively, in obese children \citep{he2000blood}. Table~\ref{Tab: obes} presents studies examining the impact of obesity on children's BP values.




\begin{table}[tb]
\caption{The results of studies investigating the effect of obesity on blood pressure values in the child population}\label{Tab: obes}
\begin{center}
\begin{tabular}{rcccccc}
\toprule
\textbf{Ref.} &\textbf{BP} & \textbf{Sex} & \multicolumn{2}{c}{\begin{tabular}{@{}c@{}} \textbf{Obese group} 
\end{tabular} }  
&  \multicolumn{2}{c}{\textbf{Non obese group}} 
\\\hline
 &  & &  N & Mean$\pm$SD  & N & Mean$\pm$SD
\\\hline
\citep{he2000blood} & SBP &  Boys &  330 & 96.0$\pm$13.3  & 331 &90.0$\pm$10.6
\\
 & SBP &  Girls &  253 & 95.0$\pm$13.2  & 251 & 90.0$\pm$11.5
\\\hline
\citep{he2000blood} & DBP &  Boys &  330 & 60.0$\pm$10.7  & 331 &60.0$\pm$9.5
\\ 
& DBP &  Girls &  253 & 60.0$\pm$11.0  & 251 & 60.0$\pm$10.0
\\\hline
 \citep{barba2006body} & SBP &  Boys &  420 & 103.3$\pm$14.8 & 1034 & 94.2$\pm$11.8 
 \\
 & SBP &  Girls &  401 & 100.7$\pm$14.1  & 1050 &93.5$\pm$11.8
\\\hline
\citep{barba2006body} & DBP &  Boys &  420 & 64.4$\pm$9.8  & 1034 & 59.6$\pm$8.7
 \\
& DBP &  Girls &  401 & 63.0$\pm$9.3  & 1050 & 58.8$\pm$9.2 
\\\hline
\citep{jena2018relationship} & SBP &  Boys &  80 & 103.0$\pm$13.0 & 143 & 98.0$\pm$11.0 
 \\
 & SBP &  Girls &  28 & 99.0$\pm$14.0 & 144 & 94.0$\pm$11.0
\\\hline
\citep{jena2018relationship} & DBP &  Boys &  80 & 57.0$\pm$9.0 & 143 & 55.0$\pm$6.0
 \\
& DBP &  Girls & 28 & 55.0$\pm$11.0 & 144 & 50.0$\pm$6.0 
 \\
\bottomrule
\end{tabular}
\end{center}
\end{table}

\subsubsection{Eating, drinking, and smoking}
Eating, drinking, and smoking habits can significantly influence our BP values \citep{Forman2009, Buckman2015-ew, Azar2016-eg}. Studies indicate that eating has both short-term and long-term effects on BP levels \citep{Appel1997}.

Postprandial (post-meal) BP initially rises due to increased activity, but later drops, particularly due to decreased total peripheral resistance from visceral vasodilation \citep{Kawano2010}. This reduction is pronounced in elderly, hypertensive individuals, and those with autonomic failure \citep{Jansen1987}. High-carbohydrate meals cause a larger BP decrease than high-fat ones. While the exact mechanism for post-meal hypotension remain unclear, impacting factors might include impaired baroreceptor reflexes, insulin-induced vasodilation, and the release of vasodilatory gastrointestinal polypeptides \citep{Sidery1993}. This BP decrease peaks around 1 hour post-meal and can last over 2 hours, influencing diurnal BP changes especially in elderly subjects with hypertension \citep{Kawano2010}. Furthermore, BP effects from alcohol, caffeine, and nicotine vary by dose and individual \citep{kallioinen2017sources}.

For long-term impacts of eating habits on BP, guidelines primarily recommend lifestyle and dietary changes to manage hypertension, emphasizing reduced salt intake. Adding potassium-rich foods like nuts, fruits and vegetables to the diet can enhance BP control \citep{burnier2019should}. Table~\ref{Tab: eating} summarizes studies on the impacts of eating, drinking, and smoking on BP.

\begin{table}[tb]
\caption{Summarized results of studies investigating the impact of eating, drinking, or smoking on blood pressure values. ``B.'' denotes before and ``A.'' denotes after each activity}\label{Tab: eating}
\begin{center}
\small{\begin{tabularx}{\textwidth}{rcccc}
\toprule
\textbf{Ref.} & \textbf{N} &\textbf{Conditions} & \textbf{SBP} & \textbf{DBP} \\
\hline
\citep{Nishiwaki2017-ud} & 11 & B. drinking AF200\textsuperscript{a} &  120.0$\pm$9.9 &  69.0$\pm$3.3 \\ 
& & 90 min. A. drinking AF200 & 123$\pm$6.6 & 74.0$\pm$3.3 
\\\hline
\citep{Nishiwaki2017-ud} & 11 & B. drinking B350\textsuperscript{d} &  123.0$\pm$6.6 &  71.0$\pm$6.6 \\ 
& & 90 min. A. drinking B350 & 123.0$\pm$9.9 & 76.0$\pm$13.2 
\\\hline
\citep{McMullen2011-qu} & 12 & B. drinking placebo &  133.5$\pm$14.1 &  86.4$\pm$8.7 \\ 
& & A. drinking placebo & 131.5$\pm$11.8 & 82.9$\pm$8.4 
\\\hline
\citep{McMullen2011-qu} & 12 & B. drinking C67\textsuperscript{f} &  127.6$\pm$9.1 &  81.9$\pm$6.7 \\ 
& & A. drinking C67 & 135.6$\pm$10.1 & 84.7$\pm$6.0 
\\\hline
\citep{McMullen2011-qu} & 12 & B. drinking C133\textsuperscript{g} &  126.9$\pm$11.1 &  81.4$\pm$7.7 \\ 
& & A. drinking C133 & 137.6$\pm$14.1 & 86.5$\pm$8.2 
\\\hline
\citep{McMullen2011-qu} & 12 & B. drinking C200\textsuperscript{h} &  127.5$\pm$10.2 &  81.1$\pm$5.5 \\ 
& & A. drinking C200 & 132.7$\pm$10.7 & 83.5$\pm$8.2 
\\\hline
\citep{Carter2011-aw} & 15 & Pre-treatment of alcohol &  120.0$\pm$11.6 &  64.0$\pm$7.7 \\ 
& & Post-treatment of alcohol & 124.0$\pm$15.5 & 69.0$\pm$ 7.7 
\\\hline
\citep{Carter2011-aw} & 15 & Pre-treatment of placebo &  117.0$\pm$7.7 &  64.0$\pm$11.6 \\ 
& & Post-treatment of placebo & 123.0$\pm$7.7 & 71.0$\pm$7.7 
\\\hline
\citep{Fantin2016-bt} & 18 & B. drinking alcohol\textsuperscript{e} & 110.3$\pm$12.0 &  80.0$\pm$8.0 \\ 
& & A. drinking alcohol & 109.5$\pm$11.4 & 76.2$\pm$7.1 
\\\hline
\citep{Nowak2019-si} & 22 & B. drinking Noni juice & 119.6$\pm$8.3 &  77.0$\pm$6.6 \\ 
& & A. drinking Noni juice & 113.6$\pm$8.5 & 72.0$\pm$4.8
\\\hline
\citep{Nowak2019-si} & 22 & B. drinking chokeberry juice & 125.6$\pm$14 & 84.0$\pm$9.8 \\ 
& & A. drinking chokeberry juice & 124.3$\pm$16.1 &81.0$\pm$9.9
\\\hline
\citep{Nowak2019-si} & 22 & B. consuming energy drink & 119.2$\pm$14.8 & 73.9$\pm$8.4 \\ 
& & A. consuming energy drink & 124.8$\pm$14.1 & 84.8$\pm$9.9
\\\hline
\citep{Nowak2019-si} & 22 & B. drinking water & 124.3$\pm$13.5 & 77.7$\pm$9.2 \\ 
& & A. drinking water & 124.0$\pm$11.4 & 75.8$\pm$8.0
\\\hline
\citep{luqmanexperimental} & 35 & B. drinking STING\textsuperscript{j} & 123.0$\pm$14.9 & 78.7$\pm$10.5 \\ 
& & A. drinking STING & 123.7$\pm$14.5 & 78.2$\pm$9.8
\\\hline
\citep{olatunji2011water} & 37 & B. drinking 50\,ml water & 119.6$\pm$11.5 & 74.1$\pm$10.1 \\ 
& & A. drinking 50\,ml water & 122.5$\pm$11.6 & 77.3$\pm$7.7 
\\\hline
\citep{olatunji2011water} & 37 & B. drinking 500\,ml water & 116.9$\pm$8.6 & 73.8$\pm$10.0 \\ 
& & A. drinking 500\,ml water & 125.8$\pm$8.8 & 76.8$\pm$10.7
\\\hline
\citep{kho2006acute} & 39 & B. non-tobacco smoking & 120.0$\pm$13.5 & 78.9$\pm$10.1 \\ 
& & 65 min. A. non-tobacco smoking & 125.8$\pm$8.8 & 76.6$\pm$6.9
\\\hline
\citep{kho2006acute} & 39 & B. tobacco smoking & 118.6$\pm$12.8 & 79.7$\pm$9.2 \\ 
& & 65 min. A. tobacco smoking & 116.9$\pm$12.4 & 80.0$\pm$8.9
\\\hline
\citep{Papakonstantinou2016-mj} & 40 & B. drinking 200\,ml cold espresso & 116.7$\pm$9.7 & 75.3$\pm$7.1 \\ 
& & A. drinking 200\,ml cold espresso & 120.0$\pm$11.1 & 79.5$\pm$9.1
\\\hline
\citep{Papakonstantinou2016-mj} & 40 & B. drinking 200\,ml filter coffee & 118.2$\pm$12.3 & 77.1$\pm$8.5 \\ 
& & A. drinking 200\,ml filter coffee & 121.2$\pm$10.6 & 79.1$\pm$6.7
\\\hline
\citep{Papakonstantinou2016-mj} & 40 & B. drinking 200\,ml cold inst. coffee & 116.7$\pm$12.3 & 77.3$\pm$8.5 \\ 
& & A. drinking 200\,ml cold inst. coffee & 121.3$\pm$11.4 & 79.6$\pm$7.3
\\\hline
\citep{Papakonstantinou2016-mj} & 40 & B. drinking 200\,ml hot inst. coffee & 118.5$\pm$10.5 & 78.2$\pm$9.3 \\ 
& & A. drinking 200\,ml hot inst. coffee & 122.6$\pm$11.8 & 80.2$\pm$8.7
\\\hline
\citep{luqmanexperimental} & 60 & B. drinking STING\textsuperscript{i} & 121.2$\pm$14.3 & 77.4$\pm$9.6 \\ 
& & A. drinking STING & 126.5$\pm$14.1 & 81.0$\pm$9.0
\\\hline
\citep{Buckman2015-ew} & 72 & Pre beverage of juice & 117.0$\pm$13.1 & 79.8$\pm$10.1 \\ 
& & Post beverage of juice & 125.9$\pm$13.2 & 85.4$\pm$9.6
\\\hline
\citep{Buckman2015-ew} & 72 & Pre beverage of placebo & 126.2$\pm$19.2 & 83.5$\pm$13.9 \\ 
& & Post beverage of placebo & 130.7$\pm$18.2 & 85.7$\pm$12.9
\\\hline
\citep{Buckman2015-ew} & 72 & Pre beverage of alcohol & 116.9$\pm$13.5 & 80.1$\pm$8.7 \\ 
& & Post beverage of alcohol & 113.2$\pm$12.6 & 79.9$\pm$9.7
\\\hline
\citep{Azar2016-eg} & 194 & B. water-pipe smoking & 120.3$\pm$15.8 & 76.4$\pm$11.3 \\ 
& & 15 min. A. water-pipe smoking & 121.1$\pm$16.1 & 77.1$\pm$10.8
\\
\bottomrule
\end{tabularx}}
\end{center}
\noindent{\footnotesize{
\textsuperscript{a} 200\,ml of alcohol-free beer; \textsuperscript{b} 350\,ml of alcohol-free beer; \textsuperscript{c} 200\,ml of beer; \textsuperscript{d} 350\,ml of beer; \textsuperscript{e} 2 glasses (2 × 125\,ml) of red wine (12\% of ethanol); \textsuperscript{f} 67\,mg of caffeine; \textsuperscript{g} 133\,mg of caffeine; \textsuperscript{h} 200\,mg of caffeine; \textsuperscript{i} a single dose (500\,ml) of energy drink; \textsuperscript{j} a 2-glasses (500\,ml) of plain water
}}
\end{table}
\subsubsection{Circadian rhythm}
The circadian clock significantly affects CVD risk factors like heart rate and BP \citep{Morris2012}. It has been linked to two mechanisms: the central clock in the hypothalamic suprachiasmatic nucleus (SCN) and the peripheral clock in most body tissues and organs \citep{Richards2012}. Environmental factors, like physical exercise, can also impact and align circadian rhythms, especially in skeletal muscle \citep{hower2018circadian}.

BP displays daily rhythmic fluctuations, peaking in the early morning and dipping around midnight. Boivin et al.\ \citep{Boivin2014-ss} examined BP changes in the morning and evening among 52 controlled hypertensive patients. BP was measured six times both in the morning and evening, three times before and after resting. Each session was nine minutes with a minute between measurements and five minutes of rest. The study found significant BP differences between morning and evening, around 5\,mmHg for DBP and 8\,mmHg for SBP.

Psychological and physical activities also contribute to BP fluctuations, with higher values commonly observed during work hours and lower values at home. While several neurohormonal systems follow a circadian rhythm with a morning peak, the sympathetic nervous system appears to be the primary determinant of these BP circadian variations. However, this principle can be reversed for individuals with specific job roles, such as shift workers \citep{hassler2005circadian}.

Table \ref{Tab: cardiac} provides a comparative analysis of studies exploring the effects of the circadian clock on BP values.

\begin{table}[tb]
\caption{Results of studies investigating the impact of the circadian clock on blood pressure values}\label{Tab: cardiac}
\begin{center}
\begin{tabular}{rccc}
\toprule
\textbf{Ref.} & \textbf{N} & \textbf{Mean daytime BP}  & \textbf{Mean nighttime BP}\\
& & SBP / DBP & SBP / DBP
\\
\hline
\text{\citep{mayet1998ethnic}}&46&(149.0$\pm$18.3)/(95.0$\pm$10.7)&(132.0$\pm$21.7)/(81.0$\pm$13.5)\\
\hline
\text{\citep{mayet1998ethnic}}&46&(145.0$\pm$14.9)/(95.0$\pm$11.5)& (136.0$\pm$17.6)/(86.0$\pm$11.5)\\
\hline
\text{\citep{krzesinski2016diagnostic}}&280&(144.7$\pm$11.9)/(91.0$\pm$8.6)&(128.2$\pm$12.9)/(77.8$\pm$9.0)\\
\hline
\text{\citep{Wang2006-jk}}&312&(119.5$\pm$8.8)/(72.5$\pm$6.6)&(108.7$\pm$9.3)/(60.4$\pm$7.2)\\
\hline
\text{\citep{Wang2006-jk}}&351&(117.7$\pm$8.1)/(70.9$\pm$6.4)&(105.9$\pm$8.4)/(57.7$\pm$6.1)\\
\hline
\text{\citep{Li2019-ax}}&17921&(129.3$\pm$15.1)/(78.8$\pm$9.3)&(112.9$\pm$15.6)/(65.1$\pm$9.6)\\
\bottomrule
\end{tabular}
\end{center}
\end{table}

\subsection{Biases related to the acquisition session}
There are specific guidelines for BP measurement, violation of which can lead to inaccurate readings. We subsequently explore often-overlooked factors during BP acquisition that can introduce bias and compromise measurement reliability.

\subsubsection{Seasonal variations and ambient temperature} 

Seasonal BP fluctuations, with peaks in winter and troughs in summer, have been reported in both home and clinic measurements \citep{Park2019, Sega1998}. Although the direct link between BP and ambient temperature has not been conclusively established, consistent seasonal patterns suggest a correlation \citep{Jansen2001-xl}. Yang et al.'s research on 23,040 individuals in China found BP values to be highest in winter (December--February) and lowest in summer (July--August) \citep{Yang2015-rc}. Table~\ref{Tab: TEM} summarizes studies comparing BP values by environmental temperature. The short-term impact of temperature on BP underscores the importance of protective measures in colder weather to maintain stable BP and reduce BP-related disease risks \citep{Xu2019-rx}.

\begin{table}[tb]
\caption{Results of studies investigating the impact of ambient temperature on blood pressure values}\label{Tab: TEM}
\begin{center}
\begin{tabular}{rcccc}
\toprule
\textbf{Ref.} & \textbf{N} & \textbf{Temp. ($^{\circ}$C)} &\textbf{SBP} &\textbf{DBP} \\ 
\hline
\citep{Jansen2001-xl} & 19 & 7.5$\pm$0.7 &118.0$\pm$7.8 & 65.0$\pm$6.1 \\ 
\hline
\citep{Jansen2001-xl} & 20 & 14.8$\pm$1.3 & 116.0$\pm$6.3 & 64.0$\pm$5.4 \\  
\hline
\citep{Jansen2001-xl} & 20 & 2.0$\pm$0.4& 121.0$\pm$7.6 &  65.0$\pm$6.3 \\  
\hline
\citep{Jansen2001-xl} & 20 & -3.4$\pm$3.0& 125.0$\pm$18.0 &  67.0$\pm$5.8 \\ 
\hline
\citep{Wu2015-dj} & 39 & 25.0 &117.0&65.0 \\  
\hline
\citep{Wu2015-dj} & 39 & 17.6 & 117.0& 66.0 \\  
\hline
\citep{Wu2015-dj} & 39 & 22.7 & 122.0& 65.0 \\  
\hline
\citep{Wu2015-dj} & 39 & 21.6 & 119.0& 65.0 \\
\hline
\citep{Xu2019-rx} & 100 & 15.7$\pm$8.7 &132.9$\pm$16.5 &80.0$\pm$10.4 \\ 
\hline
\citep{Kim2012-oi} & 327 & 31.5$\pm$1.0 &133.7$\pm$24.5 &81.7$\pm$15.4 \\
\hline
\citep{Widlansky2007-ih} & 500 & 25$\pm$1 &123.0$\pm$17.0 & 70.0$\pm$11.0 \\
\hline
\citep{Widlansky2007-ih} & 599 & 24$\pm$1 &128.0$\pm$18.0 & 72.0$\pm$12.0 \\  
\hline
\citep{Widlansky2007-ih} & 733 & 26$\pm$1 &122.0$\pm$17.0 & 69.0$\pm$11.0 \\  
\hline
\citep{Widlansky2007-ih} & 755 & 25$\pm$1 &124.0$\pm$18.0 & 71.0$\pm$11.0 \\  
\bottomrule
\end{tabular}
\end{center}
\end{table}


\subsubsection{Cuff position} 
\begin{figure}[tb]
\centering
\includegraphics[trim={8cm 4cm 15cm 3cm},clip,width=.5\columnwidth]{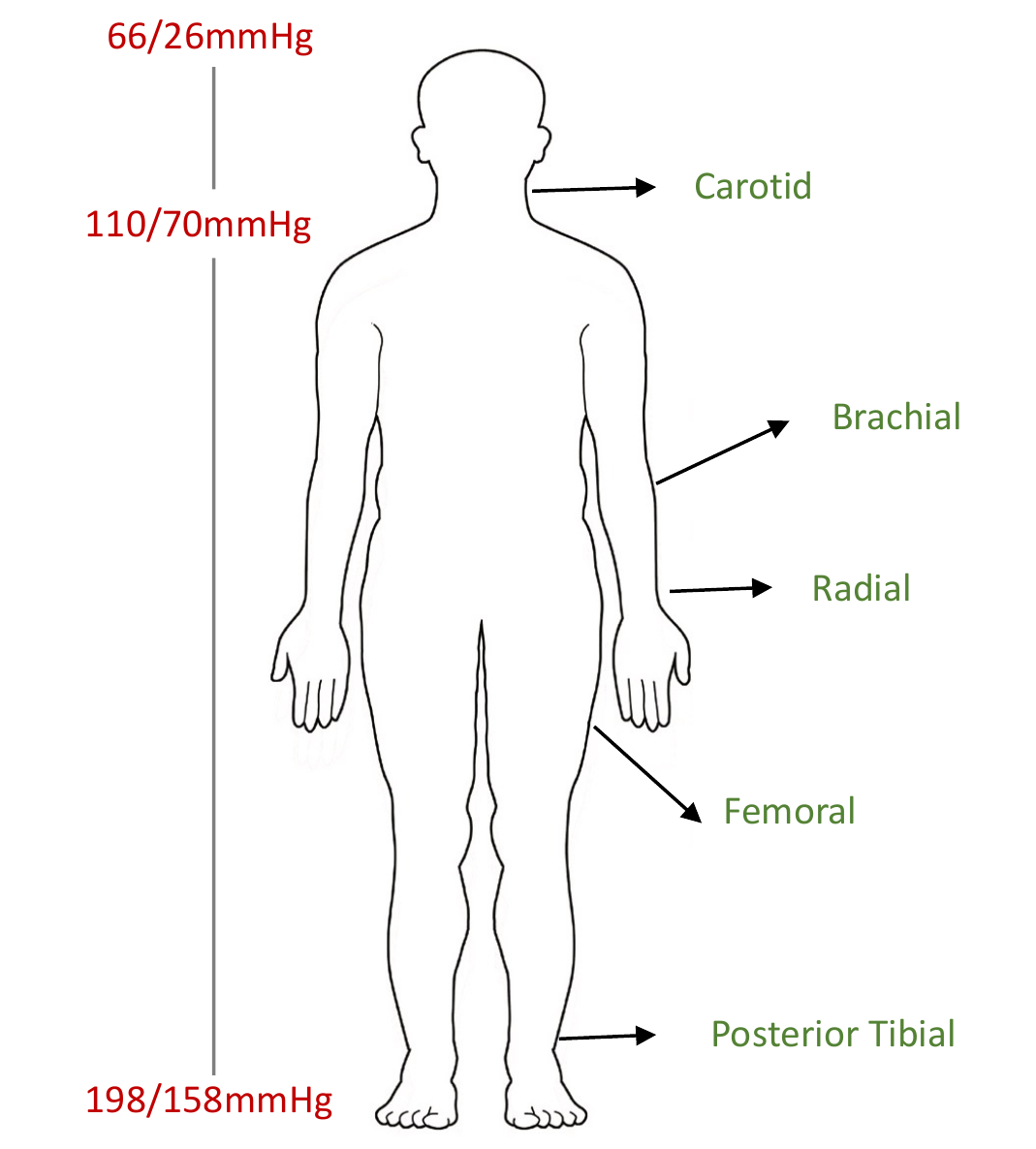}
\caption{Schematic showing blood pressure variations across arterial sites (carotid, brachial, radial, femoral, posterior tibial) influenced by distance from the heart, artery type, and gravity in a standing position \citep{Hinghofer-Szalkay2011-hu, Al-Qatatsheh2020-fh}.}\label{Fig: BPpoints}
\end{figure}

The brachial artery is commonly used for BP measurement \citep{muntner2019measurement, Bilo2017}. While wrist and finger monitors have become popular, SBP and DBP values differ across the arterial tree \citep{ogedegbe2010principles}. Fig.~\ref{Fig: BPpoints} depicts the influence of various body points on BP as blood flows through different arteries when upright. Table~\ref{Tab: place} summarizes studies comparing BP across different body points.

\begin{table}[tb]
\caption{Summary of studies investigating the impact of body points on recorded blood pressure values}\label{Tab: place}
\begin{center}
\begin{tabular}{rcccc}
\toprule
\textbf{Ref.} & \textbf{N}  &\textbf{Measuring place} & \textbf{SBP} & \textbf{DBP} \\
\hline
\citep{Kayrak2010-lk} & 45 & Upper Arm & 174.0$\pm$14.1 & 95.8$\pm$11.5 \\
 & & Wrist &  163.8$\pm$25.4  & 94.4$\pm$11.5 
\\\hline
\citep{Kayrak2010-lk} & 70 & Upper Arm & 168.3$\pm$18.4 & 83.4$\pm$9.4 \\
 & & Wrist &  159.2$\pm$18.5  & 83.2$\pm$10.5 
\\\hline
\citep{sareen2012comparison} & 250 & Arm & 127.7$\pm$15.7 & 80.7$\pm$11.2 \\
 & & Leg &  143.0$\pm$22.2  & 75.7$\pm$11.9 
\\
\bottomrule
\end{tabular}
\end{center}
\end{table}
\subsubsection{Body position} 

Body position significantly affects BP readings. Guidelines suggest measuring BP in a seated position with back support \citep{mancia20132013}. Research indicates BP values are higher when seated compared to lying down \citep{krzesinski2016diagnostic, privvsek2018epidemiological, Netea2003}. Specifically, sitting upright can increase DBP by up to 6.5\,mmHg compared to leaning back \citep{ogedegbe2010principles}. For accurate measurements, the BP cuff should be at the level of the patient's right atrium, regardless of position \citep{muntner2019measurement}. Table~\ref{Tab: bodyposition} summarizes studies on BP values in various body positions.

\begin{table}[tb]
\caption{Results of studies investigating blood pressure values in different positions of the subject during measurement}\label{Tab: bodyposition}
\begin{center}
\begin{tabular}{rcccc}
\toprule
\textbf{Ref.} & \textbf{N} &\textbf{Body position} & \textbf{SBP} & \textbf{DBP} \\
\hline
\citep{McMullen2011-qu} & 12  & Supine & 116.2$\pm$11.7 &  68.1$\pm$6.6 \\                                      & & Upright & 133.5$\pm$14.1 & 86.4 $\pm$8.7
\\\hline
\citep{Netea2003} & 57 & Sitting & 135.7$\pm$24.8 &  79.5$\pm$9.7 \\                                      & & Supine & 141.3$\pm$25.5 & 84.6 $\pm$10.5
\\\hline
\citep{ecser2007effect} & 157 & Sitting & 102.8$\pm$11.4 &  65.7$\pm$8.2 \\                                      & & Standing & 99.9$\pm$10.2 & 66.0$\pm$8.7\\                                      & & Supine & 107.9$\pm$10.7 & 66.9$\pm$9.6\\                                      & & Supine; legs crossed & 107.0$\pm$8.6 & 66.7$\pm$7.3
\\\hline
\citep{Chachula2020-nl} & 229 &Supine & 129.8$\pm$27.5 & 72.5$\pm$14.5 \\ 

& &Beach chair & 114.6$\pm$24.8 &  64.6$\pm$11.2  
\\\hline
\citep{netea1998does} & 245 & Sitting & 136.7$\pm$21.9 &
86.0$\pm$14
\\ & & Supine &135.5$\pm$20.3  &83.5$\pm$12.5 
\\\hline
\citep{cicolini2011differences} & 250 & Supine & 139.3$\pm$14.0 & 80.1$\pm$9.1 \\ 
& &Fowler’s & 138.1$\pm$13.8 & 81.9$\pm$9.4\\ 
& & Sitting & 137.2$\pm$13.7 & 83.0$\pm$9.6\\\hline
\citep{privvsek2018epidemiological} & 1298 & Sitting & 125.9$\pm$14.9 & 76.1$\pm$10.2
\\ & & Supine &  124.7$\pm$14.1 & 71.7$\pm$ 9.0
\\
\bottomrule
\end{tabular}
\end{center}
\end{table}

\subsubsection{Arm position}

Proper arm positioning during BP measurements is vital, with the arm ideally supported at heart level on a flat surface \citep{Netea2003, ogedegbe2010principles, Adiyaman2006}. Shifting the arm from horizontal to vertical can raise the pressure by 5-6\,mmHg due to hydrostatic changes \citep{ogedegbe2010principles}. Mariotti et al.\ explored the effects of arm positioning and postural hypotension during BP assessments \citep{Mariotti1987}, finding that incorrect arm positioning while standing resulted in overestimated BP readings. Table~\ref{Tab: Arm position} summarizes studies on BP values considering different arm positions.

\begin{table}[tb]
\caption{Results of studies investigating the impact of arm position on blood pressure values}\label{Tab: Arm position}
\begin{center}
\begin{tabular}{rcccc}
\toprule
\textbf{Ref.} & \textbf{N} & \textbf{Arm position} & \textbf{SBP} & \textbf{DBP} \\
\hline
\citep{Netea2003} & 57  & Arm high (at heart level) & 
137.4$\pm$29.0 &  78.2$\pm$14.4\\
                 &     & Arm low (on the bed)      &
142.1$\pm$28.0 & 82.1$\pm$13.4\\
\hline
\citep{netea1999arm} & 69  & Arm high (at heart level) & 
133.3$\pm$20.7 &  77.7$\pm$9.9\\
                 &     & Arm low (on chair arm-rest)  &
143.0$\pm$19.9 &  88.6$\pm$9.1\\
\bottomrule
\end{tabular}
\end{center}
\end{table}

\subsubsection{Leg position}
Proper leg positioning during BP measurements is crucial according to guidelines \citep{FosterFitzpatrick1999}. Studies show that crossed-leg positions yield higher BP readings than uncrossed legs or sitting with feet flat \citep{adiyaman2007effect, kallioinen2017sources}. Table~\ref{Tab: Leg position} provides a summary of studies on BP values with varying leg positions, emphasizing the importance of standardized leg positioning for accurate measurements.

\begin{table}[tb]
\caption{Results of studies investigating the impact of leg position on blood pressure values }\label{Tab: Leg position}
\begin{center}
\begin{tabular}{rcccc}
\toprule
\textbf{Ref.} & \textbf{N} &\textbf{Leg position} & \textbf{SBP} & \textbf{DBP} \\
\hline
\citep{Foster-Fitzpatrick1999-us} & 100 & Uncrossed & 146.5$\pm$18.6 & 80.9$\pm$11.2
\\
& & Crossed & 155.6$\pm$19.3 & 84.9$\pm$11.6  
\\\hline
\citep{Pinar2004-nq} & 238  & Uncrossed & 145.3$\pm$20.3 & 86.4$\pm$10.8
\\ & & Crossed & 153.6$\pm$20.2 & 92.1$\pm$11.2 
\\
\bottomrule
\end{tabular}
\end{center}
\end{table}

\subsubsection{Left vs right arm}
Clinical settings often show BP reading differences between the left and right arms \citep{fred2013accurate}. Guidelines suggest measuring BP in both arms on the first visit, using the arm with higher readings thereafter \citep{muntner2019measurement}. Conditions like coarctation of the aorta or upper-extremity arterial obstruction can cause significant BP variations between arms \citep{muntner2019measurement}. Table~\ref{Tab: LRarm} summarizes studies comparing BP measurements in both arms.

\begin{table}[tb]
\caption{Results of studies investigating the difference blood pressure values between right and left arms}\label{Tab: LRarm}
\begin{center}
\begin{tabular}{rccc}
\toprule
\textbf{Ref.} & \textbf{N} &\textbf{Right arm} & \textbf{Left arm}\\
\hline
\citep{Netea2003} & 57 &  SBP: 138.3$\pm$29.2 & SBP: 137.4$\pm$29.0
\\ & & DBP: 77.8$\pm$13.7 & DBP:78.2$\pm$14.4 
\\\hline
\citep{netea1999arm} & 69  &SBP: 133.3$\pm$20.7 & SBP: 131.8$\pm$19.1\\ && DBP: 77.7$\pm$9.9 & DBP: 78.0$\pm$9.9 
\\\hline
\citep{lane2002inter} & 400  &  SBP: 131.2$\pm$21.0 & SBP: 129.4$\pm$21.2
\\ && DBP: 76.8$\pm$11.9 & DBP: 77.1$\pm$12.6  
\\
\bottomrule
\end{tabular}
\end{center}
\end{table}

\subsubsection{Cuff size and tightness}
Using the correct cuff size is vital for accurate BP measurement \citep{bur2000accuracy}. Small cuffs can overestimate pressure, which is a frequent error \citep{Palatini2018, ogedegbe2010principles}. Larger cuffs tend to report lower BP values \citep{kallioinen2017sources}. For children, cuff size selection is challenging. The BHS suggests three sizes: 4$\times$13\,cm, 8$\times$18\,cm, and 12$\times$35\,cm (adult cuff) based on arm circumference \citep{ogedegbe2010principles}.

In obese individuals, selecting the right cuff size is essential to ensure the brachial artery is accurately compressed, yielding reliable BP readings \citep{muntner2019measurement}. Many obese individuals have tronco-conical arms, further complicating accurate BP measurement \citep{Palatini2018}. For these patients, conical cuffs can offer more accurate BP readings \citep{muntner2019measurement}. Table~\ref{Tab: Cuff SIZE} presents studies exploring the influence of cuff size on BP readings.

When measuring BP, the cuff should be securely placed around the upper arm without any clothing interference, ensuring even snugness from top to bottom. The tightness can be gauged by fitting one finger easily and two fingers with comfort between the cuff's top and bottom. Achieving the correct cuff tightness is vital for reliable and consistent BP readings \citep{muntner2019measurement}.

\begin{table}[tb]
\caption{Results of studies investigating the effect of arm position on blood pressure values}\label{Tab: Cuff SIZE}
\begin{center}
\begin{tabular}{rcccc}
\toprule
\textbf{Ref.} & \textbf{N} &\textbf{Cuff size (cm)} & \textbf{SBP} & \textbf{DBP} \\
\hline
\citep{Bakx1997-pf} & 130  &  13 $\times$ 36 & 125.1$\pm$19.2 & 75.4$\pm$12.4\\ 
& &  16 $\times$ 23 & 123.7$\pm$19.7 & 74.4$\pm$13.2\\ 
&  &  13 $\times$ 23 & 127.2$\pm$19.2 & 77.0$\pm$12.8
\\
\bottomrule
\end{tabular}
\end{center}
\end{table}

\subsubsection{Rest period before measuring BP}
Studies have investigated the effects of rest durations before taking BP measurements. Their findings indicate that not resting sufficiently before a measurement can lead to elevated SBP and DBP readings. A rest period ranging from 10 to 16 minutes showed a modest decrease in SBP and a slight drop in DBP \citep{Nikolic2013, SALA2006}. However, the exact duration of rest needed to compensate the effects of prior physical activity remains uncertain. More research is needed to specify the ideal rest period for precise BP readings \citep{kallioinen2017sources}. A summary of studies exploring the influence of rest durations on BP measurements can be found in Table~\ref{Tab: rest}.

\begin{table}[tb]
\caption{Results of studies investigating the effect of resting before measuring BP}\label{Tab: rest}
\begin{center}
\begin{tabular}{rcccc}
\toprule
\textbf{Ref.} & \textbf{N} & \textbf{Before resting} & \textbf{After resting} & \textbf{Resting Time}\\
\hline
\citep{Boivin2014-ss} & 52 & SBP: 127.9$\pm$12.0 & SBP: 121.5$\pm$10.9 & 5 minute\\ & & DBP: 78.0$\pm$8.7 & DBP: 76.0$\pm$9.0 & 
\\
\bottomrule
\end{tabular}
\end{center}
\end{table}

\subsubsection{Number of measurements}
For many individuals, the first BP reading taken in a clinical setting tends to be higher than the readings that follow. Research has explored this pattern by conducting three consecutive BP measurements \citep{muntner2019measurement}. Findings indicated that when only the first measurement was considered, approximately 35\% of adults exhibited an SBP range of 140--159\,mmHg and a DBP range of 90--99\,mmHg. However, when the average of all three measurements was taken into account, most participants registered SBP/DBP values below the 140/90\,mmHg benchmark \citep{muntner2019measurement}. Thus, relying exclusively on the initial reading can result in over-diagnosing hypertension, emphasizing the necessity of multiple readings to ensure an accurate diagnosis \citep{MANCIA1983}.

\subsubsection{Clothing}
Healthcare professionals are advised to measure BP by fully exposing the cuff on the upper arm. However, it is a common practice to measure BP by rolling up the sleeve or placing the cuff over sleeves. Table~\ref{Tab: Clothing} presents the findings from various studies that have examined the impact of clothing on BP values, demonstrating a bias in the BP readings due to clothing. 
\begin{table}[tb]
\caption{Results of studies investigating the impact of wearing clothing on the arm during blood pressure measurements on reported values}\label{Tab: Clothing}
\begin{center}
\begin{tabular}{rclcc}
\toprule
\textbf{Ref.} & \textbf{N}  &\textbf{Measuring place} & \textbf{SBP} & \textbf{DBP} \\
\hline
\citep{Ki2013-ai} & 141 & Sleeved & 128.5$\pm$10.6 & 80.7$\pm$6.3 \\
& & Rolled sleeves & 128.3$\pm$11.1 & 80.9$\pm$6.3 \\
& & Bare arm & 128.4$\pm$10.8 & 80.8$\pm$6.0 
\\
\bottomrule
\end{tabular}
\end{center}
\end{table}

\section{Cuff-less blood pressure technologies}
Significant efforts have been made to develop novel and versatile methods for measuring BP. Traditional BP measurement techniques involving cuffs have several inherent limitations \citep{peter2014review}, such as: (1) time-consuming procedures; (2) patient discomfort due to frequent cuff inflation, as seen in ABPM devices used for recording BP during the day and night \citep{Pickering2008}; (3) impracticality for continuous BP measurement in specific medical settings such as heart surgical units and acute burns cases; (4) the necessity of allowing sufficient time between successive BP measurements to enable the blood vessels beneath the cuff to return to their baseline state and to prevent vessel collapse due to cuff pressure. To address these limitations, cuff-less BP measurement methods have been introduced, which can be categorized into three groups \citep{mousavi2018designing, Sharma2017-ng, rastegar2020non}: tonometry, volume clamps, and pulse wave velocity. These innovative approaches offer the potential to overcome the drawbacks associated with traditional cuff-based BP measurement techniques, paving the way for more efficient and patient-friendly BP monitoring solutions.

The pulse wave velocity (PWV) method of measuring BP was invented by Moens and Korteweg \citep{Cole2007-rx}. They defined a fundamental relationship between vascular elasticity and pulse wave velocity in the artery, known as the Moens-Korteweg equation\citep{geddes2013handbook,peter2014review}:
\begin{equation}\label{Eq: Moens-Korteweg}
\text{PWV} = \sqrt{\frac{E\cdot h}{\rho\cdot D}}
\end{equation}
where $h$ represents the thickness of the vessel wall, $\rho$ is the blood density, and $D$ is the vessel's inner diameter. The parameter $E$ denotes Young's modulus of elasticity, which indicates the vessel wall's elasticity. Geddes empirically defined $E$ as follows \citep{peter2014review}:
\begin{equation}\label{Eq: Geddes}
E = E_{0}e^{\alpha \cdot \text{BP}}
\end{equation}
where $E_{0}$ represents the modulus of elasticity at a pressure of 0\,mmHg, $\alpha$ is a constant related to the vessel (typically ranging from 0.016 to 0.018 mmHg$^{-1}$), $e$ is the Euler number, and $\text{BP}$ is the blood pressure. The Moens-Korteweg equation and Geddes' formulation are fundamental in enabling the utilization of PWV as a valuable and non-invasive method for assessing BP and vascular elasticity. Combining \eqref{Eq: Moens-Korteweg} and \eqref{Eq: Geddes} leads to a compact relation between $\text{BP}$ and $\text{PWV}$:
\begin{equation}\label{Eq: M-K3}
\text{BP}= \frac{1}{\alpha}\ln\frac{\rho \cdot D \cdot \text{PWV}^2}{h\cdot E_{0}}
\end{equation}

Several methods are used to measure PWV. Pulse transit time $\text{(PTT)}$ is the most well-known indirect method of calculating PWV \citep{mukkamala2015toward}. The relationship between PTT and PWV is defined as follows \citep{peter2014review}:

\begin{equation}\label{Eq: PPTandPWV}
\text{PWV}=\displaystyle\frac{d}{\text{PTT}}
\end{equation}
$d$ is the distance between the heart and a specific location where the blood flows, and $\text{PTT}$ is the time it takes for the blood pulse to propagate the distance $d$. $\text{PTT}$ is calculated using different sensors and bio-signals, including \citep{Mukkamala2018-fi}: photoplethysmography (PPG) and the output signal of the Hall sensor \citep{Nam2013-ez}; PPG signal and modulated magnetic signature of the blood \citep{zhang2016mechanism}; PPG and ballistocardiography \citep{chen2013noninvasive}; PPG and impedance plethysmography \citep{liu2017cuffless}; PPG and electrocardiography \citep{chen2019non}; One or two PPG recordings \citep{mousavi2019blood}.

The parameter $d$ is not easy to find (accurately) in practice. Therefore $\text{PWV}$ and subsequently $\text{BP}$ in \eqref{Eq: PPTandPWV} are not accurately found. To address this challenge, cuff-less BP monitoring devices are increasingly integrating artificial-intelligence (AI) algorithms to learn crucial and complex features of the cardiovascular system \citep{ding2016continuous, rastegar2020non}. Ongoing research focuses on exploring how to extract the most relevant features from signals like PTT to model the cardiovascular system and to predict BP values effectively. However, it is essential to acknowledge that these studies are still in the prototype stage and must be validated on large cohorts to meet medical standards, before being used in clinical settings \citep{rastegar2020non}. Nonetheless, advances in this domain are promising for enhancing BP monitoring and improving patient care.


When developing cuff-less methods, cuff-based BP databases are typically used as references to demonstrate their ``substantial equivalency'' with an existing technology, as required for regulatory clearances like FDA 510(k) premarket approval \citep{510k,alpert2014public, donawa2010continuing,alpert2017can}. If standard requirements are not met during cuff-based BP acquisition, the calibration models for new technologies will train with biased recordings. This affects the long-term reliability of these technologies, potentially causing deviations between predicted and actual BP values. It is essential to address these biases for accurate measurements with cuff-less technologies, warranting further research to enhance their performance and reliability.

\section{Future perspectives: Using machine learning for bias removal and individualized BP level risk assessment}\label{sec: Future}
In our study, we investigated BP and various measurement and subject-wise factors that impact the accurate collection and interpretation of BP. However, given the vast BP literature, it is unfeasible to comprehensively examine every potential source of bias. To note, as of July 31, 2023, PubMed alone reports 676,855 incidences of the keyword ``blood pressure'' in its databases (obtained using the Entrez Direct (EDirect) Linux tools from the NCBI \citep{EntrezDirect}). In addition, there are billions of encounters of BP records on electronic health records (EHRs) worldwide. The one-by-one investigation of these rich resources is infeasible. We propose that future research can leverage machine learning and natural language processing techniques (especially the recent advances in large language models) to mine these massive data- and literature-driven resources in a more systematic manner. This can result in more conclusive findings regarding individual-wise BP-related risk factors across different demographic groups. Open source codes for BP-related studies is another requirement. Schwenck et al.\ recently released a versatile open-source BP analysis and visualization toolkit in R \citep{Schwenck2022}. Beyond visualization tools, rigorous statistical frameworks and models are required for building machine learning pipelines. In the sequel, we propose a stochastic framework that can be used in future research to use machine learning algorithms in BP-related studies.

\subsection{A machine learning framework for BP bias analysis}
The problem of BP measurement can be formulated as
\begin{equation}
    y_k = \text{BP}_k + e_k
\label{eq:BP_data_model}    
\end{equation}
where $y_k$ are the reported BP values in different trials/sessions, $\text{BP}_k$ is the `true' BP value, and $e_k$ denotes the total measurement errors attributed to nonstandard devices or measurement errors. Both the BP and its error can be considered random variables, with presumed (time-variant) distributions. With this data model, the problem of accurate BP measurement and bias removal can be formulated as a classical estimation problem that could be addressed using standard techniques such as least squares, maximum likelihood or Bayesian estimation, where the latter two benefit from prior distributions of the BP and measurement errors.

$e_k$ is device and technology dependent. Complying with standard BP measurement techniques and averaging over multiple trials ($k=1, \ldots, K$) can mitigate the impact of this term, resulting in a more accurate measurement of the BP.

Even with accurate measurements ($e_k\approx 0$), $\text{BP}_k$ is a random variable, which fluctuates over time and across trial. A comprehensive machine learning framework should benefit from the demographic-specific distributions of $\text{BP}_k$, namely:
\begin{equation}
    f(\text{BP}_k|\mathbf{p})
\end{equation}
which is the conditional distribution of the BP (SBP, DBP, or both) given all the demographic and comorbid factors parameterized by the vector $\mathbf{p}$ (sex, race, age, background medical conditions, comorbidities, etc.).

\begin{figure}[tb]
  \begin{adjustwidth}{-\extralength}{0cm}
  \centering
  \hfill
  \begin{subfigure}{0.25\columnwidth}
   \includegraphics[width=1\columnwidth]{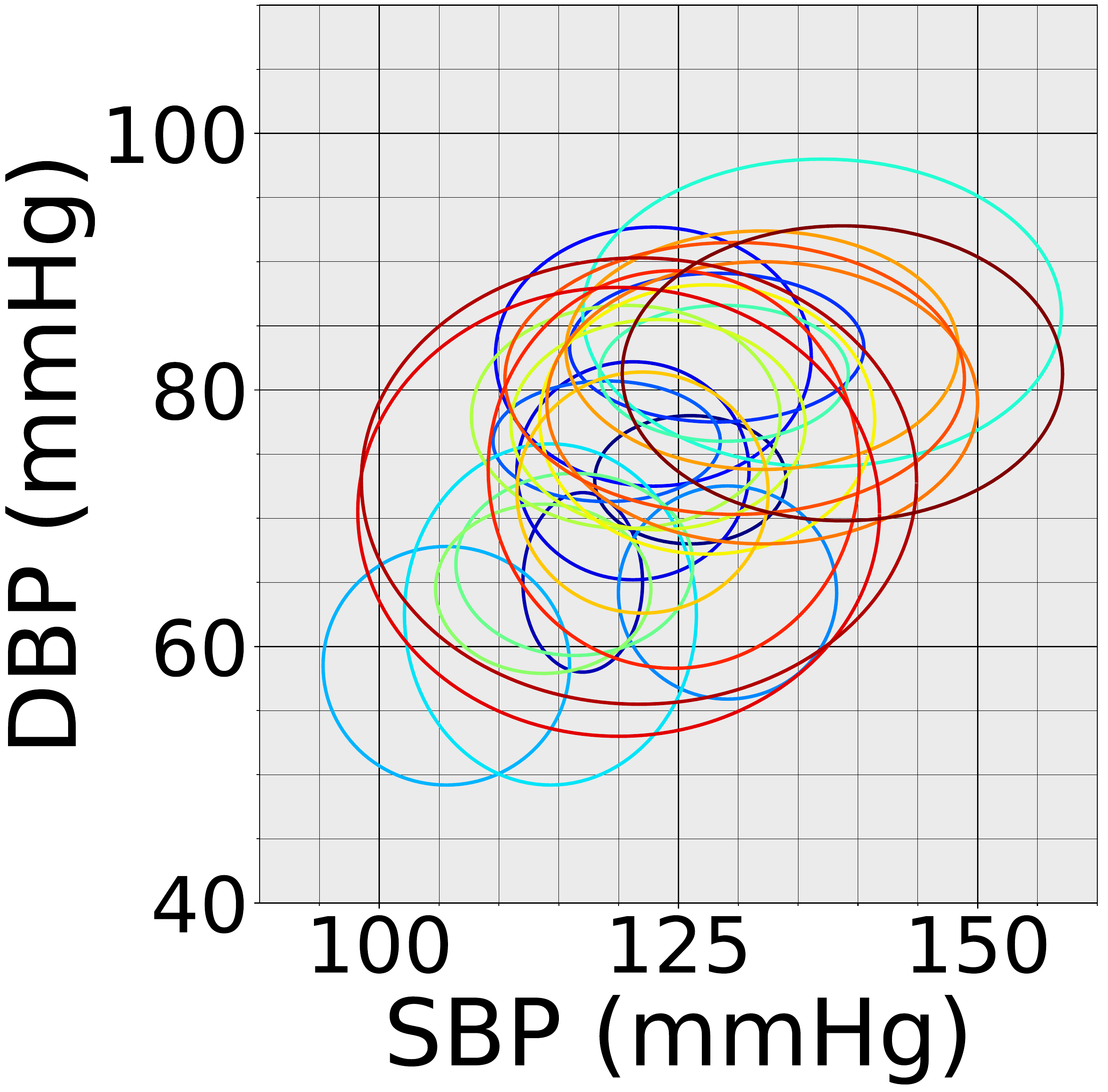}
    \caption{Male}
    \label{Fig:MaleandBP}
  \end{subfigure}%
  \hfill
  \begin{subfigure}{0.25\columnwidth}
   \includegraphics[width=1\columnwidth]{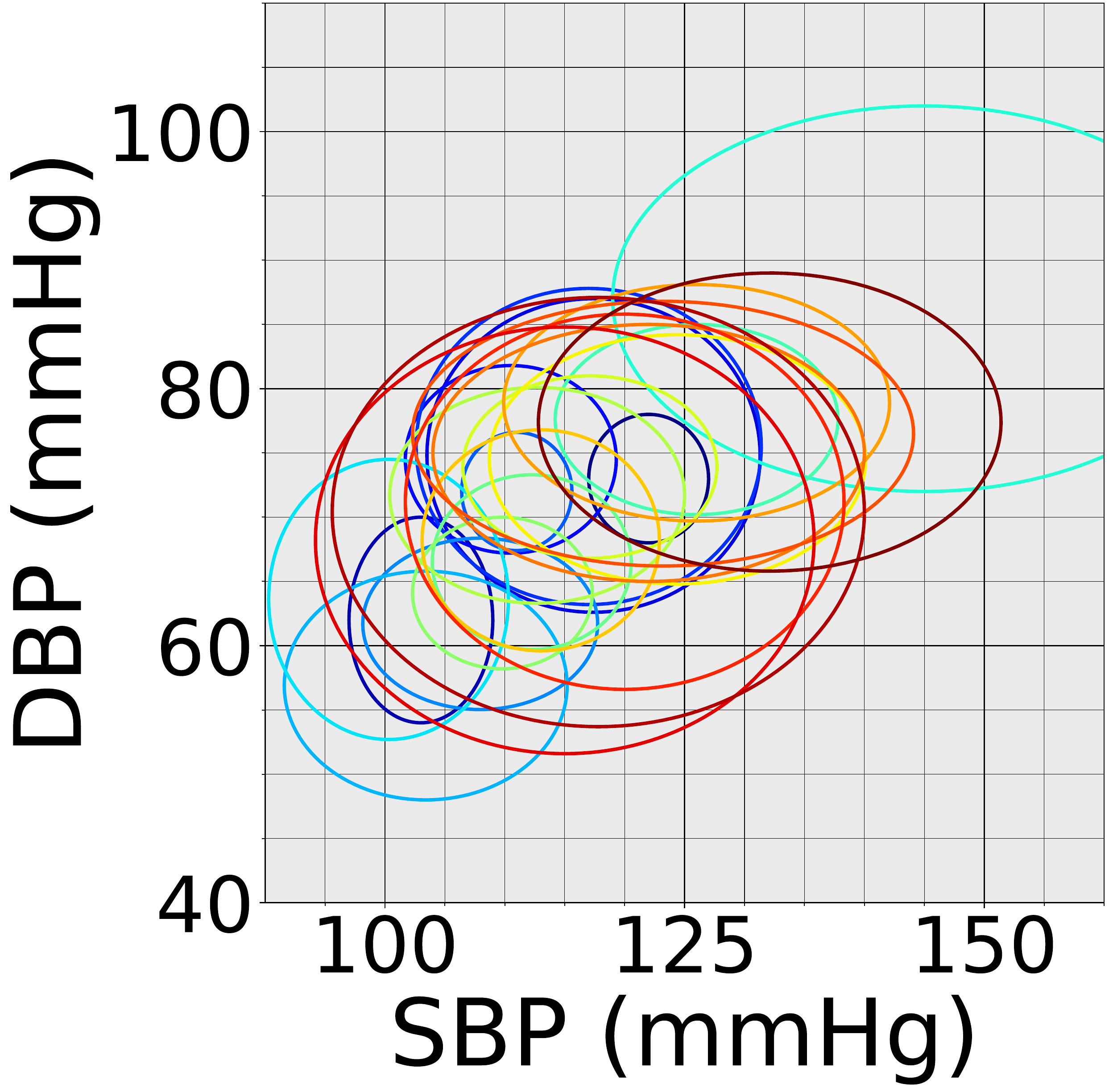}
    \caption{Female}
    \label{Fig:FemaleandBP}
  \end{subfigure}%
  \hfill
  \begin{subfigure}{0.25\columnwidth}
   \includegraphics[width=1\columnwidth]{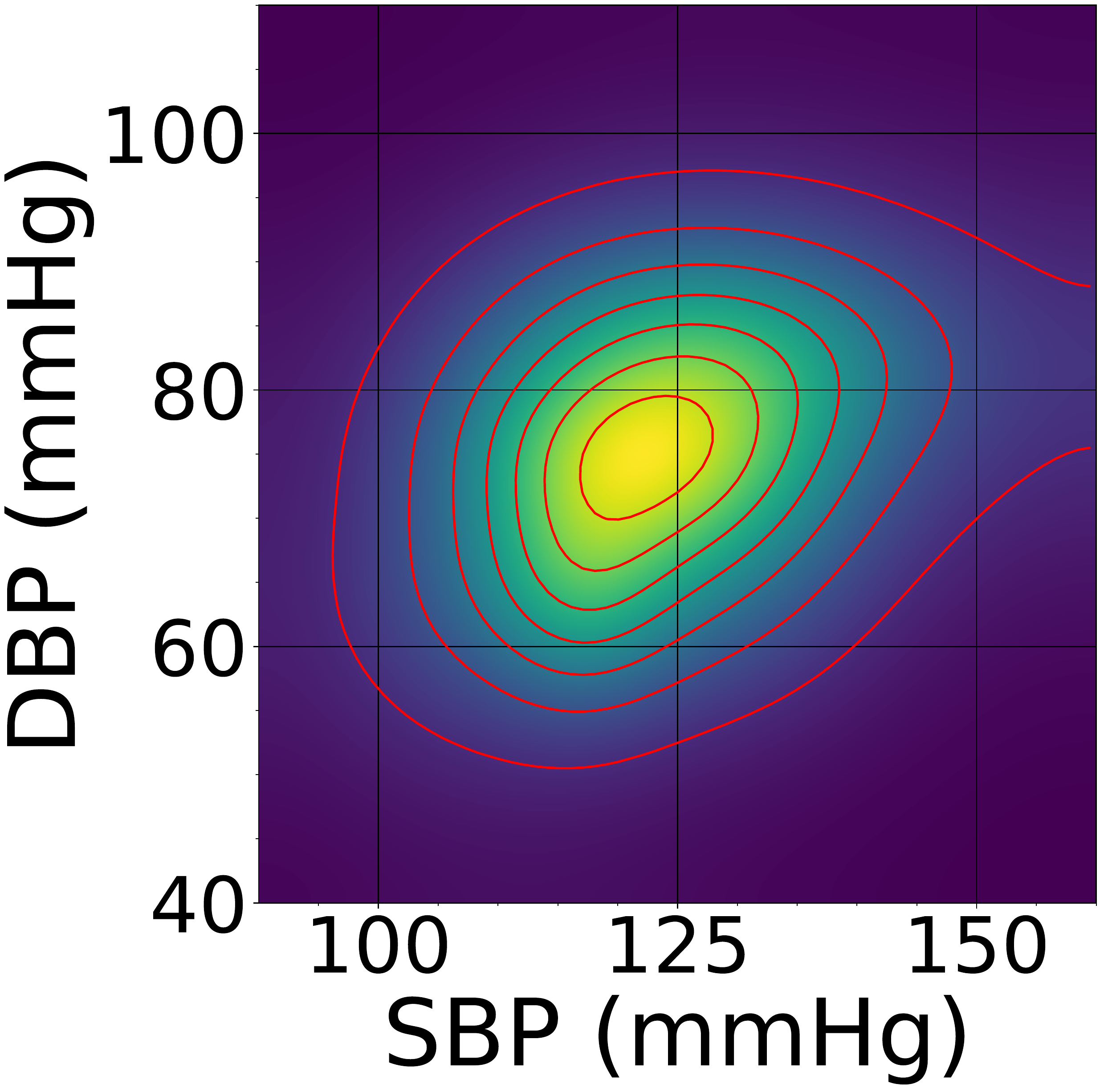}
    \caption{Male GMM}
    \label{Fig:MaleGMM}
  \end{subfigure}%
  \hfill
  \begin{subfigure}{0.25\columnwidth}
   \includegraphics[width=1\columnwidth]{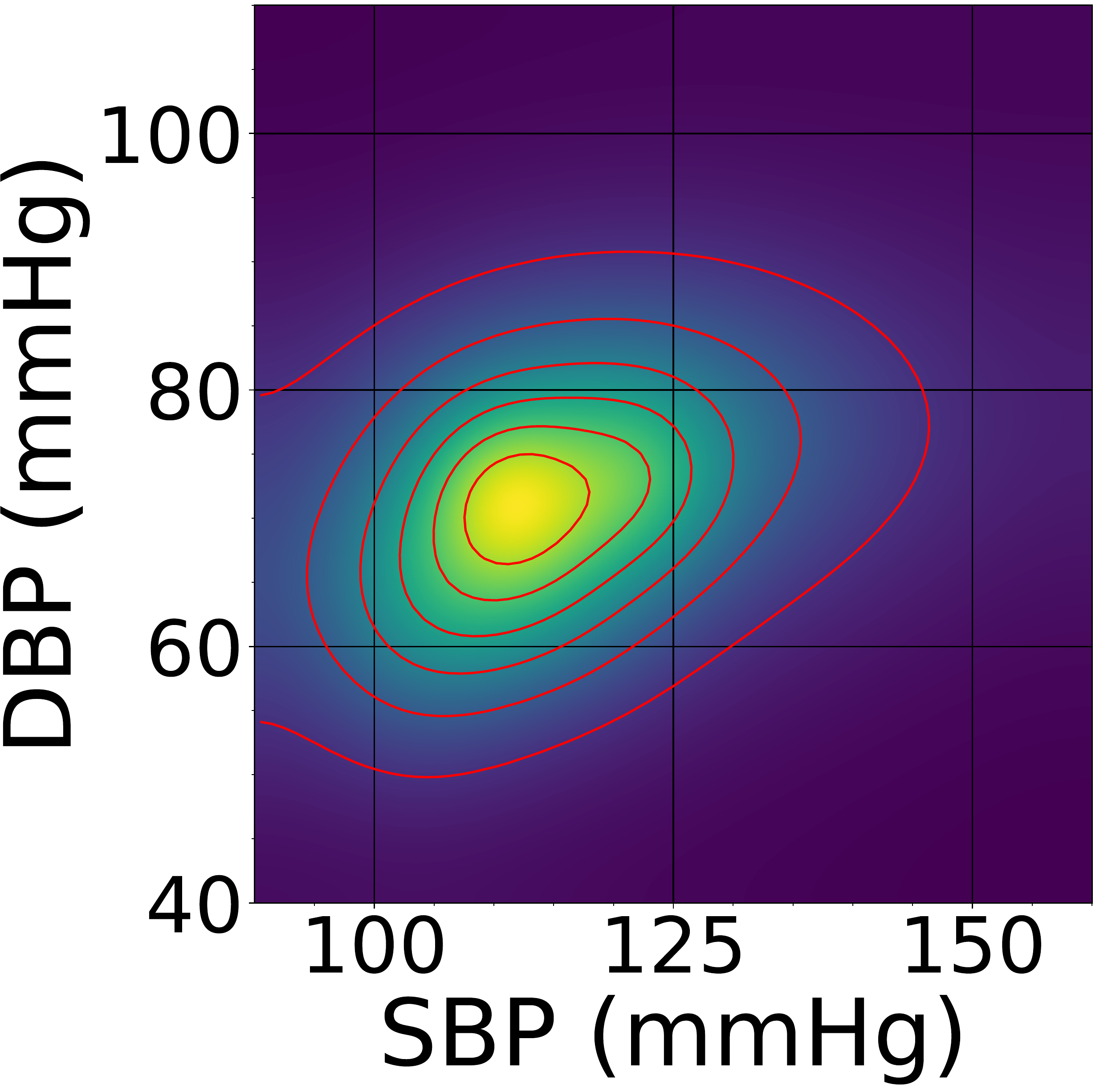}
    \caption{Female GMM}
    \label{Fig:FemaleGMM}
  \end{subfigure}
  \hfill
  \caption{Comparison of male (Fig.~\ref{Fig:MaleandBP}) vs. female (Fig.~\ref{Fig:FemaleandBP}) blood pressure values from various studies. Each ellipse represents a study, centered on the mean SBP and DBP with horizontal and vertical radii as the corresponding standard deviations. Gaussian mixture model (GMM) distributions are fitted over these reports in Figs.~\ref{Fig:MaleGMM} and \ref{Fig:FemaleGMM}.}
  \label{Fig:GenderandBP}
  \end{adjustwidth}
\end{figure}

\begin{figure}[tb]
\begin{adjustwidth}{-\extralength}{0cm}
\centering
  \hfill
\begin{subfigure}{0.25\columnwidth}
\includegraphics[width=1\columnwidth]{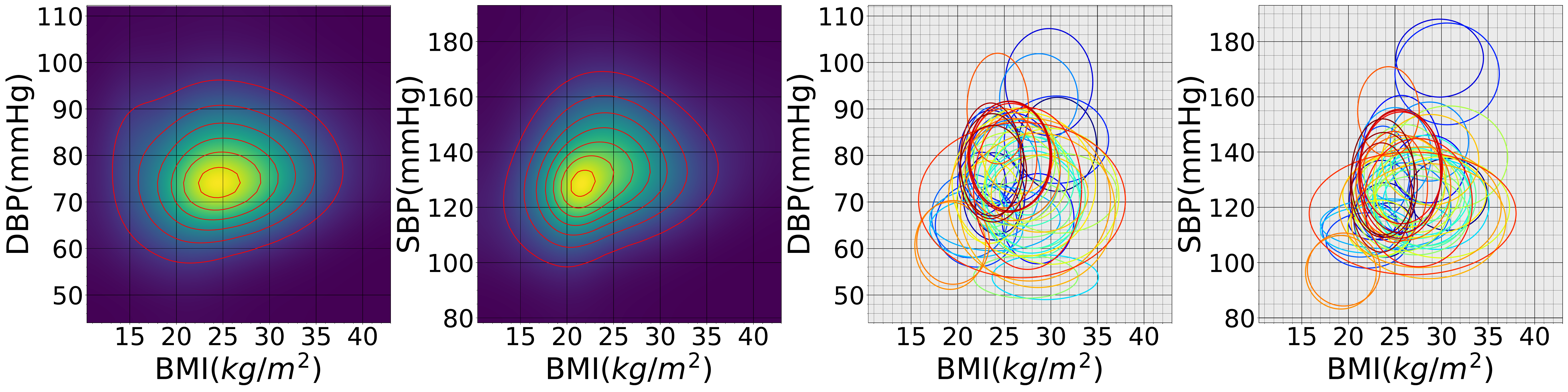}
  \caption{DBP vs BMI}
  \label{Fig: DBP_vs_BMI}
\end{subfigure}%
  \hfill
\begin{subfigure}{0.25\columnwidth}
  \includegraphics[width=1\columnwidth]{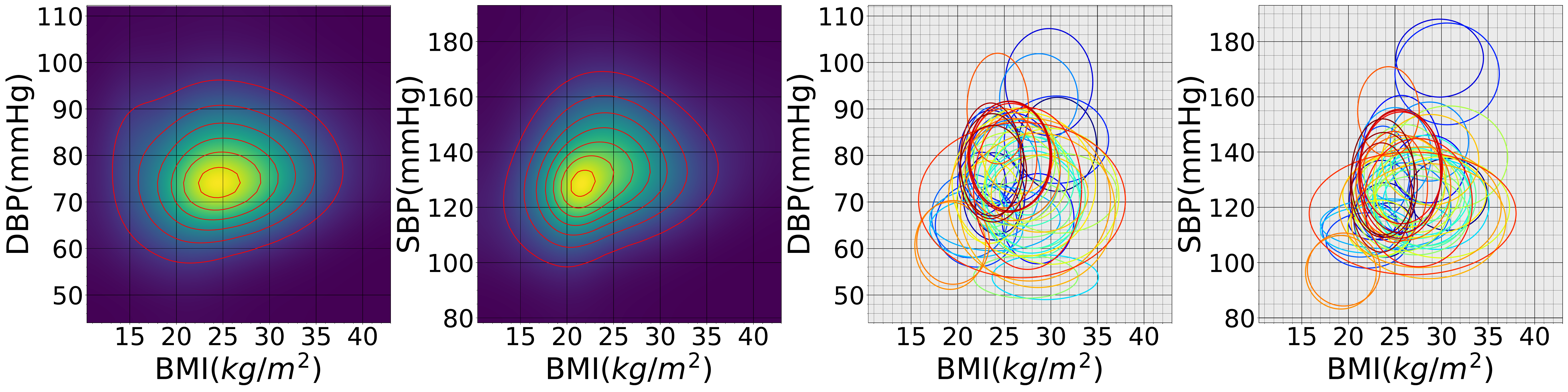}
  \caption{SBP vs BMI}
  \label{Fig: SBP_vs_BMI}
\end{subfigure}
  \hfill
\begin{subfigure}{0.25\columnwidth}
  \includegraphics[width=1\columnwidth]{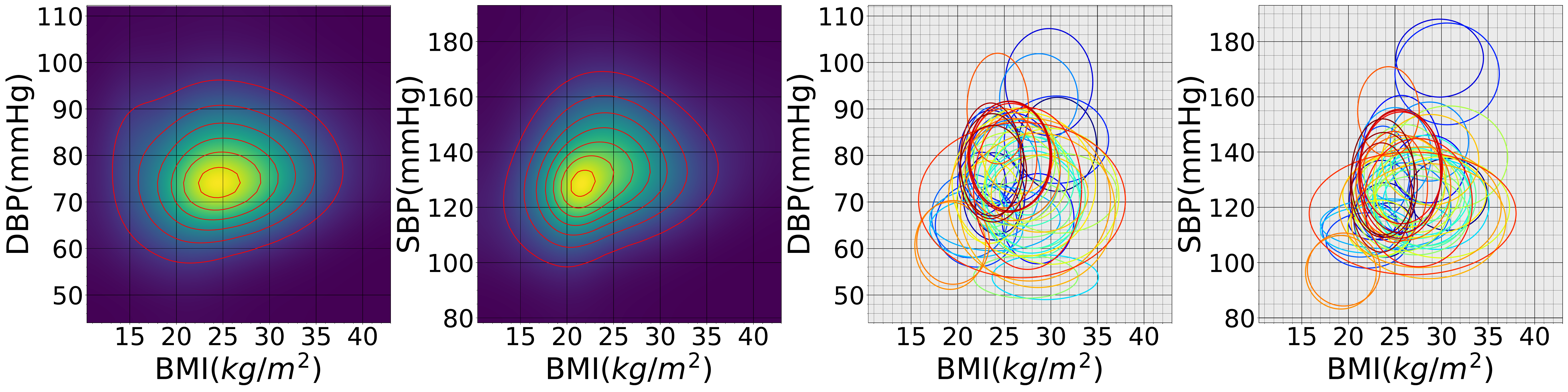}
  \caption{DBP vs BMI}
  \label{Fig: HEATMAP_DBP_BMI}
\end{subfigure}%
  \hfill
\begin{subfigure}{0.25\columnwidth}
  \includegraphics[width=1\columnwidth]{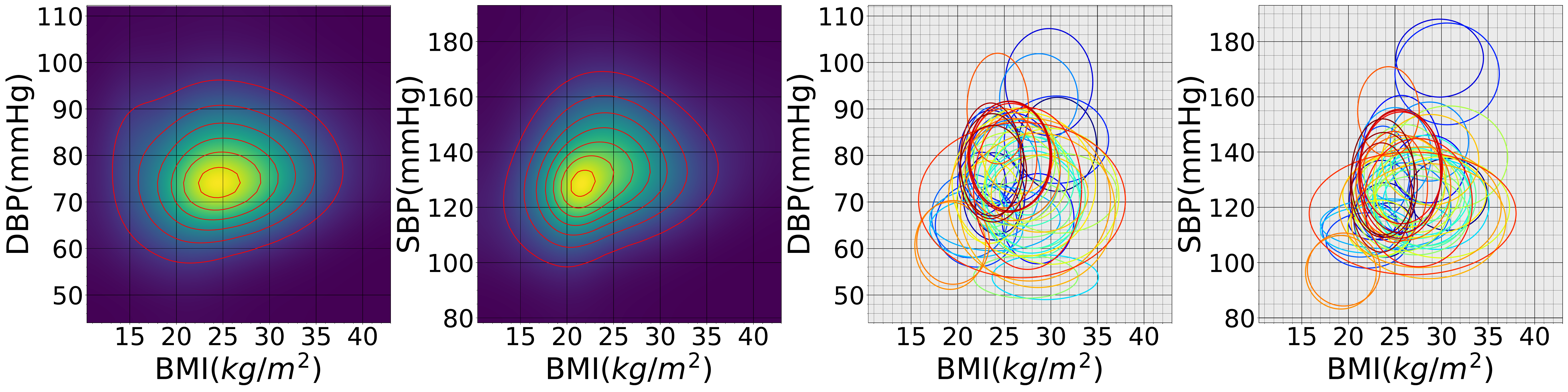}\caption{SBP vs BMI}
  \label{Fig: HEATMAP_SBP_BMI}
\end{subfigure}
  \hfill
\caption{Comparison of blood pressure values by BMI from various studies. Each ellipse represents a study, centered on mean BMI and BP with horizontal and vertical radii as their respective standard deviations.}
\label{Fig: BMIandBP}
\end{adjustwidth}
\end{figure}

We propose that our current survey of the BP literature --- and more systematic future surveys that could benefit from massive EHR registries combined with NLP technologies for BP literature survey --- can be used to estimate $f(\text{BP}_k|\mathbf{p})$ across different demographic factors. As proof of concept, graphical representations of the BP data across sex and BMI, from Tables~\ref{Tab: sex} and \ref{Tab: BMI}, are shown in Fig.~\ref{Fig:GenderandBP} and Fig.~\ref{Fig: BMIandBP}, respectively. In Fig.~\ref{Fig:GenderandBP}, each ellipse corresponds to an individual study from the studied literature, where each ellipse is centered at the mean SBP and DBP, and the horizontal and vertical radii of the ellipses equal the standard deviation of the reported SBP and DBP, respectively. Similar elliptic shapes can be constructed from massive EHR data, while controlling over individualized and demographic factors. For multi-site studies this information can be used to construct multi-modal distributions of $f(\text{BP}_k|\mathbf{p})$. For instance, assuming that the reported BP values follows a Normal distribution (which is an accurate assumption for population-wide studies), the multi-site data can be effectively modeled using Gaussian mixture models (GMMs) to capture underlying clusters or sub-populations \citep{reynolds2009gaussian}. The heatmaps in Figs.~\ref{Fig:MaleandBP}, and \ref{Fig:FemaleandBP}, \ref{Fig: HEATMAP_SBP_BMI} and \ref{Fig: HEATMAP_DBP_BMI} correspond to GMM distributions fitted over the ensemble of the hereby studied BP reports across sex and BMI.

An interesting observation in these figures is that while the literature on BP merely report the mean and standard deviation of SBP and DBP as independent random variables, there is a trend of correlation between SBP and DBP, which is neglected in the BP literature. BP data from EHR can additionally be used to identify the correlations between SBP and DBP.

\begin{figure}[tb]
\begin{adjustwidth}{-\extralength}{0cm}
\centering
\includegraphics[trim={1cm 2cm 1cm 1cm},clip,width=1.3\textwidth]{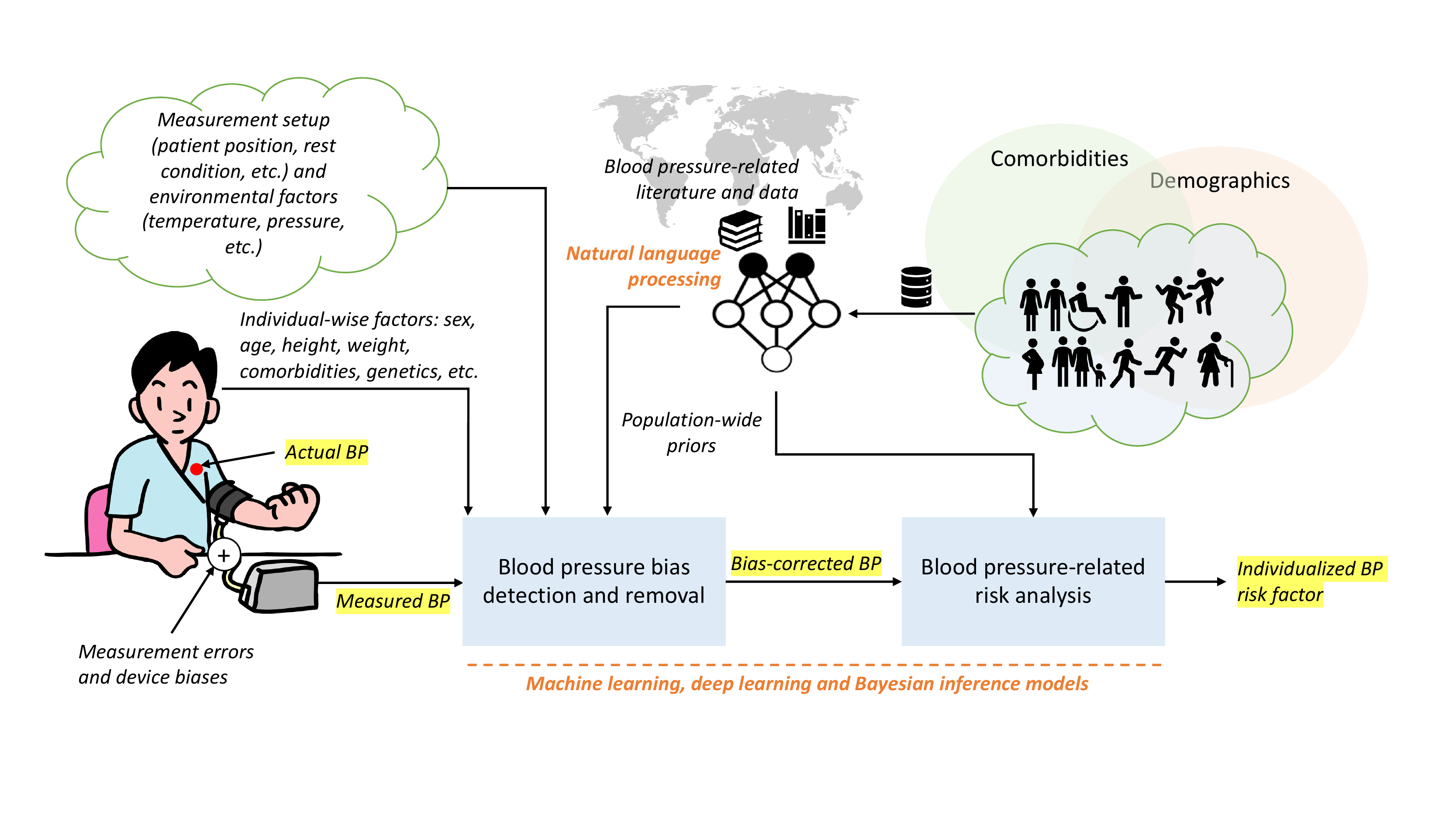}
\caption{The future of blood pressure (BP) monitoring. AI-assisted and ML-based algorithms are anticipated to be integrated in the new generation of cuff-based BP measurement devices to detect bias, correct BP measurements, and provide individualized BP-related risk-factors.}\label{Fig: CorrectedBP}
\end{adjustwidth}
\end{figure}

Future research can integrate these distributions and the data model in \eqref{eq:BP_data_model} to provide Bayesian estimates of the BP on a subject-wise bases. These distributions can also be integrated with EHR data and clinical outcomes to train machine learning technologies that quantify the risk of hypertension and BP-related complications of individuals across different demographics (see Fig.~\ref{Fig: CorrectedBP}), resulting in more accurate assessment of BP-based diagnoses \citep{Praveen2018}. With such a technology, we anticipate that we will be able to phrase a patient's BP-associated risk with such terms: \textit{``A 55-yr-old Hispanic female with a body mass index of 34.2, a consistent in-clinic cuff-based systolic BP $>$ 129\,mmHg, a diastolic BP $>$ 85\,mmHg while sitting, and a history of diabetes is 15\% at risk of stroke in the next 12 years (p-value $<$ 0.05)''}.

\section{Conclusion}\label{sec: conclusion}
BP is a crucial vital sign for monitoring health and clinical decision making. In recent years, the adoption of cuff-based BP technologies has surged, primarily driven by the growing popularity of ambulatory and home-based BP monitoring and its widespread use in medical centers and hospitals. Accurate BP values are essential, as any discrepancies between reported and actual BP measurements can lead to misinterpretations and mis-treatments. In this study, we highlighted the notion of ``bias'' and the factors that may result in bias in cuff-based BP monitoring across various social groups. The survey demonstrated how reported BP values can be diverse due to individualized and demographic factors, or become inaccurate due to a failure to adhere to BP measurement standards. Given these limitations, the development of a new generation of cuff-based BP devices supported by artificial-intelligence (AI) and machine learning (ML) techniques is highly demanded. Integrating AI and ML techniques into these devices is a promising approach to identify and correct bias and to customize normal and abnormal BP ranges on an individualized level; thereby improving the accuracy of BP measurements and clinical decision-making. In conclusion, addressing the issue of bias in cuff-based BP devices is essential for advancing patient care and ensuring reliable BP data for medical decision-making. Developing ML-based devices that can detect and correct bias will undoubtedly be a valuable contribution to the field, enhancing the overall effectiveness and reliability of BP monitoring systems. By focusing on improving BP monitoring precision, we can significantly improve patient outcomes.

\section*{Author Contributions}

\section*{Funding}
This research received no external funding. 

\section*{Institutional Review Board Statement}
The Emory University IRB has approved this study under the study protocol number STUDY00006568.

\section*{Informed Consent Statement}
Not applicable.

\section*{Data Availability Statement}
Not applicable.

\section*{Conflicts of Interest}
The authors declare no conflict of interest.

\reftitle{References}

\bibliography{References}

\begin{thebibliography}{999}

\bibitem[{World Health Organization}(2021)]{who}
{World Health Organization}.
\newblock {Cardiovascular Diseases (CVDs)}.
\newblock
  \href{https://www.who.int/news-room/fact-sheets/detail/cardiovascular-diseases-(cvds)}{https://www.who.int/news-room/fact-sheets/detail/cardiovascular-diseases-({cvds})},
   2021.
\newblock Accessed: 2021-06-21.

\bibitem[Ahmad and Anderson(2021)]{ahmad2021leading}
Ahmad, F.B.; Anderson, R.N.
\newblock {The leading causes of death in the US for 2020}.
\newblock {\em Jama} {\bf 2021}, {\em 325},~1829--1830.
\newblock {\url{https://doi.org/10.1001/jama.2021.5469}}.

\bibitem[Leonardi-Bee et~al.(2002)Leonardi-Bee, Bath, Phillips, and
  Sandercock]{LeonardiBee2002}
Leonardi-Bee, J.; Bath, P.M.; Phillips, S.J.; Sandercock, P.A.
\newblock Blood Pressure and Clinical Outcomes in the International Stroke
  Trial.
\newblock {\em Stroke} {\bf 2002}, {\em 33},~1315--1320.
\newblock {\url{https://doi.org/10.1161/01.str.0000014509.11540.66}}.

\bibitem[Lawes et~al.(2004)Lawes, Bennett, Feigin, and Rodgers]{Lawes2004}
Lawes, C.M.; Bennett, D.A.; Feigin, V.L.; Rodgers, A.
\newblock Blood Pressure and Stroke.
\newblock {\em Stroke} {\bf 2004}, {\em 35},~776--785.
\newblock {\url{https://doi.org/10.1161/01.str.0000116869.64771.5a}}.

\bibitem[Vasan et~al.(2001)Vasan, Larson, Leip, Evans,
  O{\textquotesingle}Donnell, Kannel, and Levy]{Vasan2001}
Vasan, R.S.; Larson, M.G.; Leip, E.P.; Evans, J.C.; O{\textquotesingle}Donnell,
  C.J.; Kannel, W.B.; Levy, D.
\newblock {Impact of High-Normal Blood Pressure on the Risk of Cardiovascular
  Disease}.
\newblock {\em New England Journal of Medicine} {\bf 2001}, {\em
  345},~1291--1297.
\newblock {\url{https://doi.org/10.1056/nejmoa003417}}.

\bibitem[Zhou et~al.(2017)Zhou, Bentham, Cesare, Bixby, Danaei, Cowan,
  Paciorek, Singh, Hajifathalian, Bennett, Taddei, Bilano, Carrillo-Larco,
  Djalalinia, Khatibzadeh, Lugero, Peykari, Zhang, Lu, Stevens, Riley, Bovet,
  Elliott, Gu, Ikeda, Jackson, Joffres, Kengne, Laatikainen, Lam, Laxmaiah,
  Liu, Miranda, Mondo, Neuhauser, Sundstr\"{o}m, Smeeth, Soric, Woodward,
  Ezzati, Abarca-G{\'{o}}mez, Abdeen, Rahim, Abu-Rmeileh, Acosta-Cazares,
  Adams, Aekplakorn, Afsana, Aguilar-Salinas, Agyemang, Ahmadvand, Ahrens,
  Raddadi, Woyatan, Ali, Alkerwi, Aly, Amouyel, Amuzu, Andersen, Anderssen,
  \"{A}ngquist, Anjana, Ansong, Aounallah-Skhiri, Ara{\'{u}}jo, Ariansen, Aris,
  Arlappa, Aryal, Arveiler, Assah, Assun{\c{c}}{\~{a}}o, Avdicov{\'{a}},
  Azevedo, Azizi, Babu, Bahijri, Balakrishna, Bandosz, Banegas, Barbagallo,
  Barcel{\'{o}}, Barkat, Barros, Barros, Bata, Batieha, Baur, Beaglehole,
  Romdhane, Benet, Benson, Bernabe-Ortiz, Bernotiene, Bettiol, Bhagyalaxmi,
  Bharadwaj, Bhargava, Bi, Bikbov, Bjerregaard, Bjertness, Bj\"{o}rkelund,
  Blokstra, Bo, Bobak, Boeing, Boggia, Boissonnet, Bongard, Braeckman,
  Brajkovich, Branca, Breckenkamp, Brenner, Brewster, Bruno, de~Mesquita,
  Bugge, Burns, Bursztyn, de~Le{\'{o}}n, Cacciottolo, Cameron, Can,
  C{\^{a}}ndido, Capuano, Cardoso, Carlsson, Carvalho, Casanueva, Casas,
  Caserta, Chamukuttan, Chan, Chan, Chaturvedi, Chaturvedi, Chen, Chen, Chen,
  Chen, Chen, Cheng, Dekkaki, Chetrit, Chiolero, Chiou, Chirita-Emandi, Cho,
  Cho, Chudek, Cifkova, Claessens, Clays, Concin, Cooper, Cooper, Coppinger,
  Costanzo, Cottel, Cowell, Craig, Crujeiras, Cruz, D{\textquotesingle}Arrigo,
  d{\textquotesingle}Orsi, Dallongeville, Damasceno, Dankner, Dantoft, Dauchet,
  Backer, Bacquer, de~Gaetano, Henauw, Smedt, Deepa, Dehghan, Delisle,
  Deschamps, Dhana, Castelnuovo, da~Costa, Diaz, Dickerson, Do, Dobson,
  Donfrancesco, Donoso, D\"{o}ring, Doua, Drygas, Dulskiene, D{\v{z}}akula,
  Dzerve, Dziankowska-Zaborszczyk, Eggertsen, Ekelund, Ati, Ellert, Elliott,
  Elosua, Erasmus, Erem, Eriksen, de~la Pe{\~{n}}a, Evans, Faeh, Fall,
  Farzadfar, Felix-Redondo, Ferguson, Fern{\'{a}}ndez-Berg{\'{e}}s, Ferrante,
  Ferrari, Ferreccio, Ferrieres, Finn, Fischer, F\"{o}ger, Foo, Forslund,
  Forsner, Fortmann, Fouad, Francis, do~Carmo~Franco, Franco, Frontera, Fuchs,
  Fuchs, Fujita, Furusawa, Gaciong, Gareta, Garnett, Gaspoz, Gasull, Gates,
  Gavrila, Geleijnse, Ghasemian, Ghimire, Giampaoli, Gianfagna, Giovannelli,
  Goldsmith, Gon{\c{c}}alves, Gross, Rivas, Gottrand, Graff-Iversen,
  Grafnetter, Grajda, Gregor, Grodzicki, Gr{\o}ntved, Gruden, Grujic, Gu, Guan,
  Gudnason, Guerrero, Guessous, Guimaraes, Gulliford, Gunnlaugsdottir, Gunter,
  Gupta, Gureje, Gurzkowska, Gutierrez, Gutzwiller, Hadaegh, Halkj{\ae}r,
  Hambleton, Hardy, Harikumar, Hata, Hayes, He, Hendriks, Henriques, Cadena,
  Herrala, Heshmat, Hihtaniemi, Ho, Ho, Hobbs, Hofman, Dinc, Hormiga, Horta,
  Houti, Howitt, Htay, Htet, Hu, Huerta, Husseini, Huybrechts, Hwalla,
  Iacoviello, Iannone, Ibrahim, Ikram, Irazola, Islam, Ivkovic, Iwasaki,
  Jackson, Jacobs, Jafar, Jamrozik, Janszky, Jasienska, Jelakovic, Jiang,
  Joffres, Johansson, Jonas, J{\o}rgensen, Joshi, Juolevi, Jurak, Jure{\v{s}}a,
  Kaaks, Kafatos, Kalter-Leibovici, Kamaruddin, Kasaeian, Katz, Kauhanen, Kaur,
  Kavousi, Kazakbaeva, Keil, Boker, Kein\"{a}nen-Kiukaanniemi, Kelishadi,
  Kemper, Kengne, Kersting, Key, Khader, Khalili, Khang, Khaw, Kiechl, Killewo,
  Kim, Klumbiene, Kolle, Kolsteren, Korrovits, Koskinen, Kouda, Koziel,
  Kristensen, Krokstad, Kromhout, Kruger, Kubinova, Kuciene, Kuh, Kujala, Kula,
  Kulaga, Kumar, Kurjata, Kusuma, Kuulasmaa, Kyobutungi, Laatikainen, Lachat,
  Lam, Landrove, Lanska, Lappas, Larijani, Laugsand, Laxmaiah, Bao, Le,
  Leclercq, Lee, Lee, Lehtim\"{a}ki, Lekhraj, Le{\'{o}}n-Mu{\~{n}}oz, Levitt,
  Li, Lilly, Lim, Lima-Costa, Lin, Lin, Linneberg, Lissner, Litwin, Lorbeer,
  Lotufo, Lozano, Luksiene, Lundqvist, Lunet, Lytsy, Ma, Ma, Machado-Coelho,
  Machi, Maggi, Magliano, Majer, Makdisse, Malekzadeh, Malhotra, Rao,
  Malyutina, Manios, Mann, Manzato, Margozzini, Marques-Vidal, Marrugat,
  Martorell, Mathiesen, Matijasevich, Matsha, Mbanya, Posso, McFarlane,
  McGarvey, McLachlan, McLean, McNulty, Khir, Mediene-Benchekor, Medzioniene,
  Meirhaeghe, Meisinger, Menezes, Menon, Meshram, Metspalu, Mi, Mikkel, Miller,
  Miquel, Mi{\v{s}}igoj-Durakovic, Mohamed, Mohammad, Mohammadifard, Mohan,
  Yusoff, M{\o}ller, Moln{\'{a}}r, Momenan, Mondo, Monyeki, Moreira, Morejon,
  Moreno, Morgan, Moschonis, Mossakowska, Mostafa, Mota, Motlagh, Motta,
  Muiesan, M\"{u}ller-Nurasyid, Murphy, Mursu, Musil, Nagel, Naidu, Nakamura,
  N{\'{a}}me{\v{s}}n{\'{a}}, Nang, Nangia, Narake, Navarrete-Mu{\~{n}}oz,
  Ndiaye, Neal, Nenko, Nervi, Nguyen, Nguyen, Nieto-Mart{\'{\i}}nez, Niiranen,
  Ning, Ninomiya, Nishtar, Noale, Noboa, Noorbala, Noorbala, Noto, Nsour,
  O{\textquotesingle}Reilly, Oh, Olinto, Oliveira, Omar, Onat, Ordunez, Osmond,
  Ostojic, Otero, Overvad, Owusu-Dabo, Paccaud, Padez, Pahomova, Pajak, Palli,
  Palmieri, Panda-Jonas, Panza, Papandreou, Parnell, Parsaeian, Pecin,
  Pednekar, Peer, Peeters, Peixoto, Pelletier, Peltonen, Pereira, P{\'{e}}rez,
  Peters, Petkeviciene, Pham, Pigeot, Pikhart, Pilav, Pilotto, Pitakaka,
  Plans-Rubi{\'{o}}, Polakowska, Pola{\v{s}}ek, Porta, Portegies, Pourshams,
  Pradeepa, Prashant, Price, Puiu, Punab, Qasrawi, Qorbani, Radic, Radisauskas,
  Rahman, Raitakari, Raj, Rao, Ramachandran, Ramos, Rampal, Reina, Rasmussen,
  Redon, Reganit, Ribeiro, Riboli, Rigo, de~Wit, Ritti-Dias, Robinson,
  Robitaille, Rodr{\'{\i}}guez-Artalejo, del Cristo,
  Rodr{\'{\i}}guez-Villamizar, Rojas-Martinez, Rosengren, Rubinstein, Rui,
  Ruiz-Betancourt, Horimoto, Rutkowski, Sabanayagam, Sachdev, Saidi, Sakarya,
  Salanave, Martinez, Salmer{\'{o}}n, Salomaa, Salonen, Salvetti,
  S{\'{a}}nchez-Abanto, Sans, Santos, Santos, dos Santos, Santos, Saramies,
  Sardinha, Margolis, Sarrafzadegan, Saum, Savva, Scazufca, Schargrodsky,
  Schneider, Schultsz, Schutte, Sen, Senbanjo, Sepanlou, Sharma, Shaw, Shibuya,
  Shin, Shin, Siantar, Sibai, Silva, Simon, Simons, Simons, Sj\"{o}str\"{o}m,
  Skovbjerg, Slowikowska-Hilczer, Slusarczyk, Smeeth, Smith, Snijder, So,
  Sobngwi, S\"{o}derberg, Solfrizzi, Sonestedt, Song, S{\o}rensen,
  J{\'{e}}rome, Soumare, Staessen, Starc, Stathopoulou, Stavreski,
  Steene-Johannessen, Stehle, Stein, Stergiou, Stessman, Stieber, St\"{o}ckl,
  Stocks, Stokwiszewski, Stronks, Strufaldi, Sun, Sundstr\"{o}m, Sung,
  Suriyawongpaisal, Sy, Tai, Tammesoo, Tamosiunas, Tang, Tang, Tanser, Tao,
  Tarawneh, Tarqui-Mamani, Taylor, Theobald, Thijs, Thuesen, Tjonneland,
  Tolonen, Tolstrup, Topbas, Top{\'{o}}r-Madry, Tormo, Torrent, Traissac,
  Trichopoulos, Trichopoulou, Trinh, Trivedi, Tshepo, Tulloch-Reid, Tuomainen,
  Tuomilehto, Turley, Tynelius, Tzourio, Ueda, Ugel, Ulmer, Uusitalo, Valdivia,
  Valvi, van~der Schouw, Herck, van Rossem, van Valkengoed, Vanderschueren,
  Vanuzzo, Vatten, Vega, Velasquez-Melendez, Veronesi, Verschuren, Verstraeten,
  Victora, Viet, Viikari-Juntura, Vineis, Vioque, Virtanen, Visvikis-Siest,
  Viswanathan, Vollenweider, Voutilainen, Vrdoljak, Vrijheid, Wade, Wagner,
  Walton, Mohamud, Wang, Wang, Wang, Wannamethee, Wareham, Wedderkopp,
  Weerasekera, Whincup, Widhalm, Widyahening, Wiecek, Wijga, Wilks, Willeit,
  Willeit, Williams, Wilsgaard, Wojtyniak, Wong, Wong-McClure, Woo, Woodward,
  Wu, Wu, Wu, Xu, Yan, Yang, Ye, Yiallouros, Yoshihara, Younger-Coleman,
  Yusoff, Yusoff, Zambon, Zdrojewski, Zeng, Zhao, Zhao, Zheng, Zhu, Zimmermann,
  and Cisneros]{Zhou2017}
Zhou, B.; Bentham, J.; Cesare, M.D.; Bixby, H.; Danaei, G.; Cowan, M.J.;
  Paciorek, C.J.; Singh, G.; Hajifathalian, K.; Bennett, J.E.;  et~al.
\newblock Worldwide trends in blood pressure from 1975 to 2015: a pooled
  analysis of 1479 population-based measurement studies with 19$\cdotp$1
  million participants.
\newblock {\em The Lancet} {\bf 2017}, {\em 389},~37--55.
\newblock {\url{https://doi.org/10.1016/s0140-6736(16)31919-5}}.

\bibitem[Guyenet(2006)]{Guyenet2006}
Guyenet, P.G.
\newblock The sympathetic control of blood pressure.
\newblock {\em Nature Reviews Neuroscience} {\bf 2006}, {\em 7},~335--346.
\newblock {\url{https://doi.org/10.1038/nrn1902}}.

\bibitem[Brenner et~al.(1988)Brenner, Garcia, and Anderson]{Brenner1988}
Brenner, B.M.; Garcia, D.L.; Anderson, S.
\newblock Glomeruli and Blood Pressure: Less of One, More the Other?
\newblock {\em American Journal of Hypertension} {\bf 1988}, {\em 1},~335--347.
\newblock {\url{https://doi.org/10.1093/ajh/1.4.335}}.

\bibitem[Mukkamala et~al.(2015)Mukkamala, Hahn, Inan, Mestha, Kim, T{\"o}reyin,
  and Kyal]{mukkamala2015toward}
Mukkamala, R.; Hahn, J.O.; Inan, O.T.; Mestha, L.K.; Kim, C.S.; T{\"o}reyin,
  H.; Kyal, S.
\newblock {Toward ubiquitous blood pressure monitoring via pulse transit time:
  theory and practice}.
\newblock {\em IEEE Transactions on Biomedical Engineering} {\bf 2015}, {\em
  62},~1879--1901.
\newblock {\url{https://doi.org/10.1109/TBME.2015.2441951}}.

\bibitem[Kalehoff and Oparil(2020)]{kalehoff2020story}
Kalehoff, J.P.; Oparil, S.
\newblock The story of the silent killer.
\newblock {\em Current Hypertension Reports} {\bf 2020}, {\em 22},~1--14.
\newblock
  {\url{https://doi.org/https://link.springer.com/article/10.1007/s11906-020-01077-7}}.

\bibitem[Rastegar et~al.(2020)Rastegar, GholamHosseini, and
  Lowe]{rastegar2020non}
Rastegar, S.; GholamHosseini, H.; Lowe, A.
\newblock Non-invasive continuous blood pressure monitoring systems: current
  and proposed technology issues and challenges.
\newblock {\em Physical and Engineering Sciences in Medicine} {\bf 2020}, {\em
  43},~11--28.
\newblock
  {\url{https://doi.org/https://link.springer.com/article/10.1007/s13246-019-00813-x}}.

\bibitem[Tholl et~al.(2004)Tholl, Forstner, and Anlauf]{Tholl2004}
Tholl, U.; Forstner, K.; Anlauf, M.
\newblock Measuring blood pressure: pitfalls and recommendations.
\newblock {\em Nephrology Dialysis Transplantation} {\bf 2004}, {\em
  19},~766--770.
\newblock {\url{https://doi.org/10.1093/ndt/gfg602}}.

\bibitem[Organization et~al.(2020)]{worldtechnical}
Organization, W.H.;  et~al.
\newblock WHO technical specifications for automated non-invasive blood
  pressure measuring devices with cuff. 2020.
\newblock {\em World Health Organization} {\bf 2020}, pp. 1--68.

\bibitem[Van Den~Heuvel et~al.(2019)Van Den~Heuvel, Lely, Franx, and
  Bekker]{van2019validation}
Van Den~Heuvel, J.F.; Lely, A.T.; Franx, A.; Bekker, M.N.
\newblock {Validation of the iHealth Track and Omron HEM-9210T automated blood
  pressure devices for use in pregnancy}.
\newblock {\em Pregnancy Hypertension} {\bf 2019}, {\em 15},~37--41.
\newblock {\url{https://doi.org/10.1016/j.preghy.2018.10.008}}.

\bibitem[Pickering et~al.(2006)Pickering, Shimbo, and Haas]{Pickering2006}
Pickering, T.G.; Shimbo, D.; Haas, D.
\newblock Ambulatory Blood-Pressure Monitoring.
\newblock {\em New England Journal of Medicine} {\bf 2006}, {\em
  354},~2368--2374.
\newblock {\url{https://doi.org/10.1056/nejmra060433}}.

\bibitem[Ding et~al.(2016)Ding, Zhao, Yang, Pettigrew, Lo, Miao, Li, Liu, and
  Zhang]{ding2016continuous}
Ding, X.R.; Zhao, N.; Yang, G.Z.; Pettigrew, R.I.; Lo, B.; Miao, F.; Li, Y.;
  Liu, J.; Zhang, Y.T.
\newblock {Continuous blood pressure measurement from invasive to unobtrusive:
  Celebration of 200th birth anniversary of Carl Ludwig}.
\newblock {\em IEEE journal of biomedical and health informatics} {\bf 2016},
  {\em 20},~1455--1465.
\newblock {\url{https://doi.org/10.1109/JBHI.2016.2620995}}.

\bibitem[Sewell et~al.(2016)Sewell, Halanych, Russell, Andreae, Cherrington,
  Martin, Pisu, and Safford]{Sewell2016-lp}
Sewell, K.; Halanych, J.H.; Russell, L.B.; Andreae, S.J.; Cherrington, A.L.;
  Martin, M.Y.; Pisu, M.; Safford, M.M.
\newblock {Blood pressure measurement biases in clinical settings, Alabama,
  2010-2011}.
\newblock {\em Prev. Chronic Dis.} {\bf 2016}, {\em 13},~E01.
\newblock {\url{https://doi.org/10.5888/pcd13.150348}}.

\bibitem[Hoffman et~al.(2016)Hoffman, Trawalter, Axt, and
  Oliver]{Hoffman2016-km}
Hoffman, K.M.; Trawalter, S.; Axt, J.R.; Oliver, M.N.
\newblock Racial bias in pain assessment and treatment recommendations, and
  false beliefs about biological differences between blacks and whites.
\newblock {\em Proc. Natl. Acad. Sci. U. S. A.} {\bf 2016}, {\em
  113},~4296--4301.
\newblock {\url{https://doi.org/10.1073/pnas.1516047113}}.

\bibitem[Lee et~al.(2019)Lee, Le~Saux, Siegel, Goyal, Chen, Ma, and
  Meltzer]{Lee2019-fm}
Lee, P.; Le~Saux, M.; Siegel, R.; Goyal, M.; Chen, C.; Ma, Y.; Meltzer, A.C.
\newblock {Racial and ethnic disparities in the management of acute pain in US
  emergency departments: Meta-analysis and systematic review}.
\newblock {\em Am. J. Emerg. Med.} {\bf 2019}, {\em 37},~1770--1777.
\newblock {\url{https://doi.org/10.1016/j.ajem.2019.06.014}}.

\bibitem[Valbuena et~al.(2022)Valbuena, Merchant, and Hough]{Valbuena2022-gw}
Valbuena, V.S.M.; Merchant, R.M.; Hough, C.L.
\newblock Racial and ethnic bias in pulse oximetry and clinical outcomes.
\newblock {\em JAMA Intern. Med.} {\bf 2022}, {\em 182},~699--700.
\newblock {\url{https://doi.org/10.1001/jamainternmed.2022.1903}}.

\bibitem[Obermeyer et~al.(2019)Obermeyer, Powers, Vogeli, and
  Mullainathan]{Obermeyer2019-cb}
Obermeyer, Z.; Powers, B.; Vogeli, C.; Mullainathan, S.
\newblock Dissecting racial bias in an algorithm used to manage the health of
  populations.
\newblock {\em Science} {\bf 2019}, {\em 366},~447--453.
\newblock {\url{https://doi.org/10.1126/science.aax2342}}.

\bibitem[Reyna et~al.(2022)Reyna, Nsoesie, and Clifford]{Reyna2022-on}
Reyna, M.A.; Nsoesie, E.O.; Clifford, G.D.
\newblock Rethinking algorithm performance metrics for artificial intelligence
  in diagnostic medicine.
\newblock {\em JAMA} {\bf 2022}, {\em 328},~329--330.
\newblock {\url{https://doi.org/10.1001/jama.2022.10561}}.

\bibitem[Vokinger et~al.(2021)Vokinger, Feuerriegel, and
  Kesselheim]{Vokinger2021-td}
Vokinger, K.N.; Feuerriegel, S.; Kesselheim, A.S.
\newblock Mitigating bias in machine learning for medicine.
\newblock {\em Commun Med (Lond)} {\bf 2021}, {\em 1},~25.
\newblock {\url{https://doi.org/10.1038/s43856-021-00028-w}}.

\bibitem[Magder(2018)]{Magder2018-vy}
Magder, S.
\newblock {The meaning of blood pressure}.
\newblock {\em Crit. Care} {\bf 2018}, {\em 22},~257.

\bibitem[Iaizzo(2010)]{M9}
Iaizzo, P.A.
\newblock {\em {Handbook of cardiac anatomy, physiology, and devices}};
  Springer Science \& Business Media,  2010.

\bibitem[Mousavi et~al.(2018)Mousavi, Charmi, Firouzmand, Hemmati, Moghadam,
  and Ghorbani]{mousavi2018designing}
Mousavi, S.S.; Charmi, M.; Firouzmand, M.; Hemmati, M.; Moghadam, M.; Ghorbani,
  Y.
\newblock Designing and Manufacturing a Mobile-Based Ambulatory Monitoring of
  Blood Pressure Using Electrocardiogram and Photoplethysmography Signals.
\newblock {\em Zanjan University} {\bf 2018}.

\bibitem[Lowry and Ashelford(2015)]{M5}
Lowry, M.; Ashelford, S.
\newblock {Assessing the pulse rate in adult patients.}
\newblock {\em Nursing Times} {\bf 2015}, {\em 111},~18--20.
\newblock
  {\url{https://doi.org/https://www.nursingtimes.net/clinical-archive/assessment-skills/assessing-the-pulse-rate-in-adult-patients-31-08-2015/}}.

\bibitem[Betts et~al.()Betts, Desaix, Johnson, Johnson, Korol, Kruse, Poe,
  Wise, Womble, and Young]{bettsanatomy}
Betts, J.; Desaix, P.; Johnson, E.; Johnson, J.; Korol, O.; Kruse, D.; Poe, B.;
  Wise, J.; Womble, M.; Young, K.
\newblock {Anatomy and Physiology. OpenStax, 2013}.

\bibitem[Vlachopoulos et~al.(2011)Vlachopoulos, O'Rourke, and
  Nichols]{Vlachopoulos2011-kp}
Vlachopoulos, C.; O'Rourke, M.; Nichols, W.W.
\newblock {\em {McDonald's} blood flow in arteries}, 6 ed.; Hodder Arnold:
  London, England,  2011.
\newblock {\url{https://doi.org//10.1201/b13568}}.

\bibitem[Lapum et~al.(2018)Lapum, Verkuyl, Garcia, St-Amant, and Tan]{M10}
Lapum, J.L.; Verkuyl, M.; Garcia, W.; St-Amant, O.; Tan, A.
\newblock {\em Vital sign measurement across the lifespan}; Ryerson university,
   2018.
\newblock
  {\url{https://doi.org/https://pressbooks.library.torontomu.ca/vitalsign/}}.

\bibitem[Sapra et~al.(2021)Sapra, Malik, and Bhandari]{M1}
Sapra, A.; Malik, A.; Bhandari, P.
\newblock {Vital sign assessment}. In {\em StatPearls [Internet]}; StatPearls
  Publishing,  2021.
\newblock
  {\url{https://doi.org/https://www.ncbi.nlm.nih.gov/books/NBK553213/}}.

\bibitem[DeSaix et~al.(2013)DeSaix, Betts, Johnson, Johnson, Oksana, Kruse,
  Poe, Wise, and Young]{desaix2013anatomy}
DeSaix, P.; Betts, G.J.; Johnson, E.; Johnson, J.E.; Oksana, K.; Kruse, D.H.;
  Poe, B.; Wise, J.A.; Young, K.A.
\newblock {Anatomy \& Physiology (OpenStax)},  2013.
\newblock
  {\url{https://doi.org/https://assets.openstax.org/oscms-prodcms/media/documents/AnatomyandPhysiology-OP.pdf}}.

\bibitem[Glasser(1997)]{Glasser1997-ll}
Glasser, S.
\newblock {Vascular compliance and cardiovascular disease A risk factor or a
  marker?}
\newblock {\em Am. J. Hypertens.} {\bf 1997}, {\em 10},~1175--1189.
\newblock {\url{https://doi.org/10.1016/s0895-7061(97)00311-7}}.

\bibitem[Bazett(1920)]{M12}
Bazett, H.
\newblock {The time relations of the blood-pressure changes after excision of
  the adrenal glands, with some observations on blood volume changes}.
\newblock {\em The Journal of Physiology} {\bf 1920}, {\em 53},~320.
\newblock {\url{https://doi.org/https://pubmed.ncbi.nlm.nih.gov/16993419/}}.

\bibitem[Letcher et~al.(1981)Letcher, Chien, Pickering, Sealey, and
  Laragh]{Letcher1981-qq}
Letcher, R.L.; Chien, S.; Pickering, T.G.; Sealey, J.E.; Laragh, J.H.
\newblock {Direct relationship between blood pressure and blood viscosity in
  normal and hypertensive subjects. Role of fibrinogen and concentration}.
\newblock {\em Am. J. Med.} {\bf 1981}, {\em 70},~1195--1202.
\newblock {\url{https://doi.org/10.1016/0002-9343(81)90827-5}}.

\bibitem[Jeppesen et~al.(2007)Jeppesen, Sanye-Hajari, and Bek]{Jeppesen2007-az}
Jeppesen, P.; Sanye-Hajari, J.; Bek, T.
\newblock Increased blood pressure induces a diameter response of retinal
  arterioles that increases with decreasing arteriolar diameter.
\newblock {\em Invest. Ophthalmol. Vis. Sci.} {\bf 2007}, {\em 48},~328--331.
\newblock {\url{https://doi.org/10.1167/iovs.06-0360}}.

\bibitem[Whelton et~al.(2018)Whelton, Carey, Aronow, Casey, Collins,
  Himmelfarb, DePalma, Gidding, Jamerson, Jones, MacLaughlin, Muntner,
  Ovbiagele, Smith, Spencer, Stafford, Taler, Thomas, Williams, Williamson, and
  Wright]{Whelton2018}
Whelton, P.K.; Carey, R.M.; Aronow, W.S.; Casey, D.E.; Collins, K.J.;
  Himmelfarb, C.D.; DePalma, S.M.; Gidding, S.; Jamerson, K.A.; Jones, D.W.;
  et~al.
\newblock 2017 {ACC}/{AHA}/{AAPA}/{ABC}/{ACPM}/
  {AGS}/{APhA}/{ASH}/{ASPC}/{NMA}/{PCNA} Guideline for the Prevention,
  Detection, Evaluation, and Management of High Blood Pressure in Adults.
\newblock {\em Journal of the American College of Cardiology} {\bf 2018}, {\em
  71},~e127--e248.
\newblock {\url{https://doi.org/10.1016/j.jacc.2017.11.006}}.

\bibitem[Azegami et~al.(2021)Azegami, Uchida, Tokumura, and
  Mori]{azegami2021blood}
Azegami, T.; Uchida, K.; Tokumura, M.; Mori, M.
\newblock Blood Pressure Tracking From Childhood to Adulthood.
\newblock {\em Frontiers in Pediatrics} {\bf 2021}, {\em 9}.
\newblock {\url{https://doi.org/10.3389/fped.2021.785356}}.

\bibitem[Cole(2007)]{Cole2007-rx}
Cole, E.
\newblock {Measuring central venous pressure}.
\newblock {\em Nurs. Stand.} {\bf 2007}, {\em 22},~40--42.
\newblock {\url{https://doi.org/10.7748/ns2007.10.22.7.40.c4624}}.

\bibitem[Webster(2009)]{M15}
Webster, J.
\newblock {\em Medical instrumentation: Application and design}; John Wiley \&
  Sons,  2009.

\bibitem[Kumar et~al.(2021)Kumar, Dubey, Zafer, Kumar, and Yadav]{Kumar2021-em}
Kumar, R.; Dubey, P.K.; Zafer, A.; Kumar, A.; Yadav, S.
\newblock Past, present and future of blood pressure measuring instruments and
  their calibration.
\newblock {\em Measurement (Lond.)} {\bf 2021}, {\em 172},~108845.
\newblock {\url{https://doi.org/10.1016/j.measurement.2020.108845}}.

\bibitem[Peter et~al.(2014)Peter, Noury, and Cerny]{peter2014review}
Peter, L.; Noury, N.; Cerny, M.
\newblock A review of methods for non-invasive and continuous blood pressure
  monitoring: Pulse transit time method is promising?
\newblock {\em Irbm} {\bf 2014}, {\em 35},~271--282.
\newblock {\url{https://doi.org/https://doi.org/10.1016/j.irbm.2014.07.002}}.

\bibitem[Lim and Kim(2022)]{lim2022blood}
Lim, S.H.; Kim, S.H.
\newblock {Blood pressure measurements and hypertension in infants, children,
  and adolescents: from the postmercury to mobile devices}.
\newblock {\em Clin Exp Pediatr} {\bf 2022}, {\em 65},~73--80.
\newblock {\url{https://doi.org/10.3345/cep.2021.00143}}.

\bibitem[Yarows and Qian(2001)]{Yarows2001}
Yarows, S.A.; Qian, K.
\newblock {Accuracy of aneroid sphygmomanometers in clinical usage: University
  of Michigan experience}.
\newblock {\em Blood Pressure Monitoring} {\bf 2001}, {\em 6},~101--106.
\newblock {\url{https://doi.org/10.1097/00126097-200104000-00006}}.

\bibitem[Canzanello et~al.(2001)Canzanello, Jensen, and
  Schwartz]{Canzanello2001}
Canzanello, V.J.; Jensen, P.L.; Schwartz, G.L.
\newblock {Are Aneroid Sphygmomanometers Accurate in Hospital and Clinic
  Settings?}
\newblock {\em Archives of Internal Medicine} {\bf 2001}, {\em 161},~729.
\newblock {\url{https://doi.org/10.1001/archinte.161.5.729}}.

\bibitem[Mion and Pierin(1998)]{Mion1998}
Mion, D.; Pierin, A.
\newblock How accurate are sphygmomanometers?
\newblock {\em Journal of Human Hypertension} {\bf 1998}, {\em 12},~245--248.
\newblock {\url{https://doi.org/10.1038/sj.jhh.1000589}}.

\bibitem[Ogedegbe and Pickering(2010)]{ogedegbe2010principles}
Ogedegbe, G.; Pickering, T.
\newblock Principles and techniques of blood pressure measurement.
\newblock {\em Cardiology clinics} {\bf 2010}, {\em 28},~571--586.
\newblock {\url{https://doi.org/10.1016/j.ccl.2010.07.006}}.

\bibitem[{Association for the Advancement of Medical Instrumentation and
  others}(1987)]{association1987american}
{Association for the Advancement of Medical Instrumentation and others}.
\newblock {American national standards for electronic or automated
  sphygmomanometers}.
\newblock {\em ANSI/AAMI SP 10-1987} {\bf 1987}.

\bibitem[Ger{\v{s}}ak et~al.(2009)Ger{\v{s}}ak, {\v{Z}}emva, and
  Drnov{\v{s}}ek]{gervsak2009procedure}
Ger{\v{s}}ak, G.; {\v{Z}}emva, A.; Drnov{\v{s}}ek, J.
\newblock A procedure for evaluation of non-invasive blood pressure simulators.
\newblock {\em Medical \& biological engineering \& computing} {\bf 2009}, {\em
  47},~1221--1228.
\newblock {\url{https://doi.org/10.1007/s11517-009-0532-2}}.

\bibitem[Stergiou et~al.(2018)Stergiou, Alpert, Mieke, Asmar, Atkins, Eckert,
  Frick, Friedman, Gra{\ss}l, Ichikawa, et~al.]{stergiou2018universal}
Stergiou, G.S.; Alpert, B.; Mieke, S.; Asmar, R.; Atkins, N.; Eckert, S.;
  Frick, G.; Friedman, B.; Gra{\ss}l, T.; Ichikawa, T.;  et~al.
\newblock A universal standard for the validation of Blood Pressure measuring
  devices: Association for the Advancement of Medical
  {Instrumentation/European} Society of {Hypertension/International}
  Organization for Standardization ({AAMI/ESH/ISO}) collaboration statement.
\newblock {\em Hypertension} {\bf 2018}, {\em 71},~368--374.
\newblock {\url{https://doi.org/10.1161/HYPERTENSIONAHA.117.10237}}.

\bibitem[ISO(2020)]{ISO81060-2:2018/Amd.1:2020}
{Non-Invasive Sphygmomanometers – Part 2: Clinical Investigation of
  Intermittent Automated Measurement - Amendment 1},  2020.
\newblock {ISO 81060-2:2018/Amd.1:2020}.

\bibitem[Hassler and Burnier(2005)]{hassler2005circadian}
Hassler, C.; Burnier, M.
\newblock Circadian variations in blood pressure.
\newblock {\em American journal of cardiovascular drugs} {\bf 2005}, {\em
  5},~7--15.
\newblock {\url{https://doi.org/10.2165/00129784-200505010-00002}}.

\bibitem[Wagner et~al.(2012)Wagner, Toftegaard, and Bertelsen]{Wagner2012-id}
Wagner, S.; Toftegaard, T.S.; Bertelsen, O.W.
\newblock Challenges in blood pressure self-measurement.
\newblock {\em Int. J. Telemed. Appl.} {\bf 2012}, {\em 2012},~437350.
\newblock {\url{https://doi.org/10.1155/2012/437350}}.

\bibitem[Stergiou et~al.(2021)Stergiou, Palatini, Parati, O'Brien, Januszewicz,
  Lurbe, Persu, Mancia, Kreutz, and {European Society of Hypertension Council
  and the European Society of Hypertension Working Group on Blood Pressure
  Monitoring and Cardiovascular Variability}]{Stergiou2021-fl}
Stergiou, G.S.; Palatini, P.; Parati, G.; O'Brien, E.; Januszewicz, A.; Lurbe,
  E.; Persu, A.; Mancia, G.; Kreutz, R.; {European Society of Hypertension
  Council and the European Society of Hypertension Working Group on Blood
  Pressure Monitoring and Cardiovascular Variability}.
\newblock {2021 European Society of Hypertension practice guidelines for office
  and out-of-office blood pressure measurement}.
\newblock {\em J. Hypertens.} {\bf 2021}, {\em 39},~1293--1302.
\newblock {\url{https://doi.org/10.1097/HJH.0000000000002843}}.

\bibitem[Liu et~al.(2022)Liu, Li, Li, Zheng, and Liu]{Liu2022-xk}
Liu, J.; Li, Y.; Li, J.; Zheng, D.; Liu, C.
\newblock {Sources of automatic office blood pressure measurement error: a
  systematic review}.
\newblock {\em Physiol. Meas.} {\bf 2022}, {\em 43},~09TR02.
\newblock {\url{https://doi.org/10.1088/1361-6579/ac890e}}.

\bibitem[Muntner et~al.(2019)Muntner, Shimbo, Carey, Charleston, Gaillard,
  Misra, Myers, Ogedegbe, Schwartz, Townsend, et~al.]{muntner2019measurement}
Muntner, P.; Shimbo, D.; Carey, R.M.; Charleston, J.B.; Gaillard, T.; Misra,
  S.; Myers, M.G.; Ogedegbe, G.; Schwartz, J.E.; Townsend, R.R.;  et~al.
\newblock {Measurement of blood pressure in humans: a scientific statement from
  the American Heart Association}.
\newblock {\em Hypertension} {\bf 2019}, {\em 73},~e35--e66.
\newblock {\url{https://doi.org/10.1161/HYP.0000000000000087}}.

\bibitem[Kallioinen et~al.(2017)Kallioinen, Hill, Horswill, Ward, and
  Watson]{kallioinen2017sources}
Kallioinen, N.; Hill, A.; Horswill, M.S.; Ward, H.E.; Watson, M.O.
\newblock Sources of inaccuracy in the measurement of adult patients’ resting
  blood pressure in clinical settings: a systematic review.
\newblock {\em Journal of hypertension} {\bf 2017}, {\em 35},~421.
\newblock {\url{https://doi.org/10.1097/HJH.0000000000001197}}.

\bibitem[{ESH/ESC Task Force for the Management of Arterial
  Hypertension}(2013)]{mancia20132013}
{ESH/ESC Task Force for the Management of Arterial Hypertension}.
\newblock 2013 Practice guidelines for the management of arterial hypertension
  of the European Society of Hypertension ({ESH}) and the European Society of
  Cardiology ({ESC)}: {ESH/ESC} Task Force for the Management of Arterial
  Hypertension.
\newblock {\em J. Hypertens.} {\bf 2013}, {\em 31},~1925--1938.
\newblock {\url{https://doi.org/10.1097/HJH.0b013e328364ca4c}}.

\bibitem[Yong and Geddes(1987)]{Yong1987-gn}
Yong, P.G.; Geddes, L.A.
\newblock The effect of cuff pressure deflation rate on accuracy in indirect
  measurement of blood pressure with the auscultatory method.
\newblock {\em J. Clin. Monit.} {\bf 1987}, {\em 3},~155--159.

\bibitem[Speechly et~al.(2007)Speechly, Bignell, and
  Turner]{speechly2007sphygmomanometer}
Speechly, C.; Bignell, N.; Turner, M.
\newblock {Sphygmomanometer calibration: why, how and how often?}
\newblock {\em Australian family physician} {\bf 2007}, {\em 36}.
\newblock
  {\url{https://doi.org/https://www.racgp.org.au/getattachment/5fcd5241-6455-4656-8476-a6bd1798cec7/200710turner.pdf}}.

\bibitem[Turner et~al.(2004)Turner, Baker, and Kam]{turner2004effects}
Turner, M.J.; Baker, A.B.; Kam, P.C.
\newblock Effects of systematic errors in blood pressure measurements on the
  diagnosis of hypertension.
\newblock {\em Blood pressure monitoring} {\bf 2004}, {\em 9},~249--253.
\newblock {\url{https://doi.org/10.1097/00126097-200410000-00004}}.

\bibitem[Yayan and Zengin(2020)]{yayan2020key}
Yayan, E.H.; Zengin, M.
\newblock A Key Point in Medical Measurements: Device Calibration and Knowledge
  Level of Healthcare Professionals.
\newblock {\em vol} {\bf 2020}, {\em 13},~1346--1354.
\newblock
  {\url{https://doi.org/http://www.internationaljournalofcaringsciences.org/docs/60_1_zengin_original_13_2.pdf}}.

\bibitem[Pickering et~al.(2005)Pickering, Hall, Appel, Falkner, Graves, Hill,
  Jones, Kurtz, Sheps, and Roccella]{pickering2005recommendations}
Pickering, T.G.; Hall, J.E.; Appel, L.J.; Falkner, B.E.; Graves, J.; Hill,
  M.N.; Jones, D.W.; Kurtz, T.; Sheps, S.G.; Roccella, E.J.
\newblock {Recommendations for blood pressure measurement in humans and
  experimental animals: part 1: blood pressure measurement in humans: a
  statement for professionals from the Subcommittee of Professional and Public
  Education of the American Heart Association Council on High Blood Pressure
  Research}.
\newblock {\em Hypertension} {\bf 2005}, {\em 45},~142--161.
\newblock {\url{https://doi.org/10.1161/01.HYP.0000150859.47929.8e}}.

\bibitem[Reckelhoff(2018)]{Reckelhoff2018}
Reckelhoff, J.F.
\newblock {Sex Differences in Regulation of Blood Pressure}. In {\em {Advances
  in Experimental Medicine and Biology}}; Springer International Publishing,
  2018; pp. 139--151.
\newblock {\url{https://doi.org/10.1007/978-3-319-77932-4_9}}.

\bibitem[Sandberg and Ji(2012)]{Sandberg2012}
Sandberg, K.; Ji, H.
\newblock Sex differences in primary hypertension.
\newblock {\em {Biology of Sex Differences}} {\bf 2012}, {\em 3},~7.
\newblock {\url{https://doi.org/10.1186/2042-6410-3-7}}.

\bibitem[Reckelhoff(2001)]{Reckelhoff2001-td}
Reckelhoff, J.F.
\newblock Gender differences in the regulation of blood pressure.
\newblock {\em Hypertension} {\bf 2001}, {\em 37},~1199--1208.
\newblock {\url{https://doi.org/10.1161/01.HYP.37.5.1199}}.

\bibitem[Ji et~al.(2020)Ji, Kim, Ebinger, Niiranen, Claggett, Merz, and
  Cheng]{Ji2020}
Ji, H.; Kim, A.; Ebinger, J.E.; Niiranen, T.J.; Claggett, B.L.; Merz, C.N.B.;
  Cheng, S.
\newblock Sex Differences in Blood Pressure Trajectories Over the Life Course.
\newblock {\em {JAMA} Cardiology} {\bf 2020}, {\em 5},~255.
\newblock {\url{https://doi.org/10.1001/jamacardio.2019.5306}}.

\bibitem[Somani et~al.(2018)Somani, Baross, Brook, Milne, McGowan, and
  Swaine]{Somani2018-cr}
Somani, Y.B.; Baross, A.W.; Brook, R.D.; Milne, K.J.; McGowan, C.L.; Swaine,
  I.L.
\newblock Acute response to a 2-minute isometric exercise test predicts the
  blood pressure-lowering efficacy of isometric resistance training in young
  adults.
\newblock {\em Am. J. Hypertens.} {\bf 2018}, {\em 31},~362--368.
\newblock {\url{https://doi.org/10.1093/ajh/hpx173}}.

\bibitem[Olatunji et~al.(2011)Olatunji, Aaron, Michael, and
  Oyeyipo]{olatunji2011water}
Olatunji, L.; Aaron, A.; Michael, O.; Oyeyipo, I.
\newblock Water ingestion affects orthostatic challenge-induced blood pressure
  and heart rate responses in young healthy subjects: gender implications.
\newblock {\em Nigerian Journal of Physiological Sciences} {\bf 2011}, {\em
  26}.
\newblock {\url{https://doi.org/10.1038/s41598-017-08446-4}}.

\bibitem[Kho et~al.(2006)Kho, Yi, Lee, and Chung]{kho2006acute}
Kho, Y.L.; Yi, S.G.; Lee, E.H.; Chung, M.H.
\newblock Acute Effects of Tobacco and Non-tobacco Cigarette Smoking on the
  Blood Pressure and Heart Rate.
\newblock {\em Journal of Environmental Health Sciences} {\bf 2006}, {\em
  32},~222--226.

\bibitem[Papakonstantinou et~al.(2016)Papakonstantinou, Kechribari,
  Sotirakoglou, Tarantilis, Gourdomichali, Michas, Kravvariti, Voumvourakis,
  and Zampelas]{Papakonstantinou2016-mj}
Papakonstantinou, E.; Kechribari, I.; Sotirakoglou, K.; Tarantilis, P.;
  Gourdomichali, T.; Michas, G.; Kravvariti, V.; Voumvourakis, K.; Zampelas, A.
\newblock Acute effects of coffee consumption on self-reported gastrointestinal
  symptoms, blood pressure and stress indices in healthy individuals.
\newblock {\em Nutr. J.} {\bf 2016}, {\em 15},~26.
\newblock {\url{https://doi.org/10.1186/s12937-016-0146-0}}.

\bibitem[Monnard and Grasser(2017)]{Monnard2017-je}
Monnard, C.R.; Grasser, E.K.
\newblock Water ingestion decreases cardiac workload time-dependent in healthy
  adults with no effect of gender.
\newblock {\em Sci. Rep.} {\bf 2017}, {\em 7},~7939.
\newblock {\url{https://doi.org/10.1038/s41598-017-08446-4}}.

\bibitem[Helfer and McCubbin(2001)]{helfer2001does}
Helfer, S.G.; McCubbin, J.A.
\newblock Does gender affect the relation between blood pressure and pain
  sensitivity?
\newblock {\em International Journal of Behavioral Medicine} {\bf 2001}, {\em
  8},~220--229.
\newblock {\url{https://doi.org/10.1207/S15327558IJBM0803_4}}.

\bibitem[Harshfield et~al.(1989)Harshfield, Alpert, Willey, Somes, Murphy, and
  Dupaul]{harshfield1989race}
Harshfield, G.A.; Alpert, B.S.; Willey, E.S.; Somes, G.W.; Murphy, J.K.;
  Dupaul, L.M.
\newblock Race and gender influence ambulatory blood pressure patterns of
  adolescents.
\newblock {\em Hypertension} {\bf 1989}, {\em 14},~598--603.
\newblock {\url{https://doi.org/10.1161/01.HYP.14.6.598}}.

\bibitem[Costa-Hong et~al.(2018)Costa-Hong, Muela, Macedo, Sales, and
  Bortolotto]{costa2018gender}
Costa-Hong, V.A.; Muela, H.C.S.; Macedo, T.A.; Sales, A.R.K.; Bortolotto, L.A.
\newblock Gender differences of aortic wave reflection and influence of
  menopause on central blood pressure in patients with arterial hypertension.
\newblock {\em BMC cardiovascular disorders} {\bf 2018}, {\em 18},~1--6.
\newblock {\url{https://doi.org/10.1186/s12872-018-0855-8}}.

\bibitem[Ki et~al.(2013)Ki, Oh, and Lee]{Ki2013-ai}
Ki, J.H.; Oh, M.K.; Lee, S.H.
\newblock Differences in blood pressure measurements obtained using an
  automatic oscillometric sphygmomanometer depending on clothes-wearing status.
\newblock {\em Korean J. Fam. Med.} {\bf 2013}, {\em 34},~145--151.
\newblock {\url{https://doi.org/10.4082/kjfm.2013.34.2.145}}.

\bibitem[Wang et~al.(2006)Wang, Poole, Treiber, Harshfield, Hanevold, and
  Snieder]{Wang2006-jk}
Wang, X.; Poole, J.C.; Treiber, F.A.; Harshfield, G.A.; Hanevold, C.D.;
  Snieder, H.
\newblock {Ethnic and gender differences in ambulatory blood pressure
  trajectories: results from a 15-year longitudinal study in youth and young
  adults}.
\newblock {\em Circulation} {\bf 2006}, {\em 114},~2780--2787.
\newblock {\url{https://doi.org/10.1161/CIRCULATIONAHA.106.643940}}.

\bibitem[Song et~al.(2016)Song, Kim, Shim, Lee, and Choi]{Song2016-ho}
Song, B.M.; Kim, H.C.; Shim, J.S.; Lee, M.H.; Choi, D.P.
\newblock {Inter-arm difference in brachial blood pressure in the general
  population of Koreans}.
\newblock {\em Korean Circ. J.} {\bf 2016}, {\em 46},~374--383.
\newblock {\url{https://doi.org/10.4070/kcj.2016.46.3.374}}.

\bibitem[Lan and Chen(2012)]{lan2012prevalence}
Lan, Y.L.; Chen, T.L.
\newblock {Prevalence of high blood pressure and its relationship with body
  weight factors among inpatients with schizophrenia in Taiwan}.
\newblock {\em Asian Nursing Research} {\bf 2012}, {\em 6},~13--18.
\newblock {\url{https://doi.org/10.1016/j.anr.2012.02.003}}.

\bibitem[Priv{\v{s}}ek et~al.(2018)Priv{\v{s}}ek, Hellgren, R{\aa}stam,
  Lindblad, and Daka]{privvsek2018epidemiological}
Priv{\v{s}}ek, E.; Hellgren, M.; R{\aa}stam, L.; Lindblad, U.; Daka, B.
\newblock Epidemiological and clinical implications of blood pressure measured
  in seated versus supine position.
\newblock {\em Medicine} {\bf 2018}, {\em 97}.
\newblock {\url{https://doi.org/10.1097/MD.0000000000011603}}.

\bibitem[Cui et~al.(2002)Cui, Hopper, and Harrap]{cui2002genes}
Cui, J.; Hopper, J.L.; Harrap, S.B.
\newblock {Genes and family environment explain correlations between blood
  pressure and body mass index}.
\newblock {\em Hypertension} {\bf 2002}, {\em 40},~7--12.
\newblock {\url{https://doi.org/10.1161/01.HYP.0000022693.11752.E9}}.

\bibitem[Vall{\'e}e et~al.(2019)Vall{\'e}e, Perrine, Deschamps, Blacher, and
  Oli{\'e}]{vallee2019relationship}
Vall{\'e}e, A.; Perrine, A.L.; Deschamps, V.; Blacher, J.; Oli{\'e}, V.
\newblock {Relationship between dynamic changes in body weight and blood
  pressure: the ESTEBAN survey}.
\newblock {\em American Journal of Hypertension} {\bf 2019}, {\em
  32},~1003--1012.
\newblock {\url{https://doi.org/10.1093/ajh/hpz096}}.

\bibitem[Giggey et~al.(2011)Giggey, Wendell, Zonderman, and
  Waldstein]{Giggey2011-ri}
Giggey, P.P.; Wendell, C.R.; Zonderman, A.B.; Waldstein, S.R.
\newblock Greater coffee intake in men is associated with steeper age-related
  increases in blood pressure.
\newblock {\em Am. J. Hypertens.} {\bf 2011}, {\em 24},~310--315.
\newblock {\url{https://doi.org/10.1038/ajh.2010.225}}.

\bibitem[Bourgeois et~al.(2017)Bourgeois, Watts, Thomas, Carmichael, Hu, Heo,
  Hall, and Heymsfield]{bourgeois2017associations}
Bourgeois, B.; Watts, K.; Thomas, D.M.; Carmichael, O.; Hu, F.B.; Heo, M.;
  Hall, J.E.; Heymsfield, S.B.
\newblock {Associations between height and blood pressure in the United States
  population}.
\newblock {\em Medicine} {\bf 2017}, {\em 96}.
\newblock {\url{https://doi.org/10.1097/MD.0000000000009233}}.

\bibitem[Pan et~al.(1986)Pan, Nanas, Dyer, Liu, Mcdonald, Schoenberger,
  Shekelle, Stamler, and Stamler]{pan1986role}
Pan, W.H.; Nanas, S.; Dyer, A.; Liu, K.; Mcdonald, A.; Schoenberger, J.A.;
  Shekelle, R.B.; Stamler, R.; Stamler, J.
\newblock The role of weight in the positive association between age and blood
  pressure.
\newblock {\em American journal of epidemiology} {\bf 1986}, {\em
  124},~612--623.
\newblock {\url{https://doi.org/10.1093/oxfordjournals.aje.a114434}}.

\bibitem[Carrico et~al.(2013)Carrico, Sun, Sima, and
  Rosner]{carrico2013predictive}
Carrico, R.J.; Sun, S.S.; Sima, A.P.; Rosner, B.
\newblock The predictive value of childhood blood pressure values for adult
  elevated blood pressure.
\newblock {\em Open journal of pediatrics} {\bf 2013}, {\em 3},~116.
\newblock {\url{https://doi.org/10.4236/ojped.2013.32022}}.

\bibitem[Jones and Hall(2006)]{Jones2006}
Jones, D.W.; Hall, J.E.
\newblock {Racial and Ethnic Differences in Blood Pressure}.
\newblock {\em Circulation} {\bf 2006}, {\em 114},~2757--2759.
\newblock {\url{https://doi.org/10.1161/circulationaha.106.668731}}.

\bibitem[Hardy et~al.(2021)Hardy, Chen, Cherrington, Moise, Jaeger, Foti,
  Sakhuja, Wozniak, Abdalla, and Muntner]{hardy2021racial}
Hardy, S.T.; Chen, L.; Cherrington, A.L.; Moise, N.; Jaeger, B.C.; Foti, K.;
  Sakhuja, S.; Wozniak, G.; Abdalla, M.; Muntner, P.
\newblock {Racial and ethnic differences in blood pressure among US adults,
  1999--2018}.
\newblock {\em Hypertension} {\bf 2021}, {\em 78},~1730--1741.
\newblock {\url{https://doi.org/10.1161/HYPERTENSIONAHA.121.18086}}.

\bibitem[Cooper and Kaufman(1998)]{Cooper1998}
Cooper, R.S.; Kaufman, J.S.
\newblock {Race and Hypertension}.
\newblock {\em Hypertension} {\bf 1998}, {\em 32},~813--816.
\newblock {\url{https://doi.org/10.1161/01.hyp.32.5.813}}.

\bibitem[Fryar et~al.(2017)Fryar, Ostchega, Hales, Zhang, and
  Kruszon-Moran]{fryar2017hypertension}
Fryar, C.D.; Ostchega, Y.; Hales, C.M.; Zhang, G.; Kruszon-Moran, D.
\newblock {Hypertension prevalence and control among adults: United States,
  2015-2016}.
\newblock {\em National Center for Health Statistics} {\bf 2017}.
\newblock
  {\url{https://doi.org/https://www.cdc.gov/nchs/products/databriefs/db289.htm}}.

\bibitem[Staessen et~al.(1997)Staessen, Bieniaszewski, O'Brien, Gosse, Hayashi,
  Imai, Kawasaki, Otsuka, Palatini, Thijs, et~al.]{staessen1997nocturnal}
Staessen, J.A.; Bieniaszewski, L.; O'Brien, E.; Gosse, P.; Hayashi, H.; Imai,
  Y.; Kawasaki, T.; Otsuka, K.; Palatini, P.; Thijs, L.;  et~al.
\newblock {Nocturnal blood pressure fall on ambulatory monitoring in a large
  international database. The ``Ad Hoc' Working Group}.
\newblock {\em Hypertension} {\bf 1997}, {\em 29},~30--39.
\newblock {\url{https://doi.org/10.1161/01.HYP.29.1.30}}.

\bibitem[Mayet et~al.(1998)Mayet, Chapman, Li, Shahi, Poulter, Sever, Foale,
  and Thom]{mayet1998ethnic}
Mayet, J.; Chapman, N.; Li, C.K.C.; Shahi, M.; Poulter, N.R.; Sever, P.S.;
  Foale, R.A.; Thom, S.A.M.
\newblock Ethnic differences in the hypertensive heart and 24-hour blood
  pressure profile.
\newblock {\em Hypertension} {\bf 1998}, {\em 31},~1190--1194.
\newblock {\url{https://doi.org/10.1161/01.HYP.31.5.1190}}.

\bibitem[He et~al.(2000)He, Ding, Fong, and Karlberg]{he2000blood}
He, Q.; Ding, Z.Y.; Fong, D.Y.T.; Karlberg, J.
\newblock Blood pressure is associated with body mass index in both normal and
  obese children.
\newblock {\em Hypertension} {\bf 2000}, {\em 36},~165--170.
\newblock {\url{https://doi.org/10.1161/01.HYP.36.2.165}}.

\bibitem[Jena et~al.(2018)Jena, Pattnaik, et~al.]{jena2018relationship}
Jena, S.K.; Pattnaik, M.;  et~al.
\newblock Relationship between body mass index and blood pressure in school
  students.
\newblock {\em CHRISMED Journal of Health and Research} {\bf 2018}, {\em
  5},~187.
\newblock
  {\url{https://doi.org/https://www.cjhr.org/temp/CHRISMEDJHealthRes53187-1339962_034319.pdf}}.

\bibitem[Neter et~al.(2003{\natexlab{a}})Neter, Stam, Kok, Grobbee, and
  Geleijnse]{neter2003influence}
Neter, J.E.; Stam, B.E.; Kok, F.J.; Grobbee, D.E.; Geleijnse, J.M.
\newblock Influence of weight reduction on blood pressure: a meta-analysis of
  randomized controlled trials.
\newblock {\em Hypertension} {\bf 2003}, {\em 42},~878--884.
\newblock {\url{https://doi.org/10.1161/01.HYP.0000094221.86888.AE}}.

\bibitem[Neter et~al.(2003{\natexlab{b}})Neter, Stam, Kok, Grobbee, and
  Geleijnse]{Neter2003}
Neter, J.E.; Stam, B.E.; Kok, F.J.; Grobbee, D.E.; Geleijnse, J.M.
\newblock Influence of Weight Reduction on Blood Pressure.
\newblock {\em Hypertension} {\bf 2003}, {\em 42},~878--884.
\newblock {\url{https://doi.org/10.1161/01.hyp.0000094221.86888.ae}}.

\bibitem[Tibana et~al.(2013)Tibana, Pereira, Navalta, Bottaro, and
  Prestes]{Tibana2013-ve}
Tibana, R.A.; Pereira, G.B.; Navalta, J.W.; Bottaro, M.; Prestes, J.
\newblock Acute effects of resistance exercise on 24-h blood pressure in middle
  aged overweight and obese women.
\newblock {\em Int. J. Sports Med.} {\bf 2013}, {\em 34},~460--464.
\newblock {\url{https://doi.org/10.1055/s-0032-1323819}}.

\bibitem[Karatzi et~al.(2013)Karatzi, Rontoyanni, Protogerou, Georgoulia,
  Xenos, Chrysou, Sfikakis, and Sidossis]{Karatzi2013-yr}
Karatzi, K.; Rontoyanni, V.G.; Protogerou, A.D.; Georgoulia, A.; Xenos, K.;
  Chrysou, J.; Sfikakis, P.P.; Sidossis, L.S.
\newblock Acute effects of beer on endothelial function and hemodynamics: a
  single-blind, crossover study in healthy volunteers.
\newblock {\em Nutrition} {\bf 2013}, {\em 29},~1122--1126.
\newblock {\url{https://doi.org/10.1016/j.nut.2013.02.016}}.

\bibitem[Fantin et~al.(2016)Fantin, Bulpitt, Zamboni, Cheek, and
  Rajkumar]{Fantin2016-bt}
Fantin, F.; Bulpitt, C.J.; Zamboni, M.; Cheek, E.; Rajkumar, C.
\newblock Arterial compliance may be reduced by ingestion of red wine.
\newblock {\em J. Hum. Hypertens.} {\bf 2016}, {\em 30},~68--72.
\newblock {\url{https://doi.org/10.1038/jhh.2015}}.

\bibitem[Kayrak et~al.(2010)Kayrak, Ulgen, Yazici, Yilmaz, Demir, Dogan, Ozhan,
  Alihanoglu, Koc, and Bodur]{Kayrak2010-lk}
Kayrak, M.; Ulgen, M.S.; Yazici, M.; Yilmaz, R.; Demir, K.; Dogan, Y.; Ozhan,
  H.; Alihanoglu, Y.; Koc, F.; Bodur, S.
\newblock {A comparison of blood pressure and pulse pressure values obtained by
  oscillometric and central measurements in hypertensive patients}.
\newblock {\em Blood Press.} {\bf 2010}, {\em 19},~98--103.
\newblock {\url{https://doi.org/10.3109/08037050903516318}}.

\bibitem[Cunha et~al.(2017)Cunha, Vila{\c{c}}a-Alves, Noleto, Silva, Costa,
  Silva, P{\'o}voa, and Lehnen]{cunha2017acute}
Cunha, R.M.; Vila{\c{c}}a-Alves, J.; Noleto, M.V.; Silva, J.S.; Costa, A.M.;
  Silva, C.N.F.; P{\'o}voa, T.I.R.; Lehnen, A.M.
\newblock Acute blood pressure response in hypertensive elderly women
  immediately after water aerobics exercise: A crossover study.
\newblock {\em Clinical and Experimental Hypertension} {\bf 2017}, {\em
  39},~17--22.
\newblock {\url{https://doi.org/10.1080/10641963.2016.1226891}}.

\bibitem[Netea et~al.(2003)Netea, Lenders, Smits, and Thien]{Netea2003}
Netea, R.T.; Lenders, J.W.M.; Smits, P.; Thien, T.
\newblock Both body and arm position significantly influence blood pressure
  measurement.
\newblock {\em Journal of Human Hypertension} {\bf 2003}, {\em 17},~459--462.
\newblock {\url{https://doi.org/10.1038/sj.jhh.1001573}}.

\bibitem[Talukder et~al.(2016)Talukder, Rutherford, Phung, Islam, and
  Chu]{talukder2016effect}
Talukder, M.R.R.; Rutherford, S.; Phung, D.; Islam, M.Z.; Chu, C.
\newblock {The effect of drinking water salinity on blood pressure in young
  adults of coastal Bangladesh}.
\newblock {\em Environmental pollution} {\bf 2016}, {\em 214},~248--254.
\newblock {\url{https://doi.org/10.1016/j.envpol.2016.03.074}}.

\bibitem[Xu et~al.(2019)Xu, Zhang, Wang, Yang, Ban, Liu, and Li]{Xu2019-rx}
Xu, D.; Zhang, Y.; Wang, B.; Yang, H.; Ban, J.; Liu, F.; Li, T.
\newblock {Acute effects of temperature exposure on blood pressure: An hourly
  level panel study}.
\newblock {\em Environ. Int.} {\bf 2019}, {\em 124},~493--500.
\newblock {\url{https://doi.org/10.1016/j.envint.2019.01.045}}.

\bibitem[Azar et~al.(2016)Azar, Frangieh, Mrou{\'e}, Bassila, Kasty, Hage, and
  Kadri]{Azar2016-eg}
Azar, R.R.; Frangieh, A.H.; Mrou{\'e}, J.; Bassila, L.; Kasty, M.; Hage, G.;
  Kadri, Z.
\newblock Acute effects of waterpipe smoking on blood pressure and heart rate:
  a real-life trial.
\newblock {\em Inhal. Toxicol.} {\bf 2016}, {\em 28},~339--342.
\newblock {\url{https://doi.org/10.3109/08958378.2016.1171934}}.

\bibitem[Krzesi{\'n}ski et~al.(2016)Krzesi{\'n}ski, Sta{\'n}czyk, Gielerak,
  Piotrowicz, Banak, and W{\'o}jcik]{krzesinski2016diagnostic}
Krzesi{\'n}ski, P.; Sta{\'n}czyk, A.; Gielerak, G.; Piotrowicz, K.; Banak, M.;
  W{\'o}jcik, A.
\newblock The diagnostic value of supine blood pressure in hypertension.
\newblock {\em Archives of Medical Science} {\bf 2016}, {\em 12},~310--318.
\newblock {\url{https://doi.org/10.5114/aoms.2016.59256}}.

\bibitem[Li et~al.(2019)Li, Thijs, Zhang, Asayama, Hansen, Boggia,
  Bj{\"o}rklund-Bodeg{\aa}rd, Yang, Niiranen, Ntineri, Wei, Kikuya, Ohkubo,
  Dolan, Hozawa, Tsuji, Stolarz-Skrzypek, Huang, Melgarejo, Tikhonoff,
  Malyutina, Casiglia, Nikitin, Lind, Sandoya, Aparicio, Barochiner,
  Gilis-Malinowska, Narkiewicz, Kawecka-Jaszcz, Maestre, Jula, Johansson,
  Kuznetsova, Filipovsk{\'y}, Stergiou, Wang, Imai, O'Brien, Staessen, and
  {International Database on Ambulatory and Home Blood Pressure in Relation to
  Cardiovascular Outcome Investigators}]{Li2019-ax}
Li, Y.; Thijs, L.; Zhang, Z.Y.; Asayama, K.; Hansen, T.W.; Boggia, J.;
  Bj{\"o}rklund-Bodeg{\aa}rd, K.; Yang, W.Y.; Niiranen, T.J.; Ntineri, A.;
  et~al.
\newblock Opposing age-related trends in absolute and relative risk of adverse
  health outcomes associated with out-of-office blood pressure.
\newblock {\em Hypertension} {\bf 2019}, {\em 74},~1333--1342.
\newblock {\url{https://doi.org/10.1161/HYPERTENSIONAHA.119.12958}}.

\bibitem[Walker et~al.(2019)Walker, Sharrett, Wu, Schneider, Albert, Lutsey,
  Bandeen-Roche, Coresh, Gross, Windham, Knopman, Power, Rawlings, Mosley, and
  Gottesman]{Walker2019-oe}
Walker, K.A.; Sharrett, A.R.; Wu, A.; Schneider, A.L.C.; Albert, M.; Lutsey,
  P.L.; Bandeen-Roche, K.; Coresh, J.; Gross, A.L.; Windham, B.G.;  et~al.
\newblock Association of midlife to late-life blood pressure patterns with
  incident dementia.
\newblock {\em JAMA} {\bf 2019}, {\em 322},~535--545.
\newblock {\url{https://doi.org/10.1001/jama.2019.10575}}.

\bibitem[Widlansky et~al.(2007)Widlansky, Vita, Keyes, Larson, Hamburg, Levy,
  Mitchell, Osypiuk, Vasan, and Benjamin]{Widlansky2007-ih}
Widlansky, M.E.; Vita, J.A.; Keyes, M.J.; Larson, M.G.; Hamburg, N.M.; Levy,
  D.; Mitchell, G.F.; Osypiuk, E.W.; Vasan, R.S.; Benjamin, E.J.
\newblock {Relation of season and temperature to endothelium-dependent
  flow-mediated vasodilation in subjects without clinical evidence of
  cardiovascular disease (from the Framingham Heart Study)}.
\newblock {\em Am. J. Cardiol.} {\bf 2007}, {\em 100},~518--523.
\newblock {\url{https://doi.org/10.1016/j.amjcard.2007.03.055}}.

\bibitem[Barba et~al.(2006)Barba, Troiano, Russo, Strazzullo, and
  Siani]{barba2006body}
Barba, G.; Troiano, E.; Russo, P.; Strazzullo, P.; Siani, A.
\newblock {Body mass, fat distribution and blood pressure in Southern Italian
  children: results of the ARCA project}.
\newblock {\em Nutrition, metabolism and cardiovascular diseases} {\bf 2006},
  {\em 16},~239--248.
\newblock {\url{https://doi.org/10.1016/j.numecd.2006.02.005}}.

\bibitem[Sano et~al.(2020)Sano, Hara, Asayama, Miyazaki, Kikuya, Imai, and
  Ohkubo]{Sano2020-zk}
Sano, H.; Hara, A.; Asayama, K.; Miyazaki, S.; Kikuya, M.; Imai, Y.; Ohkubo, T.
\newblock {Antihypertensive drug effects according to the pretreatment
  self-measured home blood pressure: the {HOMED-BP} study}.
\newblock {\em BMJ Open} {\bf 2020}, {\em 10},~e040524.
\newblock {\url{https://doi.org/10.1136/bmjopen-2020-040524.}}

\bibitem[Zheng et~al.(2021)Zheng, Wang, Cheng, Kang, Nie, Mi, Li, Jin, Zhang,
  and Bai]{Zheng2021-tq}
Zheng, S.; Wang, M.Z.; Cheng, Z.Y.; Kang, F.; Nie, Y.H.; Mi, X.Y.; Li, H.Y.;
  Jin, L.; Zhang, Y.W.; Bai, Y.N.
\newblock {Effects of outdoor temperature on blood pressure in a prospective
  cohort of northwest China}.
\newblock {\em Biomed. Environ. Sci.} {\bf 2021}, {\em 34},~89--100.
\newblock {\url{https://doi.org/10.3967/bes2021.014}}.

\bibitem[Kang et~al.(2020)Kang, Han, Guan, Wang, Xue, Chen, Jiang, Zhang,
  Zheng, Wang, Gao, and {China Hypertension Survey investigators}]{Kang2020-mq}
Kang, Y.; Han, Y.; Guan, T.; Wang, X.; Xue, T.; Chen, Z.; Jiang, L.; Zhang, L.;
  Zheng, C.; Wang, Z.;  et~al.
\newblock Clinical blood pressure responses to daily ambient temperature
  exposure in China: An analysis based on a representative nationwide
  population.
\newblock {\em Sci. Total Environ.} {\bf 2020}, {\em 705},~135762.
\newblock {\url{https://doi.org/10.1016/j.scitotenv.2019.135762}}.

\bibitem[Lewington et~al.(2012)Lewington, Li, Sherliker, Guo, Millwood, Bian,
  Whitlock, Yang, Collins, Chen, Wu, Wang, Hu, Jiang, Yang, Lacey, Peto, and
  Chen]{Lewington2012-iv}
Lewington, S.; Li, L.; Sherliker, P.; Guo, Y.; Millwood, I.; Bian, Z.;
  Whitlock, G.; Yang, L.; Collins, R.; Chen, J.;  et~al.
\newblock {Seasonal variation in blood pressure and its relationship with
  outdoor temperature in 10 diverse regions of China}.
\newblock {\em J. Hypertens.} {\bf 2012}, {\em 30},~1383--1391.
\newblock {\url{https://doi.org/10.1097/HJH.0b013e32835465b5}}.

\bibitem[Ohkubo et~al.(2002)Ohkubo, Hozawa, Yamaguchi, Kikuya, Ohmori,
  Michimata, Matsubara, Hashimoto, Hoshi, Araki, Tsuji, Satoh, Hisamichi, and
  Imai]{Ohkubo2002}
Ohkubo, T.; Hozawa, A.; Yamaguchi, J.; Kikuya, M.; Ohmori, K.; Michimata, M.;
  Matsubara, M.; Hashimoto, J.; Hoshi, H.; Araki, T.;  et~al.
\newblock Prognostic significance of the nocturnal decline in blood pressure in
  individuals with and without high 24-h blood pressure.
\newblock {\em Journal of Hypertension} {\bf 2002}, {\em 20},~2183--2189.
\newblock {\url{https://doi.org/10.1097/00004872-200211000-00017}}.

\bibitem[Routledge and Mc~Fetridge-Durdle(2007)]{routledge2007nondipping}
Routledge, F.; Mc~Fetridge-Durdle, J.
\newblock Nondipping blood pressure patterns among individuals with essential
  hypertension: a review of the literature.
\newblock {\em European Journal of Cardiovascular Nursing} {\bf 2007}, {\em
  6},~9--26.
\newblock {\url{https://doi.org/10.1016/j.ejcnurse.2006.05.001}}.

\bibitem[Pickering and Kario(2001)]{Pickering2001}
Pickering, T.G.; Kario, K.
\newblock Nocturnal non-dipping: what does it augur?
\newblock {\em {Current Opinion in Nephrology and Hypertension}} {\bf 2001},
  {\em 10},~611--616.
\newblock {\url{https://doi.org/10.1097/00041552-200109000-00010}}.

\bibitem[Ragot et~al.(2000)Ragot, Genes, Vaur, and Herpin]{Ragot2000-ym}
Ragot, S.; Genes, N.; Vaur, L.; Herpin, D.
\newblock {Comparison of three blood pressure measurement methods for the
  evaluation of two antihypertensive drugs: feasibility, agreement, and
  reproducibility of blood pressure response}.
\newblock {\em Am. J. Hypertens.} {\bf 2000}, {\em 13},~632--639.
\newblock {\url{https://doi.org/10.1016/S0895-7061(99)00258-7}}.

\bibitem[Asayama et~al.(2022)Asayama, Ohkubo, Rakugi, Miyakawa, Mori, Katsuya,
  Ikehara, Ueda, Ohya, Tsuchihashi, Kario, Miura, Ito, Umemura, and {Japanese
  Society of Hypertension Working Group on the COmparison of Self-measured
  home, Automated unattended office, Conventional attended office blood
  pressure (COSAC) study}]{Asayama2022-ft}
Asayama, K.; Ohkubo, T.; Rakugi, H.; Miyakawa, M.; Mori, H.; Katsuya, T.;
  Ikehara, Y.; Ueda, S.; Ohya, Y.; Tsuchihashi, T.;  et~al.
\newblock Direct comparison of the reproducibility of in-office and
  self-measured home blood pressures.
\newblock {\em J. Hypertens.} {\bf 2022}, {\em 40},~398--407.
\newblock {\url{https://doi.org/10.1097/HJH.0000000000003026}}.

\bibitem[Katebi and Clifford(2022)]{Katebi2022}
Katebi, N.; Clifford, G.D.
\newblock Deep sequence learning for assessing hypertension in pregnancy from
  Doppler signals.

\bibitem[Katebi et~al.(2023)Katebi, Bremer, Nguyen, Phan, Jeff, Armstrong,
  Phabian-Millbrook, Platner, Carroll, Shoai, Rohloff, Boulet, Franklin, and
  Clifford]{Katebi2023}
Katebi, N.; Bremer, W.; Nguyen, T.; Phan, D.; Jeff, J.; Armstrong, K.;
  Phabian-Millbrook, P.; Platner, M.; Carroll, K.; Shoai, B.;  et~al.
\newblock Automated image transcription for perinatal blood pressure monitoring
  using mobile health technology.

\bibitem[Forman(2009)]{Forman2009}
Forman, J.P.
\newblock {Diet and Lifestyle Risk Factors Associated With Incident
  Hypertension in Women}.
\newblock {\em {JAMA}} {\bf 2009}, {\em 302},~401.
\newblock {\url{https://doi.org/10.1001/jama.2009.1060}}.

\bibitem[Buckman et~al.(2015)Buckman, Eddie, Vaschillo, Vaschillo, Garcia, and
  Bates]{Buckman2015-ew}
Buckman, J.F.; Eddie, D.; Vaschillo, E.G.; Vaschillo, B.; Garcia, A.; Bates,
  M.E.
\newblock Immediate and complex cardiovascular adaptation to an acute alcohol
  dose.
\newblock {\em Alcohol. Clin. Exp. Res.} {\bf 2015}, {\em 39},~2334--2344.
\newblock {\url{https://doi.org/10.1111/acer.12912}}.

\bibitem[Appel et~al.(1997)Appel, Moore, Obarzanek, Vollmer, Svetkey, Sacks,
  Bray, Vogt, Cutler, Windhauser, Lin, Karanja, Simons-Morton, McCullough,
  Swain, Steele, Evans, Miller, and Harsha]{Appel1997}
Appel, L.J.; Moore, T.J.; Obarzanek, E.; Vollmer, W.M.; Svetkey, L.P.; Sacks,
  F.M.; Bray, G.A.; Vogt, T.M.; Cutler, J.A.; Windhauser, M.M.;  et~al.
\newblock {A Clinical Trial of the Effects of Dietary Patterns on Blood
  Pressure}.
\newblock {\em New England Journal of Medicine} {\bf 1997}, {\em
  336},~1117--1124.
\newblock {\url{https://doi.org/10.1056/nejm199704173361601}}.

\bibitem[Kawano(2010)]{Kawano2010}
Kawano, Y.
\newblock Diurnal blood pressure variation and related behavioral factors.
\newblock {\em Hypertension Research} {\bf 2010}, {\em 34},~281--285.
\newblock {\url{https://doi.org/10.1038/hr.2010.241}}.

\bibitem[Jansen et~al.(1987)Jansen, Penterman, Lier, and
  Hoefnagels]{Jansen1987}
Jansen, R.W.; Penterman, B.J.; Lier, H.J.V.; Hoefnagels, W.H.
\newblock Blood pressure reduction after oral glucose loading and its relation
  to age, blood pressure and insulin.
\newblock {\em The American Journal of Cardiology} {\bf 1987}, {\em
  60},~1087--1091.
\newblock {\url{https://doi.org/10.1016/0002-9149(87)90358-4}}.

\bibitem[Sidery et~al.(1993)Sidery, Cowley, and MacDonald]{Sidery1993}
Sidery, M.B.; Cowley, A.J.; MacDonald, I.A.
\newblock Cardiovascular responses to a high-fat and a high-carbohydrate meal
  in healthy elderly subjects.
\newblock {\em {Clinical Science}} {\bf 1993}, {\em 84},~263--270.
\newblock {\url{https://doi.org/10.1042/cs0840263}}.

\bibitem[Burnier(2019)]{burnier2019should}
Burnier, M.
\newblock Should we eat more potassium to better control blood pressure in
  hypertension?
\newblock {\em Nephrology Dialysis Transplantation} {\bf 2019}, {\em
  34},~184--193.
\newblock {\url{https://doi.org/10.1093/ndt/gfx340}}.

\bibitem[Nishiwaki et~al.(2017)Nishiwaki, Kora, and
  Matsumoto]{Nishiwaki2017-ud}
Nishiwaki, M.; Kora, N.; Matsumoto, N.
\newblock Ingesting a small amount of beer reduces arterial stiffness in
  healthy humans.
\newblock {\em Physiol. Rep.} {\bf 2017}, {\em 5}.
\newblock {\url{https://doi.org/10.14814/phy2.13381}}.

\bibitem[McMullen et~al.(2011)McMullen, Whitehouse, Shine, and
  Towell]{McMullen2011-qu}
McMullen, M.K.; Whitehouse, J.M.; Shine, G.; Towell, A.
\newblock Habitual coffee and tea drinkers experienced increases in blood
  pressure after consuming low to moderate doses of caffeine; these increases
  were larger upright than in the supine posture.
\newblock {\em Food Funct.} {\bf 2011}, {\em 2},~197--203.
\newblock {\url{https://doi.org/10.1039/C0FO00166J}}.

\bibitem[Carter et~al.(2011)Carter, Stream, Durocher, and
  Larson]{Carter2011-aw}
Carter, J.R.; Stream, S.F.; Durocher, J.J.; Larson, R.A.
\newblock Influence of acute alcohol ingestion on sympathetic neural responses
  to orthostatic stress in humans.
\newblock {\em Am. J. Physiol. Endocrinol. Metab.} {\bf 2011}, {\em
  300},~E771--8.
\newblock {\url{https://doi.org/10.1152/ajpendo.00674.2010}}.

\bibitem[Nowak et~al.(2019)Nowak, Go{\'s}li{\'n}ski, Weso{\l}owska, Berenda,
  and Pop{\l}awski]{Nowak2019-si}
Nowak, D.; Go{\'s}li{\'n}ski, M.; Weso{\l}owska, A.; Berenda, K.; Pop{\l}awski,
  C.
\newblock {Effects of acute consumption of Noni and chokeberry juices vs.
  Energy drinks on blood pressure, heart rate, and blood glucose in young
  adults}.
\newblock {\em Evid. Based. Complement. Alternat. Med.} {\bf 2019}, {\em
  2019},~6076751.
\newblock {\url{https://doi.org/10.1155/2019/6076751}}.

\bibitem[Luqman and Khan(2019)]{luqmanexperimental}
Luqman, A.; Khan, M.
\newblock An Experimental Study of Short-term Physiological Effects of a Single
  Dose of Energy Drink in Healthy Male Medical Students.
\newblock {\em Pakistan Journal of Medical and Health Sciences} {\bf 2019},
  {\em 13},~685–9.

\bibitem[Morris et~al.(2012)Morris, Yang, and Scheer]{Morris2012}
Morris, C.J.; Yang, J.N.; Scheer, F.A.
\newblock The impact of the circadian timing system on cardiovascular and
  metabolic function. In {\em {Progress in Brain Research}}; Elsevier,  2012;
  pp. 337--358.
\newblock {\url{https://doi.org/10.1016/b978-0-444-59427-3.00019-8}}.

\bibitem[Richards and Gumz(2012)]{Richards2012}
Richards, J.; Gumz, M.L.
\newblock Advances in understanding the peripheral circadian clocks.
\newblock {\em {The {FASEB} Journal}} {\bf 2012}, {\em 26},~3602--3613.
\newblock {\url{https://doi.org/10.1096/fj.12-203554}}.

\bibitem[Hower et~al.(2018)Hower, Harper, and Buford]{hower2018circadian}
Hower, I.M.; Harper, S.A.; Buford, T.W.
\newblock Circadian rhythms, exercise, and cardiovascular health.
\newblock {\em Journal of circadian rhythms} {\bf 2018}, {\em 16}.
\newblock {\url{https://doi.org/10.5334/jcr.164}}.

\bibitem[Boivin et~al.(2014)Boivin, Boutte, Fay, Rossignol, and
  Zannad]{Boivin2014-ss}
Boivin, J.M.; Boutte, E.; Fay, R.; Rossignol, P.; Zannad, F.
\newblock {Home blood pressure monitoring: a few minutes of rest before
  measurement may not be appropriate}.
\newblock {\em Am. J. Hypertens.} {\bf 2014}, {\em 27},~932--938.
\newblock {\url{https://doi.org/10.1093/ajh/hpu001}}.

\bibitem[Park et~al.(2019)Park, Kario, Chia, Turana, Chen, Buranakitjaroen,
  Nailes, Hoshide, Siddique, Sison, Soenarta, Sogunuru, Tay, Teo, Zhang, Shin,
  Minh, Tomitani, Kabutoya, Sukonthasarn, Verma, Wang, and and]{Park2019}
Park, S.; Kario, K.; Chia, Y.C.; Turana, Y.; Chen, C.H.; Buranakitjaroen, P.;
  Nailes, J.; Hoshide, S.; Siddique, S.; Sison, J.;  et~al.
\newblock {The influence of the ambient temperature on blood pressure and how
  it will affect the epidemiology of hypertension in Asia}.
\newblock {\em The Journal of Clinical Hypertension} {\bf 2019}, {\em
  22},~438--444.
\newblock {\url{https://doi.org/10.1111/jch.13762}}.

\bibitem[Sega et~al.(1998)Sega, Cesana, Bombelli, Grassi, Stella, Zanchetti,
  and Mancia]{Sega1998}
Sega, R.; Cesana, G.; Bombelli, M.; Grassi, G.; Stella, M.L.; Zanchetti, A.;
  Mancia, G.
\newblock Seasonal variations in home and ambulatory blood pressure in the
  {PAMELA} population.
\newblock {\em Journal of Hypertension} {\bf 1998}, {\em 16},~1585--1592.
\newblock {\url{https://doi.org/10.1097/00004872-199816110-00004}}.

\bibitem[Jansen et~al.(2001)Jansen, Leineweber, and Thien]{Jansen2001-xl}
Jansen, P.M.; Leineweber, M.J.; Thien, T.
\newblock The effect of a change in ambient temperature on blood pressure in
  normotensives.
\newblock {\em J. Hum. Hypertens.} {\bf 2001}, {\em 15},~113--117.
\newblock {\url{https://doi.org/10.1038/sj.jhh.1001134}}.

\bibitem[Yang et~al.(2015)Yang, Li, Lewington, Guo, Sherliker, Bian, Collins,
  Peto, Liu, Yang, Zhang, Li, Liu, Chen, and {China Kadoorie Biobank Study
  Collaboration}]{Yang2015-rc}
Yang, L.; Li, L.; Lewington, S.; Guo, Y.; Sherliker, P.; Bian, Z.; Collins, R.;
  Peto, R.; Liu, Y.; Yang, R.;  et~al.
\newblock {Outdoor temperature, blood pressure, and cardiovascular disease
  mortality among 23000 individuals with diagnosed cardiovascular diseases from
  China}.
\newblock {\em Eur. Heart J.} {\bf 2015}, {\em 36},~1178--1185.
\newblock {\url{https://doi.org/10.1093/eurheartj/ehv023}}.

\bibitem[Wu et~al.(2015)Wu, Deng, Huang, Wang, Qin, Zheng, Wei, Shima, and
  Guo]{Wu2015-dj}
Wu, S.; Deng, F.; Huang, J.; Wang, X.; Qin, Y.; Zheng, C.; Wei, H.; Shima, M.;
  Guo, X.
\newblock {Does ambient temperature interact with air pollution to alter blood
  pressure? A repeated-measure study in healthy adults}.
\newblock {\em J. Hypertens.} {\bf 2015}, {\em 33},~2414--2421.
\newblock {\url{https://doi.org/10.1097/HJH.0000000000000738}}.

\bibitem[Kim et~al.(2012)Kim, Kim, Cheong, Ahn, and Choi]{Kim2012-oi}
Kim, Y.M.; Kim, S.; Cheong, H.K.; Ahn, B.; Choi, K.
\newblock Effects of heat wave on body temperature and blood pressure in the
  poor and elderly.
\newblock {\em Environ. Health Toxicol.} {\bf 2012}, {\em 27},~e2012013.
\newblock {\url{https://doi.org/10.5620/eht.2012.27.e2012013}}.

\bibitem[Bilo et~al.(2017)Bilo, Sala, Perego, Faini, Gao, G{\l}uszewska, Ochoa,
  Pellegrini, Lonati, and Parati]{Bilo2017}
Bilo, G.; Sala, O.; Perego, C.; Faini, A.; Gao, L.; G{\l}uszewska, A.; Ochoa,
  J.E.; Pellegrini, D.; Lonati, L.M.; Parati, G.
\newblock Impact of cuff positioning on blood pressure measurement accuracy:
  may a specially designed cuff make a difference?
\newblock {\em Hypertension Research} {\bf 2017}, {\em 40},~573--580.
\newblock {\url{https://doi.org/10.1038/hr.2016.184}}.

\bibitem[Hinghofer-Szalkay(2011)]{Hinghofer-Szalkay2011-hu}
Hinghofer-Szalkay, H.
\newblock Gravity, the hydrostatic indifference concept and the cardiovascular
  system.
\newblock {\em Eur. J. Appl. Physiol.} {\bf 2011}, {\em 111},~163--174.
\newblock {\url{https://doi.org/10.1007/s00421-010-1646-9}}.

\bibitem[Al-Qatatsheh et~al.(2020)Al-Qatatsheh, Morsi, Zavabeti, Zolfagharian,
  Salim, Z~Kouzani, Mosadegh, and Gharaie]{Al-Qatatsheh2020-fh}
Al-Qatatsheh, A.; Morsi, Y.; Zavabeti, A.; Zolfagharian, A.; Salim, N.;
  Z~Kouzani, A.; Mosadegh, B.; Gharaie, S.
\newblock Blood pressure sensors: Materials, fabrication methods, performance
  evaluations and future perspectives.
\newblock {\em Sensors (Basel)} {\bf 2020}, {\em 20},~4484.
\newblock {\url{https://doi.org/10.3390/s20164484}}.

\bibitem[Sareen et~al.(2012)Sareen, Saxena, Sareen, and
  Taneja]{sareen2012comparison}
Sareen, P.; Saxena, K.; Sareen, B.; Taneja, B.
\newblock Comparison of arm and calf blood pressure.
\newblock {\em Indian Journal of Anaesthesia} {\bf 2012}, {\em 56},~83.
\newblock {\url{https://doi.org/10.4103/0019-5049.93354}}.

\bibitem[E{\c{s}}er et~al.(2007)E{\c{s}}er, Khorshid, Yapucu~G{\"u}ne{\c{s}},
  and Demir]{ecser2007effect}
E{\c{s}}er, {\.I}.; Khorshid, L.; Yapucu~G{\"u}ne{\c{s}}, {\"U}.; Demir, Y.
\newblock The effect of different body positions on blood pressure.
\newblock {\em Journal of Clinical Nursing} {\bf 2007}, {\em 16},~137--140.
\newblock {\url{https://doi.org/10.1111/j.1365-2702.2005.01494.x}}.

\bibitem[Chachula et~al.(2020)Chachula, Lieb, Hess, Welter, Graf, and
  Dullenkopf]{Chachula2020-nl}
Chachula, K.; Lieb, F.; Hess, F.; Welter, J.; Graf, N.; Dullenkopf, A.
\newblock {Non-invasive continuous blood pressure monitoring ({ClearSight™}
  system) during shoulder surgery in the beach chair position: a prospective
  self-controlled study}.
\newblock {\em BMC Anesthesiol.} {\bf 2020}, {\em 20},~271.
\newblock {\url{https://doi.org/10.1186/s12871-020-01185-6}}.

\bibitem[Netea et~al.(1998)Netea, Smits, Lenders, and Thien]{netea1998does}
Netea, R.T.; Smits, P.; Lenders, J.W.; Thien, T.
\newblock Does it matter whether blood pressure measurements are taken with
  subjects sitting or supine?
\newblock {\em Journal of hypertension} {\bf 1998}, {\em 16},~263--268.
\newblock
  {\url{https://doi.org/https://journals.lww.com/jhypertension/Abstract/1998/16030/Does_it_matter_whether_blood_pressure_measurements.2.aspx}}.

\bibitem[Cicolini et~al.(2011)Cicolini, Pizzi, Palma, Bucci, Schioppa,
  Mezzetti, and Manzoli]{cicolini2011differences}
Cicolini, G.; Pizzi, C.; Palma, E.; Bucci, M.; Schioppa, F.; Mezzetti, A.;
  Manzoli, L.
\newblock Differences in blood pressure by body position (supine, Fowler's, and
  sitting) in hypertensive subjects.
\newblock {\em American journal of hypertension} {\bf 2011}, {\em
  24},~1073--1079.
\newblock {\url{https://doi.org/10.1038/ajh.2011.106}}.

\bibitem[Adiyaman et~al.(2006)Adiyaman, Verhoeff, Lenders, Deinum, and
  Thien]{Adiyaman2006}
Adiyaman, A.; Verhoeff, R.; Lenders, J.W.; Deinum, J.; Thien, T.
\newblock The position of the arm during blood pressure measurement in sitting
  position.
\newblock {\em Blood Pressure Monitoring} {\bf 2006}, {\em 11},~309--313.
\newblock {\url{https://doi.org/10.1097/01.mbp.0000218007.57957.56}}.

\bibitem[Mariotti et~al.(1987)Mariotti, Alli, Avanzini, Canciani, Tullio,
  Manzini, Salmoirago, Taioli, Zussino, and Radice]{Mariotti1987}
Mariotti, G.; Alli, C.; Avanzini, F.; Canciani, C.; Tullio, M.D.; Manzini, M.;
  Salmoirago, E.; Taioli, E.; Zussino, A.; Radice, M.
\newblock Arm position as a source of error in blood pressure measurement.
\newblock {\em Clinical Cardiology} {\bf 1987}, {\em 10},~591--593.
\newblock {\url{https://doi.org/10.1002/clc.4960101016}}.

\bibitem[Netea et~al.(1999)Netea, Lenders, Smits, and Thien]{netea1999arm}
Netea, R.; Lenders, J.; Smits, P.; Thien, T.
\newblock Arm position is important for blood pressure measurement.
\newblock {\em Journal of human hypertension} {\bf 1999}, {\em 13},~105--109.
\newblock {\url{https://doi.org/https://doi.org/10.1038/sj.jhh.1000720}}.

\bibitem[Foster-Fitzpatrick et~al.(1999)Foster-Fitzpatrick, Ortiz, Sibilano,
  Marcantonio, and Braun]{FosterFitzpatrick1999}
Foster-Fitzpatrick, L.; Ortiz, A.; Sibilano, H.; Marcantonio, R.; Braun, L.T.
\newblock {The Effects of Crossed Leg on Blood Pressure Measurement}.
\newblock {\em Nursing Research} {\bf 1999}, {\em 48},~105--108.
\newblock {\url{https://doi.org/10.1097/00006199-199903000-00009}}.

\bibitem[Adiyaman et~al.(2007)Adiyaman, Tosun, Elving, Deinum, Lenders, and
  Thien]{adiyaman2007effect}
Adiyaman, A.; Tosun, N.; Elving, L.D.; Deinum, J.; Lenders, J.W.; Thien, T.
\newblock The effect of crossing legs on blood pressure.
\newblock {\em Blood pressure monitoring} {\bf 2007}, {\em 12},~189--193.

\bibitem[Foster-Fitzpatrick et~al.(1999)Foster-Fitzpatrick, Ortiz, Sibilano,
  Marcantonio, and Braun]{Foster-Fitzpatrick1999-us}
Foster-Fitzpatrick, L.; Ortiz, A.; Sibilano, H.; Marcantonio, R.; Braun, L.T.
\newblock {The effects of crossed leg on blood pressure measurement}.
\newblock {\em Nurs. Res.} {\bf 1999}, {\em 48},~105--108.
\newblock {\url{https://doi.org/10.1097/00006199-199903000-00009}}.

\bibitem[Pinar et~al.(2004)Pinar, Sabuncu, and Oksay]{Pinar2004-nq}
Pinar, R.; Sabuncu, N.; Oksay, A.
\newblock Effects of crossed leg on blood pressure.
\newblock {\em Blood Press.} {\bf 2004}, {\em 13},~252--254.
\newblock {\url{https://doi.org/10.1080/08037050410000903}}.

\bibitem[Fred(2013)]{fred2013accurate}
Fred, H.L.
\newblock Accurate blood pressure measurements and the other arm: the doctor is
  ultimately responsible.
\newblock {\em Texas Heart Institute Journal} {\bf 2013}, {\em 40},~217.

\bibitem[Lane et~al.(2002)Lane, Beevers, Barnes, Bourne, John, Malins, and
  Beevers]{lane2002inter}
Lane, D.; Beevers, M.; Barnes, N.; Bourne, J.; John, A.; Malins, S.; Beevers,
  D.G.
\newblock {Inter-arm differences in blood pressure: when are they clinically
  significant}.
\newblock {\em Journal of hypertension} {\bf 2002}, {\em 20},~1089--1095.
\newblock
  {\url{https://doi.org/https://journals.lww.com/jhypertension/Abstract/2002/06000/Inter_arm_differences_in_blood_pressure__when_are.19.aspx}}.

\bibitem[Bur et~al.(2000)Bur, Hirschl, Herkner, Oschatz, Kofler,
  Woisetschl{\"a}ger, and Laggner]{bur2000accuracy}
Bur, A.; Hirschl, M.M.; Herkner, H.; Oschatz, E.; Kofler, J.;
  Woisetschl{\"a}ger, C.; Laggner, A.N.
\newblock Accuracy of oscillometric blood pressure measurement according to the
  relation between cuff size and upper-arm circumference in critically ill
  patients.
\newblock {\em Critical care medicine} {\bf 2000}, {\em 28},~371--376.

\bibitem[Palatini and Asmar(2018)]{Palatini2018}
Palatini, P.; Asmar, R.
\newblock Cuff challenges in blood pressure measurement.
\newblock {\em The Journal of Clinical Hypertension} {\bf 2018}, {\em
  20},~1100--1103.
\newblock {\url{https://doi.org/10.1111/jch.13301}}.

\bibitem[Bakx et~al.(1997)Bakx, Oerlemans, van~den Hoogen, van Weel, and
  Thien]{Bakx1997-pf}
Bakx, C.; Oerlemans, G.; van~den Hoogen, H.; van Weel, C.; Thien, T.
\newblock The influence of cuff size on blood pressure measurement.
\newblock {\em J. Hum. Hypertens.} {\bf 1997}, {\em 11},~439--445.
\newblock {\url{https://doi.org/10.1038/sj.jhh.1000470}}.

\bibitem[Nikolic et~al.(2013)Nikolic, Abhayaratna, Leano, Stowasser, and
  Sharman]{Nikolic2013}
Nikolic, S.B.; Abhayaratna, W.P.; Leano, R.; Stowasser, M.; Sharman, J.E.
\newblock Waiting a few extra minutes before measuring blood pressure has
  potentially important clinical and research ramifications.
\newblock {\em Journal of Human Hypertension} {\bf 2013}, {\em 28},~56--61.
\newblock {\url{https://doi.org/10.1038/jhh.2013.38}}.

\bibitem[Sala et~al.(2006)Sala, Santin, Rescaldani, and Magrini]{SALA2006}
Sala, C.; Santin, E.; Rescaldani, M.; Magrini, F.
\newblock {How Long Shall the Patient Rest Before Clinic Blood Pressure
  Measurement?}
\newblock {\em American Journal of Hypertension} {\bf 2006}, {\em
  19},~713--717.
\newblock {\url{https://doi.org/10.1016/j.amjhyper.2005.08.021}}.

\bibitem[Mancia(1983)]{MANCIA1983}
Mancia, G.
\newblock Effects of blood-pressure measurement by the doctor on patient's
  blood pressure and heart rate.
\newblock {\em The Lancet} {\bf 1983}, {\em 322},~695--698.
\newblock {\url{https://doi.org/10.1016/s0140-6736(83)92244-4}}.

\bibitem[Pickering and White(2008)]{Pickering2008}
Pickering, T.G.; White, W.B.
\newblock When and how to use self (home) and ambulatory blood pressure
  monitoring.
\newblock {\em Journal of the American Society of Hypertension} {\bf 2008},
  {\em 2},~119--124.
\newblock {\url{https://doi.org/10.1016/j.jash.2008.04.002}}.

\bibitem[Sharma et~al.(2017)Sharma, Barbosa, Ho, Griggs, Ghirmai, Krishnan,
  Hsiai, Chiao, and Cao]{Sharma2017-ng}
Sharma, M.; Barbosa, K.; Ho, V.; Griggs, D.; Ghirmai, T.; Krishnan, S.; Hsiai,
  T.; Chiao, J.C.; Cao, H.
\newblock "Cuff-less and continuous blood pressure monitoring: A methodological
  review".
\newblock {\em Technologies (Basel)} {\bf 2017}, {\em 5},~21.
\newblock {\url{https://doi.org/10.3390/technologies5020021}}.

\bibitem[Geddes(2013)]{geddes2013handbook}
Geddes, L.A.
\newblock {\em Handbook of blood pressure measurement}; Springer Science \&
  Business Media,  2013.

\bibitem[Mukkamala and Hahn(2018)]{Mukkamala2018-fi}
Mukkamala, R.; Hahn, J.O.
\newblock Toward ubiquitous blood pressure monitoring via pulse transit time:
  Predictions on maximum calibration period and acceptable error limits.
\newblock {\em IEEE Trans. Biomed. Eng.} {\bf 2018}, {\em 65},~1410--1420.
\newblock {\url{https://doi.org/10.1109/TBME.2017.2756018}}.

\bibitem[Nam et~al.(2013)Nam, Lee, Hong, and Lee]{Nam2013-ez}
Nam, D.H.; Lee, W.B.; Hong, Y.S.; Lee, S.S.
\newblock Measurement of spatial pulse wave velocity by using a clip-type
  pulsimeter equipped with a Hall sensor and photoplethysmography.
\newblock {\em Sensors} {\bf 2013}, {\em 13},~4714--4723.
\newblock {\url{https://doi.org/10.3390/s130404714}}.

\bibitem[Zhang et~al.(2016)Zhang, Li, Chen, and Deng]{zhang2016mechanism}
Zhang, Y.; Li, Y.; Chen, X.; Deng, N.
\newblock Mechanism of magnetic pulse wave signal for blood pressure
  measurement.
\newblock {\em Journal of Biomedical Science and Engineering} {\bf 2016}, {\em
  9},~29--36.
\newblock {\url{https://doi.org/10.4236/jbise.2016.910B004}}.

\bibitem[Chen et~al.(2013)Chen, Yang, Teo, and Ng]{chen2013noninvasive}
Chen, Z.; Yang, X.; Teo, J.T.; Ng, S.H.
\newblock Noninvasive monitoring of blood pressure using optical
  Ballistocardiography and Photoplethysmograph approaches.
\newblock In Proceedings of the 2013 35th annual international conference of
  the IEEE engineering in medicine and biology society (EMBC). IEEE,  2013, pp.
  2425--2428.
\newblock {\url{https://doi.org/10.1109/EMBC.2013.6610029}}.

\bibitem[Liu et~al.(2017)Liu, Cheng, and Su]{liu2017cuffless}
Liu, S.H.; Cheng, D.C.; Su, C.H.
\newblock A cuffless blood pressure measurement based on the impedance
  plethysmography technique.
\newblock {\em Sensors} {\bf 2017}, {\em 17},~1176.
\newblock {\url{https://doi.org/10.3390/s17051176}}.

\bibitem[Chen et~al.(2019)Chen, Ji, Wu, and Xu]{chen2019non}
Chen, S.; Ji, Z.; Wu, H.; Xu, Y.
\newblock A non-invasive continuous blood pressure estimation approach based on
  machine learning.
\newblock {\em Sensors} {\bf 2019}, {\em 19},~2585.
\newblock {\url{https://doi.org/10.3390/s19112585}}.

\bibitem[Mousavi et~al.(2019)Mousavi, Firouzmand, Charmi, Hemmati, Moghadam,
  and Ghorbani]{mousavi2019blood}
Mousavi, S.S.; Firouzmand, M.; Charmi, M.; Hemmati, M.; Moghadam, M.; Ghorbani,
  Y.
\newblock Blood pressure estimation from appropriate and inappropriate PPG
  signals using A whole-based method.
\newblock {\em Biomedical Signal Processing and Control} {\bf 2019}, {\em
  47},~196--206.
\newblock {\url{https://doi.org/10.1016/j.bspc.2018.08.022}}.

\bibitem[{Division of Small Manufacturers Assistance, Office of Training and
  Assistance}(1990)]{510k}
{Division of Small Manufacturers Assistance, Office of Training and
  Assistance}.
\newblock {\em Premarket notification, 510(k) : regulatory requirements for
  medical devices}; Rockville, MD: U.S. Dept. of Health and Human Services,
  Public Health Service, Food and Drug Administration, Center for Devices and
  Radiological Health; Washington, D.C.,  1990.

\bibitem[Alpert et~al.(2014)Alpert, Dart, and Sica]{alpert2014public}
Alpert, B.S.; Dart, R.A.; Sica, D.A.
\newblock Public-use blood pressure measurement: the kiosk quandary.
\newblock {\em {Journal of the American Society of Hypertension}} {\bf 2014},
  {\em 8},~739--742.
\newblock {\url{https://doi.org/http://dx.doi.org/10.1016/j.jash.2014.07.034}}.

\bibitem[Donawa(2010)]{donawa2010continuing}
Donawa, M.E.
\newblock {Continuing Evolution of the US FDA 510 (k) Process}.
\newblock {\em European medical device technology} {\bf 2010}, {\em 1},~10--13.

\bibitem[Alpert(2017)]{alpert2017can}
Alpert, B.S.
\newblock {Can ‘FDA-cleared’ blood pressure devices be trusted? A call to
  action}.
\newblock {\em Blood Pressure Monitoring} {\bf 2017}, {\em 22},~179--181.
\newblock {\url{https://doi.org/10.1097/MBP.0000000000000279}}.

\bibitem[Kans(2013)]{EntrezDirect}
Kans, J.
\newblock {\em {Entrez Direct: E-utilities on the Unix Command Line}},  2013.
\newblock [Updated 2023 Jun 30].

\bibitem[Schwenck et~al.(2022)Schwenck, Punjabi, and Gaynanova]{Schwenck2022}
Schwenck, J.; Punjabi, N.M.; Gaynanova, I.
\newblock {bp: Blood pressure analysis in R}.
\newblock {\em {PLOS} {ONE}} {\bf 2022}, {\em 17},~e0268934.
\newblock {\url{https://doi.org/10.1371/journal.pone.0268934}}.

\bibitem[Reynolds et~al.(2009)]{reynolds2009gaussian}
Reynolds, D.A.;  et~al.
\newblock Gaussian mixture models.
\newblock {\em Encyclopedia of biometrics} {\bf 2009}, {\em 741}.

\bibitem[Praveen et~al.(2018)Praveen, Peiris, MacMahon, Mogulluru, Raghu,
  Rodgers, Chilappagari, Prabhakaran, Clifford, Maulik, Atkins, Joshi,
  Heritier, Jan, and Patel]{Praveen2018}
Praveen, D.; Peiris, D.; MacMahon, S.; Mogulluru, K.; Raghu, A.; Rodgers, A.;
  Chilappagari, S.; Prabhakaran, D.; Clifford, G.D.; Maulik, P.K.;  et~al.
\newblock {Cardiovascular disease risk and comparison of different strategies
  for blood pressure management in rural India}.
\newblock {\em {BMC} Public Health} {\bf 2018}, {\em 18}.
\newblock {\url{https://doi.org/10.1186/s12889-018-6142-x}}.

\end{thebibliography}
\end{document}